# Asymptotic, second-order homogenization of linear elastic beam networks

Yang Ye
Laboratoire de Mécanique des Solides
CNRS, Institut Polytechnique de Paris
91120 Palaiseau, France

B. Audoly
Laboratoire de Mécanique des Solides
CNRS, Institut Polytechnique de Paris
91120 Palaiseau, France

C. Lestringant
Institut Jean Le Rond d'Alembert,
Sorbonne Université, CNRS,
75005 Paris, France

*April 17, 2024*

We propose a general approach to the higher-order homogenization of discrete elastic networks made up of linear elastic beams or springs in dimension 2 or 3. The network may be *nearly* (rather than *exactly*) periodic: its elastic and geometric properties are allowed to vary slowly in space, in addition to being periodic at the scale of the unit cell. The reference configuration may be prestressed. A homogenized strain energy depending on both the macroscopic strain $\varepsilon$ and its gradient $\nabla \varepsilon$ is obtained by means of a two-scale expansion. The homogenized energy is asymptotically exact two orders beyond that obtained by classical homogenization. The homogenization method is implemented in a symbolic calculation language and applied to various types of networks, such as a 2D honeycomb, a 2D Kagome lattice, a 3D truss and a 1D pantograph. It is validated by comparing the predictions of the microscopic displacement to that obtained by full, discrete simulations. This second-order method remains highly accurate even when the strain gradient effects are significant, such as near the lips of a crack tip or in regions where a gradient of pre-strain is imposed.

**Keywords** Elastic lattices, Asymptotic homogenization, Second-order homogenization, Energy methods.

## 1. INTRODUCTION

Man-made materials have evolved over time, achieving a constant improvement in properties such as strength, stiffness, lightweight or wave transmission. Some of the more recent breakthroughs were made possible by designing the architecture at small scale. Progress in additive manufacturing enables the fabrication of cellular materials whose microstructure can be precisely controlled and tailored: a wide range of new architectured materials can be fabricated, using periodic arrangements of thin beams as fundamental structural elements and ceramics, metals, polymers or composites as base materials [FDA10]. The ongoing revolution consists in integrating these materials into structures having optimized properties.

The design of such systems calls for effective *homogenized* models capable of precisely capturing the mechanical behavior of these complex periodic or quasi-periodic microstructures [AG97; CP12; VP12]. Recent work shows that the accuracy of Cauchy-type continuum obtained by classical homogenization techniques is limited. In various cases, the effective models need to include higher-order terms in order to accurately capture physical phenomena such as localization in shear-bands or domain walls [GLTK19; DYF+20], mechanisms generating long-range modes of deformation [NCH20; DLSS22; ] or symmetries that are not present in the low-order models [RA16; RA19]. Besides, higher-order terms improve the accuracy of the effective models, especially when scale separation is poor, for example when modeling a crack in a relatively coarse microstructure [RKD+15; RDK17; SCO+22].

Various strategies have been proposed to derive higher-order effective models: numerical identification procedures [VDP14; GLTK19; RDVB19a; RDVB19b; RAPG19] or analytical approaches based on formal derivations that slave local displacements to macroscopic fields through a Taylor expansion [TB93; BT94; KM04]. The two-scale expansion method offers a rigorous methodology to derive effective models without introducing such ad-hoc kinematic assumptions. It has been applied to continuous periodic composites [Bou96; SC00; Bou19] and discrete microstructures [CMR06; DO11; DG12; Dav13; LR13], including for deriving higher-order contributions [AS18b; AS18a; ASB19; AB21; DLSS22].

In a recent paper [AL23], we revisited higher-order asymptotic homogenization of linear elastic, discrete microstructures. Building upon our previous work on asymptotic dimension reduction [LA20] and following [LM18], we proposed to tackle asymptotic homogenization at the energy level. Starting from an abstract, generic energy formulation of the discrete microstructure, we introduced a series of steps that can be applied in an automated manner to derive a homogenized energy. All the steps are implemented in a symbolic calculation language and distributed as the open-source library shoal (for Second-order HOmogenization Automated in a Library) [Aud23]. This approach can handle microstructures featuring both pre-stress and spatially varying (graded) properties, two aspects that are rarely addressed in the context of high-order, asymptotic homogenization and are very useful for designing and optimizing microstructures.

The present paper addresses higher-order homogenization of periodic networks made up of slender 1D elements (typically beams or springs), further referred to as *lattice material* or simply *lattice*. The homogenization is done symbolically and in an automated way, and builds up on the general-purpose homogenization engine from our previous work [AL23]. It delivers closed-form analytical expressions of the effective elastic properties without any adjustable internal kinematic variable or parameter. By contrast with earlier work focussing on particular lattice architectures [DO11; Dav13; LR13] or 1D/2D networks of springs [MA02; MA06], our approach is designed to be general and versatile. It works in 1D, 2D or 3D and can handle arbitrary lattice topologies, the presence of prestress, slowly varying geometrical or elastic properties, a variety of member types (such as springs, beams or curved arches), as well as a large elastic contrast.

We illustrate our homogenization method on a comprehensive set of examples: a uniform 2D honeycomb, spatially varying 2D Kagome and honeycomb architectures, a 3D quasi-octet lattice and 1D pantograph. With a view of assessing the accuracy of the resulting models, we propose a numerical verification procedure that compares the microscopic fluctuations extracted from full numerical simulations to those predicted by the homogenization method. We apply this procedure to a range of structural problems including macroscopic samples with cracks, holes, as well as spatially varying pre-stress, elastic and geometric properties. Our numerical results show that the higher-order effective model very accurately captures the fine features of the local displacements and rotations away from the boundaries, where the presence of boundary layers is known to make homogenization break down.

Higher-order homogenization comes with several difficulties, and some remain beyond the scope of the present work. Most notably, we do not address the presence of boundary layers, and accept that the gradient stiffness is often negative (see Remark 5.3 below). These limitations prevent from using the homogenized energy to set up self-contained simulations so far, as discussed in further detail in the conclusion.

## 2. GENERATING THE ELASTIC LATTICE

In this section, we describe the process of generating the elastic lattices. Given a small parameter $\eta \ll 1$, we produce a lattice having cells of size $\mathcal{O}(\eta)$ whose elastic and geometric properties vary over the characteristic length $L = \mathcal{O}(1)$. This sets the stage for homogenization which captures the effective properties of the lattice in the limit $\eta \to 0$ where the scales are well separated.

We introduce a general description of lattices based on the particular example of a *curved honeycomb lattice*. Other lattice geometries are illustrated in the numerical examples in Section 7. Specifically, we consider *networks of elastic beams connected by rigid hinges*. We refer to rigid hinges as *nodes*, and to the elastic beams as *elements*: the degrees of freedom are attached to nodes and the strain energy is stored in elements, as in the finite-element method.

### 2.1. Underlying topological lattice

The lattice topology is defined by an *underlying topological lattice* which is periodic by assumption, see Figure 2.1a. It is a mathematical abstraction: the real elastic lattice may be curved, hence non-periodic, as shown in Figure 2.1b. The following notions are associated with the topological lattice:

- **nodes**　　The nodes are labelled by an index $\beta$. Their position is denoted as $\tilde{X}_\beta \in \mathbb{R}^d$, where $d$ is the dimension of the space in which the lattice is embedded. All quantities relating to the topological lattice are denoted with a tilde.

- **Bravais sub-lattices**　　The nodes can be grouped into Bravais sub-lattices indexed by an integer $b$ with $1 \leq b \leq n_b$, where $n_b$ is the number of sub-lattices. The sub-lattices are defined as the subsets of nodes that are mapped to one another through the fundamental translations of the topological lattice. For instance, the honeycomb lattice in Figure 2.1a has $n_b = 2$ Bravais sub-lattices, shown using the open vs. solid disks. The index of the Bravais sub-lattice to which a particular node $\beta$ belongs is denoted as $b(\beta)$.

- **elements**　　Elements, which are elastic beams in the present illustration, are labelled using an index $\alpha$. An orientation is assigned to every element $\alpha$, and we denote by $(\beta_\alpha^-, \beta_\alpha^+)$ its endpoints, with $\beta_\alpha^-$ as the tail and $\beta_\alpha^+$ as the head, see Figure 2.1b'.

- **element families**　　Elements are grouped into $n_\varphi$ families indexed by $\varphi$, having similar elastic and geometric properties. The honeycomb lattice in Figure 2.1a, for instance, has $n_\varphi = 3$ element families, $\varphi \in \{I, II, III\}$, corresponding to the three possible orientations of the beams. The family to which a particular element $\alpha$ belongs is denoted as $\varphi_\alpha$. The density of elements of a particular family $\varphi$ per unit surface (if the space dimension is $n = 2$) or volume (if $n = 3$) is denoted as $\tilde{\rho}_\varphi$.

The topological lattice being periodic, the vector $\tilde{X}_{\beta_\alpha^+} - \tilde{X}_{\beta_\alpha^-}$ joining the endpoints of a particular element $\alpha$ is the same for all elements $\alpha$ in a particular family $\varphi$ and can be written as

$$\tilde{X}_{\beta_\alpha^+} - \tilde{X}_{\beta_\alpha^-} = \tilde{\boldsymbol{\delta}}_{\varphi_\alpha}, \tag{2.1}$$

where $\tilde{\boldsymbol{\delta}}_\varphi$ maps a family index $\varphi$ to a vector in $\mathbb{R}^d$.

By a similar argument, the Bravais sub-lattice $b(\beta_\alpha^\pm)$ of the endpoints $\pm$ of an element $\alpha$ depends on the element *family* $\varphi_\alpha$ and not on the element $\alpha$ itself: there exists a function $b_\varphi^\pm$ mapping a family $\varphi$ and an endpoint index $\pm$ (head or tail) to a Bravais index $b$, such that

$$b(\beta_\alpha^\pm) = b_{\varphi_\alpha}^\pm. \tag{2.2}$$

The maps $\tilde{\boldsymbol{\delta}}_\varphi$ and $b_\varphi^\pm$ for the topological honeycomb lattice are given in Table 2.1.

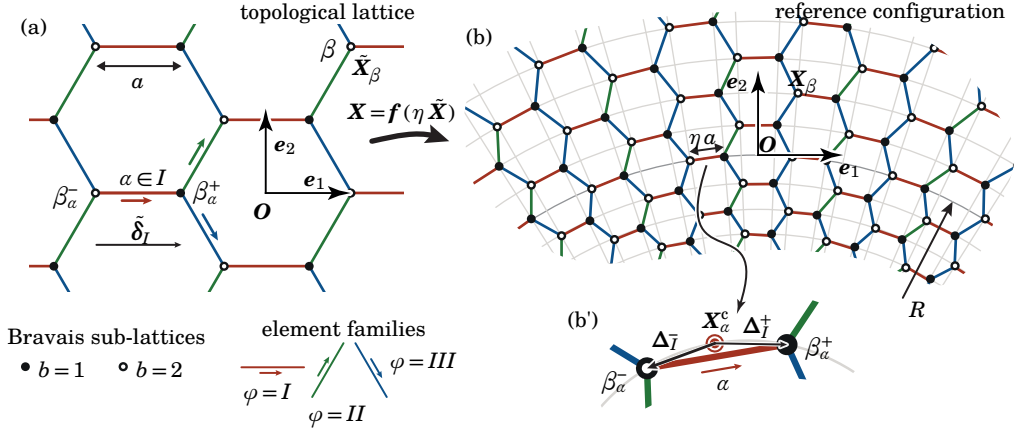

**Figure 2.1.** Generation of a lattice possessing slowly varying properties. (a) Periodic, underlying topological lattice; the conventional element orientation is indicated by the arrows. (b) The reference configuration of the lattice is produced by applying a contraction factor $\eta$ and a diffeomorphism $\boldsymbol{f}$, see (2.3). (b') Close-up view of a particular element (beam) $\alpha$, belonging to family $\varphi_\alpha = I$; note that the conventional element center $\boldsymbol{X}_\alpha^c$ is distinct from its midpoint. The figure was generated for a honeycomb lattice with the diffeomorphism in (2.4) and with parameters $a = 1$, $\eta = 0.4$ and $R = 5$.

| $d$ | $n_b$ | $n_\varphi$ | $\tilde{\rho}_\varphi$ |
|---|---|---|---|
| 2 | 2 | 3 | $\frac{2}{3\sqrt{3}a^2}$ |

| $\varphi$ | $I$ | $II$ | $III$ |
|---|---|---|---|
| $b_\varphi^-$ | 2 | 1 | 1 |
| $b_\varphi^+$ | 1 | 2 | 2 |
| $\tilde{\boldsymbol{\delta}}_\varphi$ | $a\boldsymbol{e}_1$ | $a\left(\frac{\boldsymbol{e}_1}{2} + \frac{\sqrt{3}}{2}\boldsymbol{e}_2\right)$ | $a\left(\frac{\boldsymbol{e}_1}{2} - \frac{\sqrt{3}}{2}\boldsymbol{e}_2\right)$ |

**Table 2.1.** Properties of the topological honeycomb lattice in Figure 2.1a: space dimension $d$, number $n_b$ of Bravais sub-lattices, number $n_\varphi$ of element families, density $\tilde{\rho}_\varphi$ of elements in a given family per unit area, connectivity $b_\varphi^\pm$, edge vectors $\tilde{\boldsymbol{\delta}}_\varphi$.

**Remark 2.1.** By contrast with most of the existing work on periodic homogenization, our approach does not require a unit cell to be defined: the information contained in a unit cell is entirely captured by the quantities shown in Table 2.1.

## 2.2. Curved reference configuration

The reference configuration of the elastic lattice, shown in Figure 2.1b, is produced by applying a diffeomorphism $\boldsymbol{f} : \mathbb{R}^d \to \mathbb{R}^d$ combined with a contraction with ratio $\eta$, to the topological lattice: the position in the reference configuration of the node $\beta$ is

$$\boldsymbol{X}_\beta = \boldsymbol{f}(\eta \tilde{\boldsymbol{X}}_\beta). \tag{2.3}$$

This generation procedure allows for a variety of lattice designs for a given underlying topological lattice.

By way of illustration, the curved honeycomb lattice shown in Figure 2.1b has been produced using the 'complex-exponential' diffeomorphism

$$\boldsymbol{f}(\tilde{\boldsymbol{x}}) = R \exp\left(\frac{\tilde{x}_2}{R}\right) \boldsymbol{e}_r\left(\frac{\pi}{2} - \frac{\tilde{x}_1}{R}\right) - R\boldsymbol{e}_2, \quad \text{(curved honeycomb lattice)} \tag{2.4}$$

where $\boldsymbol{e}_r(\theta) = \cos\theta \, \boldsymbol{e}_1 + \sin\theta \, \boldsymbol{e}_2$ denotes the unit radial vector and $R$ is the curvature radius of the arc of circle that is the image of the curve $\tilde{X}_2 = 0$. The case of a periodic, non-curved honeycomb lattice corresponds to $R \to \infty$, which yields $\boldsymbol{f} = \mathbf{id}_2$ (identity in $\mathbb{R}^2$).

Each element $\alpha$ is assigned a conventional *center* $\boldsymbol{X}_\alpha^c$ in reference configuration. Various definitions of the center are possible. We choose to define it as the image by the diffeomorphism of the edge's midpoint of the underlying topological lattice,

$$\boldsymbol{X}_\alpha^c = \boldsymbol{f}\left(\frac{\eta}{2}(\tilde{\boldsymbol{X}}_{\beta_\alpha^-} + \tilde{\boldsymbol{X}}_{\beta_\alpha^+})\right). \tag{2.5}$$

A particular center $\boldsymbol{X}_\alpha^c$ is shown in Figure 2.1b'.

For any particular element, we now consider the vector $\boldsymbol{X}_{\beta_a^\pm} - \boldsymbol{X}_a^c$ joining (in reference configuration) the conventional center $\boldsymbol{X}_a^c$ to either one of the endpoints $\boldsymbol{X}_{\beta_a^\pm}$. We claim that this vector can be written as

$$\boldsymbol{X}_{\beta_a^\pm} - \boldsymbol{X}_a^c = \boldsymbol{\Delta}_{\varphi_a}^\pm(\eta, \boldsymbol{X}_a^c), \quad \text{with } \boldsymbol{\Delta}_\varphi^\pm(\eta, \boldsymbol{X}) = \mathcal{O}(\eta). \tag{2.6}$$

Here, the $\boldsymbol{\Delta}_\varphi^\pm(\eta, \boldsymbol{X})$'s are *smooth* functions of their continuous arguments $(\eta, \boldsymbol{X})$ that further depend on the discrete variables in subscript (element family $\varphi$) and superscript ($\pm$ labelling endpoints). The dependence on the element $a$ in the left-hand side of (2.6) has been turned in the right-hand side into a dependence on the element *family* $\varphi_a$ and on the center position $\boldsymbol{X}_a^c$: Equation (2.6) captures the fact that $\boldsymbol{X}_{\beta_a^\pm} - \boldsymbol{X}_a^c$ varies slowly across neighboring elements belonging to the same family.

A constructive proof of (2.6) is as follows. Observe first that (2.1) and (2.5) can be combined into $\tilde{\boldsymbol{X}}_{\beta_a^\pm} = \frac{1}{2}(\tilde{\boldsymbol{X}}_{\beta_a^-} + \tilde{\boldsymbol{X}}_{\beta_a^+}) \pm \frac{1}{2}(\tilde{\boldsymbol{X}}_{\beta_a^+} - \tilde{\boldsymbol{X}}_{\beta_a^-}) = \frac{1}{\eta}\boldsymbol{f}^{-1}(\boldsymbol{X}_a^c) \pm \frac{1}{2}\tilde{\boldsymbol{\delta}}_{\varphi_a}$, so that $\boldsymbol{X}_{\beta_a^\pm} - \boldsymbol{X}_a^c = \boldsymbol{f}(\eta\,\tilde{\boldsymbol{X}}_{\beta_a^\pm}) - \boldsymbol{X}_a^c = \boldsymbol{f}(\boldsymbol{f}^{-1}(\boldsymbol{X}_a^c) \pm \frac{\eta}{2}\tilde{\boldsymbol{\delta}}_{\varphi_a}) - \boldsymbol{X}_a^c$. Identifying with (2.6), we find

$$\boldsymbol{\Delta}_\varphi^\pm(\eta, \boldsymbol{X}) = \boldsymbol{f}\left(\boldsymbol{f}^{-1}(\boldsymbol{X}) \pm \frac{\eta}{2}\tilde{\boldsymbol{\delta}}_\varphi\right) - \boldsymbol{X}. \tag{2.7}$$

The right-hand side vanishes when $\eta \to 0$ and the smoothness of $\boldsymbol{f}$ proves the estimate $\boldsymbol{\Delta}_\varphi^\pm(\eta, \boldsymbol{X}) = \mathcal{O}(\eta)$ stated in (2.6).

By way of illustration, consider elements of type $\varphi = I$ in a curved honeycomb, see Figure 2.1(b). Combining $\boldsymbol{f}^{-1}(\boldsymbol{X}) = R\left(\tan^{-1}\left(\frac{X_1}{R+X_2}\right)\boldsymbol{e}_1 + \frac{1}{R}\ln\sqrt{X_1^2 + (R+X_2)^2}\,\boldsymbol{e}_2\right)$ from (2.4) and $\tilde{\boldsymbol{\delta}}_I = a\,\boldsymbol{e}_1$, we have from (2.7)

$$\boldsymbol{\Delta}_I^\pm(\eta, \boldsymbol{X}) = \sqrt{X_1^2 + (R+X_2)^2}\,\boldsymbol{e}_r\left(\frac{\pi}{2} - \tan^{-1}\left(\frac{X_1}{R+X_2}\right) \mp \eta\,\frac{a}{2R}\right) - (X_1\,\boldsymbol{e}_1 + (R+X_2)\,\boldsymbol{e}_2) \quad \text{(curved honeycomb lattice).} \tag{2.8}$$

The two other functions $\boldsymbol{\Delta}_{II}^\pm(\eta, \boldsymbol{X})$ and $\boldsymbol{\Delta}_{III}^\pm(\eta, \boldsymbol{X})$ are obtained in a similar way.

### 2.3. Assumed power-series dependence on $\eta$

Expanding Equation (2.7) in powers of $\eta$, we get

$$\boldsymbol{\Delta}_\varphi^\pm(\eta, \boldsymbol{X}) = \overbrace{\boldsymbol{f}(\boldsymbol{f}^{-1}(\boldsymbol{X})) - \boldsymbol{X}}^{\boldsymbol{0}} \pm \frac{\eta}{2}\nabla\boldsymbol{f}(\boldsymbol{f}^{-1}(\boldsymbol{X})) \cdot \tilde{\boldsymbol{\delta}}_\varphi + \frac{\eta^2}{4}\nabla^2\boldsymbol{f}(\boldsymbol{f}^{-1}(\boldsymbol{X})) : (\tilde{\boldsymbol{\delta}}_\varphi \otimes \tilde{\boldsymbol{\delta}}_\varphi) + \cdots \tag{2.9}$$

where the cancellation at leading order leads to the estimate (2.6). The diffeomorphism $\boldsymbol{f}$ being infinitely smooth by assumption, this expansion could be pushed to any other.

More generally, we will assume that *any* function $g(\eta, \boldsymbol{X})$ depending on the expansion parameter $\eta$ can be expanded as a power series in $\eta$: when we write $g(\eta, \boldsymbol{X})$ with $g = \mathcal{O}(\eta^p)$, we imply that $g$ is of the form

$$g(\eta, \boldsymbol{X}) = \eta^p\,(g_{(p)}(\boldsymbol{X}) + \eta\,g_{(p+1)}(\boldsymbol{X}) + \cdots), \tag{2.10}$$

where the $g_{(k)}$'s are smooth functions of $\boldsymbol{X}$ that do not depend on $\eta$. Expansions of the form (2.10) are known as *regular*. For any lattice property, regular expansions can be *obtained in explicit form* as we just did for $g = \boldsymbol{\Delta}_\varphi^\pm$ in (2.9). For all the elastic fields, the regular expansions will be *assumed*, see Equation (3.2) below, as we approach homogenization based on formal expansions. At boundary layers, the regular expansions break down and the elastic fields are given by *singular* expansions instead: their analysis is beyond the scope of this paper.

### 2.4. Kinematic analysis of the beam-elements

For the moment, we limit attention to naturally straight, *2D* beam elements: this applies to the lattice shown in Figure 2.1b'. The extension to elastic *springs* is discussed later in Section 7. The extension to *3D beams* or *arches* in both 2D and 3D (*i.e.*, of beams possessing natural curvature) is straightforward and is left for future work.

The lattice is deformed by applying forces. The displacement and rotation of the node $\boldsymbol{X}_\beta$ are denoted as $\boldsymbol{v}_\beta \in \mathbb{R}^d$, and $\boldsymbol{\theta}_\beta \in \mathbb{R}^{n_r}$, respectively (in 2D, the infinitesimal rotation $\boldsymbol{\theta}_\beta = (\theta_\beta)$ is a scalar, $n_r = 1$; see Section A.2 in the Appendix for the extension to an arbitrary dimension $d$). The assumption of rigid nodes implies that the infinitesimal rotation $\boldsymbol{\theta}_\beta$ is common to the adjacent elements at any particular node $\boldsymbol{X}_\beta$. The nodal degrees of freedom are therefore

$$(\boldsymbol{v}_\beta, \boldsymbol{\theta}_\beta) \in \mathbb{R}^{n_n}, \qquad \text{where } n_n = d + n_r = \begin{cases} 1 & \text{if } d = 1 \\ 3 & \text{if } d = 2 \\ 6 & \text{if } d = 3 \end{cases}. \tag{2.11}$$

For the 2D elements discussed here, $d = 2$, $n_r = 1$ and $n_n = 3$.

Each element family $\varphi$ is characterized by a set of $n_{E_\varphi}$ element strain measures, which are collected into an *element strain vector* $\boldsymbol{E}_a \in \mathbb{R}^{n_{E_\varphi}}$. In 2D, beam elements uses $n_{E_\varphi} = 3$ strain measures, namely the *extensional strain* $\varepsilon_a$, the *bending strain* $\kappa_a$ and the *shear strain* $\tau_a$, which are defined by

$$\boldsymbol{E}_a = \left(\overbrace{\frac{(\boldsymbol{v}_{\beta_a^+} - \boldsymbol{v}_{\beta_a^-}) \cdot \boldsymbol{t}_a}{\ell_a}}^{\varepsilon_a},\ \overbrace{\theta_{\beta_a^+} - \theta_{\beta_a^-}}^{\kappa_a},\ \overbrace{\frac{\theta_{\beta_a^-} + \theta_{\beta_a^+}}{2} - \frac{(\boldsymbol{v}_{\beta_a^+} - \boldsymbol{v}_{\beta_a^-}) \cdot \boldsymbol{n}_a}{\ell_a}}^{\tau_a}\right) \in \mathbb{R}^3, \qquad n_{E_\varphi} = 3, \qquad \text{(2D beam)} \tag{2.12}$$

where $\ell_\alpha = |\boldsymbol{X}_{\beta_\alpha^+} - \boldsymbol{X}_{\beta_\alpha^-}|$ is the undeformed length, $\boldsymbol{t}_\alpha = (\boldsymbol{X}_{\beta_\alpha^+} - \boldsymbol{X}_{\beta_\alpha^-})/\ell_\alpha$ is the unit tangent, $\boldsymbol{n}_\alpha = \boldsymbol{e}_3 \times \boldsymbol{t}_\alpha$ is the undeformed unit normal and $\boldsymbol{e}_3 = \boldsymbol{e}_1 \times \boldsymbol{e}_2$ is the unit normal to the Euclidean plane, see Figure 2.2. The quantities $(\varepsilon_\alpha, \kappa_\alpha, \tau_\alpha)$ are *effective* measures characterizing the strain in a beam-element *globally*, which do not depend on the arc-length parameter. As shown in Figure 2.2(b), for instance, the shearing mode $\tau_\alpha$ defined at the element scale yields '*locally*' non-uniform curvature as a function of arc-length—this local response will captured by the element stiffness derived in the forthcoming Section 2.5.

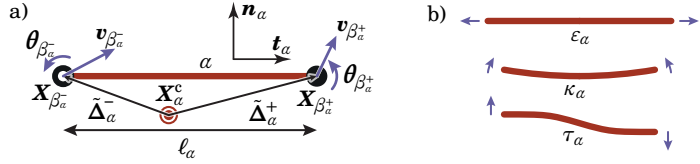

**Figure 2.2.** A 2D beam element, having index $\alpha$ and conventional center $\boldsymbol{X}_\alpha^c$: (a) geometry in reference configuration and (b) the $n_{E_\varphi} = 3$ deformation modes : extension, bending and shearing.

We observe that the undeformed element's length $\ell_\alpha$, unit tangent $\boldsymbol{t}_\alpha$ and unit normal $\boldsymbol{n}_\alpha$ can each be expressed as functions of the element family $\varphi_\alpha$, of the scale separation parameter $\eta$ and of the element center $\boldsymbol{X}_\alpha^c$. Indeed, with help of (2.6), we have $\boldsymbol{X}_{\beta_\alpha^+} - \boldsymbol{X}_{\beta_\alpha^-} = \Delta_{\varphi_\alpha}^+(\eta, \boldsymbol{X}_\alpha^c) - \Delta_{\varphi_\alpha}^-(\eta, \boldsymbol{X}_\alpha^c)$, which yields

$$\begin{aligned}
\ell_\alpha &= \hat{\ell}_{\varphi_\alpha}(\eta, \boldsymbol{X}_\alpha^c) &&\text{where } \hat{\ell}_\varphi(\eta, \boldsymbol{X}) = |\Delta_\varphi^+(\eta, \boldsymbol{X}) - \Delta_\varphi^-(\eta, \boldsymbol{X})| = \mathcal{O}(\eta^1), \\
\boldsymbol{t}_\alpha &= \hat{\boldsymbol{t}}_{\varphi_\alpha}(\eta, \boldsymbol{X}_\alpha^c) &&\text{where } \hat{\boldsymbol{t}}_\varphi(\eta, \boldsymbol{X}) = \frac{\Delta_\varphi^+(\eta, \boldsymbol{X}) - \Delta_\varphi^-(\eta, \boldsymbol{X})}{\hat{\ell}_\varphi(\eta, \boldsymbol{X})} = \mathcal{O}(\eta^0) \\
\boldsymbol{n}_\alpha &= \hat{\boldsymbol{n}}_{\varphi_\alpha}(\eta, \boldsymbol{X}_\alpha^c) &&\text{where } \hat{\boldsymbol{n}}_\varphi(\eta, \boldsymbol{X}) = \boldsymbol{e}_3 \times \hat{\boldsymbol{t}}_\varphi(\eta, \boldsymbol{X}) = \mathcal{O}(\eta^0).
\end{aligned} \quad (2.13)$$

The dependence of these function on $\boldsymbol{X}$ applies to lattices having graded geometrical properties (see §7.3). The functions $\hat{\ell}_{\varphi_\alpha}$, $\hat{\boldsymbol{t}}_{\varphi_\alpha}$ and $\hat{\boldsymbol{n}}_{\varphi_\alpha}$ have regular expansions in $\eta$ of the form (2.10) as can be checked using (2.9).

In view of this, the element strain $\boldsymbol{E}_\alpha$ in (2.12) can be rewritten in block-matrix notation as

$$\boldsymbol{E}_\alpha = \mathscr{D}_{\varphi_\alpha}(\eta, \boldsymbol{X}_\alpha^c) : \left( \begin{pmatrix} \boldsymbol{v}_{\beta_\alpha^-} \\ \theta_{\beta_\alpha^-} \end{pmatrix} \begin{pmatrix} \boldsymbol{v}_{\beta_\alpha^+} \\ \theta_{\beta_\alpha^+} \end{pmatrix} \right), \qquad (2.14)$$

where : denotes the double contraction, see Appendix A, $(\boldsymbol{v}_\beta \ \theta_\beta) \in \mathbb{R}^{n_n}$ are the nodal degrees of freedom, see (2.11), and $\mathscr{D}_\varphi \in \mathbb{T}^{(n_{E_\varphi}, 2, n_n)}$ is the displacement-to-strain tensor,

$$\mathscr{D}_\varphi(\eta, \boldsymbol{X}) = \left( \begin{pmatrix} -\hat{\boldsymbol{t}}_\varphi(\eta, \boldsymbol{X})/\hat{\ell}_\varphi(\eta, \boldsymbol{X}) & 0 \\ +\hat{\boldsymbol{t}}_\varphi(\eta, \boldsymbol{X})/\hat{\ell}_\varphi(\eta, \boldsymbol{X}) & 0 \end{pmatrix} \begin{pmatrix} \boldsymbol{0}_2 & -1 \\ \boldsymbol{0}_2 & +1 \end{pmatrix} \begin{pmatrix} +\hat{\boldsymbol{n}}_\varphi(\eta, \boldsymbol{X})/\hat{\ell}_\varphi(\eta, \boldsymbol{X}) & 1/2 \\ -\hat{\boldsymbol{n}}_\varphi(\eta, \boldsymbol{X})/\hat{\ell}_\varphi(\eta, \boldsymbol{X}) & 1/2 \end{pmatrix} \right) \in \mathbb{T}^{(3,2,3)} \quad \text{(2D beam)}. \qquad (2.15)$$

The function $\mathscr{D}_\varphi(\eta, \boldsymbol{X})$ can be expressed as a regular expansion in $\eta$ of the form (2.10), as can be checked using (2.9) and (2.13).

To keep the presentation general, we will refrain from using the specific expression (2.15) of $\mathscr{D}_\varphi$ applicable to 2D beams, and use the generic form (2.14) of the displacement-strain relation instead.

## 2.5. Strain energy of the lattice

A beam $\alpha$ made up of a Hookean material and having a circular cross-section is characterized by a stretching modulus $(EA)_\alpha$ and a bending modulus $(EI)_\alpha$. The linear beam theory yields the strain energy $w_\alpha$ of the beam as $w_\alpha = \frac{1}{2}\left((EA)_\alpha \ell_\alpha \varepsilon_\alpha^2 + \frac{(EI)_\alpha}{\ell_\alpha}(\kappa_\alpha^2 + 12\tau_\alpha^2)\right)$, which can be rewritten as

$$w_\alpha = \frac{(EA)_\alpha \ell_\alpha}{2}(\varepsilon_\alpha^2 + \chi_\alpha(\kappa_\alpha^2 + 12\tau_\alpha^2)) \qquad (2.16)$$

(no implicit sum over $\alpha$) in terms of the beam's aspect-ratio parameter

$$\chi_\alpha = \frac{(EI)_\alpha}{\ell_\alpha^2 (EA)_\alpha}. \qquad (2.17)$$

For the slender beam theory to be applicable, the beam's radius $r_\alpha$ must be much smaller than its length, $r_\alpha \ll \ell_\alpha$, implying that the aspect-ratio parameter $\chi_\alpha \sim (r_\alpha/\ell_\alpha)^2 \ll 1$ is small.

To handle the case of lattices whose elastic properties vary slowly in space, we assume that the beam's elastic constants $(EA)_\alpha$ and $\chi_\alpha$ are prescribed as smooth functions $\widehat{(EA)}_\varphi(\eta, \boldsymbol{X})$ and $\hat{\chi}_\varphi(\eta, \boldsymbol{X})$ of the scale separation parameter $\eta$ and of the element center $\boldsymbol{X}_\alpha^c$, with an additional dependence on the discrete family index $\varphi$, i.e.,

$$(EA)_\alpha = \widehat{EA}_{(\varphi_\alpha)}(\eta, \boldsymbol{X}_\alpha^c), \qquad \chi_\alpha = \hat{\chi}_{(\varphi_\alpha)}(\eta, \boldsymbol{X}_\alpha^c). \qquad (2.18)$$

In view of (2.16), we define the element stiffness matrix $\mathscr{H}_\varphi \in \mathbb{T}^{(n_{E_\varphi}, n_{E_\varphi})}$ as

$$\mathscr{H}_\varphi(\eta, \boldsymbol{X}) = \widehat{EA}_\varphi(\eta, \boldsymbol{X}) \, \hat{\ell}_\varphi(\eta, \boldsymbol{X}) \, \text{diag}(1, \hat{\chi}_\varphi(\eta, \boldsymbol{X}), 12\, \hat{\chi}_\varphi(\eta, \boldsymbol{X})) \in \mathbb{T}^{(3,3)} \qquad \text{(2D beam)}, \qquad (2.19)$$

so that the strain energy of an element takes the form

$$w_\alpha(\boldsymbol{E}_\alpha) = \frac{1}{2} \boldsymbol{E}_\alpha \cdot \mathcal{H}_{\varphi_\alpha}(\eta, \boldsymbol{X}_\alpha^c) \cdot \boldsymbol{E}_\alpha \tag{2.20}$$

(no implicit sum over $\alpha$). To keep the presentation general, we will use this expression (2.20) of the strain energy in the following, as it is applies to linearly elastic element of various kinds such as springs, beams or arches in both 2D and 3D. The element stiffness matrix $\mathcal{H}_\varphi \in \mathbb{T}^{(n_{E_\varphi}, n_{E_\varphi})}$ is always symmetric but not necessarily diagonal.

The element stiffness is assumed to obey the scaling assumption

$$\mathcal{H}_\varphi = \mathcal{O}(\eta^d), \tag{2.21}$$

where $d$ is the space dimension. This assumption is a matter of convention: applying a global factor to all elements' stiffness scales the homogenized energy by the same factor. The convention (2.21) warrants that the energy is $\Phi = \mathcal{O}(\eta^0)$, which is convenient, see Equation (3.42) below. Since $\hat{\ell}_\varphi = \mathcal{O}(\eta)$ by (2.13), the stretching modulus $\widehat{EA}_\varphi$ in (2.19) will be viewed as a quantity of order

$$EA = \mathcal{O}(\eta^{d-1}) \tag{2.22}$$

in stretching-dominated lattices.

**Remark 2.2.** As noted below Equation (2.17), the function $\hat{\chi}_\varphi(\eta, \boldsymbol{X})$ is small, $\hat{\chi}_\varphi(\eta, \boldsymbol{X}) \ll 1$. Mathematically, $\hat{\chi}_\varphi$ will be defined as a power of the aspect-ratio $\eta$. Different power laws are considered in the different examples, such as $\hat{\chi}_\varphi = \mathcal{O}(\eta)$ for the honeycomb, see (5.7), or $\hat{\chi}_\varphi = \mathcal{O}(\eta^2)$ for the pantograph, see (7.28). The only assumption we make at this stage is that $\hat{\chi}_\varphi$ is a function of $\eta$ with $\hat{\chi}_\varphi = \mathcal{O}(\eta^0)$—this includes $\hat{\chi}_\varphi = \mathcal{O}(\eta^1)$ and $\hat{\chi}_\varphi = \mathcal{O}(\eta^2)$ as particular cases. In any case, the matrix $\mathcal{H}_\varphi(\eta, \boldsymbol{X})$ appearing in (2.19) may contain terms of different orders in $\eta$ and may therefore depend on $\eta$ through more than just an overall scaling factor.

Anticipating that the homogenization will fail near the edges of the lattice due to the presence of boundary layers, we limit attention to a sub-region $\Omega \subset \mathbb{R}^d$ of the physical lattice that exclude the edges, see Figure 2.3: the strain energy of the lattice $\Phi_d$ in the region $\Omega$ is defined as the sum of the elastic energy $w_\alpha$ over all beams $\alpha$ whose center $\boldsymbol{X}_\alpha^c$ (shown as a red dot in Figure 2.3) lies inside the domain $\Omega$,

$$\Phi_d = \sum_{\alpha \text{ such that } \boldsymbol{X}_\alpha^c \in \Omega} w_\alpha(\boldsymbol{E}_\alpha). \tag{2.23}$$

Points where the loading is not smooth (such as points of application of point-like forces) are also excluded from the domain $\Omega$, along with a finite neighborhood around such points. The subscript 'd' in $\Phi_d$ stands for 'discrete': a continuous variant $\Phi$ of the lattice strain energy, bearing no subscript, will be introduced later.

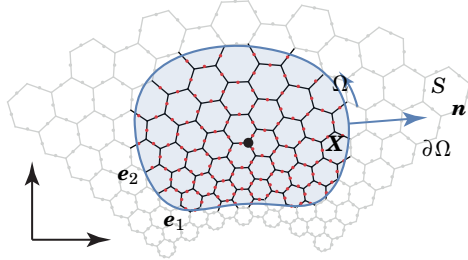

**Figure 2.3.** To avoid boundary layers, the homogenization is carried out in a sub-region $\Omega$ of the lattice (shaded blue region) that excludes the physical boundaries of the lattice.

**Remark 2.3.** In Sections 7.1, we treat the case of an elastic truss whose bending modulus $EI(\boldsymbol{X})$ varies slowly in space. In Section 7.3, we treat the case of circular arch, whose geometric properties vary slowly in space. In both case, the element stiffness matrix $\mathcal{H}_\varphi$ depends explicitly on $\boldsymbol{X}$, see Equation (7.22) for the arch.

## 3. HOMOGENIZATION PROBLEM IN CANONICAL FORM

In this Section, we cast the energy formulation of the discrete lattice problem obtained in Section 2 into a standard form, the *canonical form*, which we will be able to feed into the symbolic homogenization tool [Aud23] available from our previous work [AL23]. The following steps are involved:

- describe the discrete solution in terms of continuous functions (continualization step, §3.1),
- introduce a vector $\boldsymbol{l}(\boldsymbol{X})$ collecting the degrees of freedom which are fixed during homogenization (§3.3) and a vector $\boldsymbol{y}(\boldsymbol{X})$ collecting those that are relaxed during homogenization (§3.4),
- introduce an assembled strain vector $\boldsymbol{E}(\boldsymbol{X})$ in terms of $\boldsymbol{l}(\boldsymbol{X})$, $\boldsymbol{y}(\boldsymbol{X})$ and their gradients (§3.5),
- express the strain energy of the lattice in terms of $\boldsymbol{E}(\boldsymbol{X})$ (§3.6–3.7).

### 3.1. Parameterization of discrete solution using continuous functions

A key assumption in homogenization is that the equilibrium solution of the *discrete* lattice can be captured by the following functions of the *continuous* variable $X$:

- a macroscopic displacement $u(X) \in \mathbb{R}^d$ representing the local average of the nodal displacement over the different Bravais sub-lattices,
- for each Bravais sub-lattice $1 \leq b \leq n_b$, a microscopic displacement $\xi_b(X) \in \mathbb{R}^d$ and a microscopic rotation $\psi_b(X) \in \mathbb{R}^{n_r}$.

Their orders of magnitude are taken as (see Remark 3.1 for a justification),

$$u(X) = \mathcal{O}(\eta^0), \qquad \xi_b(X) = \mathcal{O}(\eta^1), \qquad \psi_b(X) = \mathcal{O}(\eta^0). \tag{3.1}$$

All these functions may depend on the expansion parameter $\eta \ll 1$, even though this dependence is implicit in our notation. As earlier in (2.10), we assume that this dependence takes place through regular expansions, with a leading term set by (3.1), *i.e.*,

$$\begin{array}{rcl} u(X) &=& u_{(0)}(X) + \eta\, u_{(1)}(X) + \cdots \\ \xi_b(X) &=& \eta\, (\xi_b)_{(1)}(X) + \cdots \\ \psi_b(X) &=& (\psi_b)_{(0)}(X) + \eta\, (\psi_b)_{(1)}(X) + \cdots \end{array} \tag{3.2}$$

where all the 'terms' in the series $u_{(k)}(X)$, $(\xi_b)_{(k)}(X)$ and $(\psi_b)_{(k)}(X)$ are smooth functions of $X$. Throughout this paper and in the interest of legibility, we will use the quantities $u(X)$, $\xi_b(X)$ and $\psi_b(X)$ as blocks, refraining as much as possible from exposing the underlying expansions. The only concrete consequences of (3.2) are that (*i*) we determine these fields perturbatively, see (4.3–4.4), at the successive orders of the homogenization procedure and (*ii*) the gradients scale the same way as the undifferentiated quantities, *i.e.*, $\nabla^k u = \mathcal{O}(\eta^0)$, $\nabla^k \xi_b = \mathcal{O}(\eta^1)$ and $\nabla^k \psi_b = \mathcal{O}(\eta^0)$, which we summarize by the formal rule

$$\nabla = \mathcal{O}(\eta^0). \tag{3.3}$$

As usual in continuum mechanics, the macroscopic rotation $\gamma(X) \in \mathbb{R}^{n_r}$ is a vector extracted from the antisymmetric part of $\nabla u(X)$: in our notation, this extraction is denoted as $\gamma(X) = -\frac{1}{2} \nabla u(X) : \mathcal{N} = \mathcal{O}(\eta^0)$ where $\mathcal{N}$ is a constant tensor that is antisymmetric with respect to its first pair of indices, see Equation (3.8)$_2$ below.

Following the Cauchy-Born 'hypothesis', the nodal degrees of freedom are postulated in terms of the continuous fields as

$$v_\beta = u(X_\beta) + \xi_{b(\beta)}(X_\beta), \qquad \theta_\beta = \gamma(X_\beta) + \psi_{b(\beta)}(X_\beta). \tag{3.4}$$

Here, both the nodal displacement $v_\beta$ and rotation $\theta_\beta$ are expressed as the sum of a *macroscopic* quantity (displacement $u(X)$ and rotation $\gamma(X)$) and a *microscopic* one (displacement $\xi_b(X)$ and rotation $\psi_b(X)$). The macroscopic quantities $u(X)$ and $\gamma(X)$ are functions on $X$ only, and therefore vary slowly, whereas the additional dependence of $\xi_b$ and $\psi_b$ on the Bravais index $b = b(\beta)$ of the node causes 'fast' variations at the scale of the cell. Equation (3.4) is a variant of the double scale expansion used as a starting point for the homogenization of periodic continua, see for instance [San80; Bou96], which has been modified to account for the fact that $b$ is a *discrete* variable here.

The decomposition of the nodal displacement $v_\beta$ in (3.4)$_1$ is not unique: shifting $u(X)$ by $\tilde{u}(X)$, and all the $\xi_b(X)$'s by $-\tilde{u}(X)$ leaves $v_\beta$ unchanged. To ensure unicity, we impose the constraint

$$\sum_{b=1}^{n_b} \xi_b(X) = 0, \quad \forall X. \tag{3.5}$$

By averaging both sides of (3.4)$_1$ with respect to the Bravais sub-lattices $b$, one can then confirm that $u(X)$ is the average nodal displacement as announced earlier.

**Remark 3.1.** The scaling assumptions in (3.1) imply that the element strain $E_\alpha$ in (2.12) has magnitude

$$E_\alpha = \mathcal{O}(\eta^0). \tag{3.6}$$

Considering the stretching strain, for instance, $|\varepsilon_\alpha| = \left| \frac{(v_{\beta_\alpha^+} - v_{\beta_\alpha^-}) \cdot t_\alpha}{\ell_\alpha} \right| = \left| \frac{(u(X_{\beta_\alpha^+}) + \xi_{b(\beta_\alpha^+)}(X_{\beta_\alpha^+}) - u(X_{\beta_\alpha^-}) - \xi_{b(\beta_\alpha^-)}(X_{\beta_\alpha^-}))}{\ell_\alpha} \right| \leq \frac{|\nabla u \cdot (\Delta_{\varphi_\alpha}^+ - \Delta_{\varphi_\alpha}^-)|}{|\Delta_{\varphi_\alpha}^+ - \Delta_{\varphi_\alpha}^-|} + \frac{|\xi_{b(\beta_\alpha^+)}(X_{\beta_\alpha^+})|}{|\Delta_{\varphi_\alpha}^+ - \Delta_{\varphi_\alpha}^-|} + \frac{|\xi_{b(\beta_\alpha^-)}(X_{\beta_\alpha^-})|}{|\Delta_{\varphi_\alpha}^+ - \Delta_{\varphi_\alpha}^-|}$ which yields $|\varepsilon_\alpha| = \mathcal{O}(\eta^0)$ when combined with (3.1)$_{1,2}$ and (2.6). A similar reasoning holds for $\kappa_\alpha = \mathcal{O}(\eta^0)$ and $\tau_\alpha = \mathcal{O}(\eta^0)$. In situations where the strain scales differently than in (3.6), one can simply shift the orders in the scaling assumptions (3.1).

**Remark 3.2.** Given an equilibrium configuration of the lattice, the continuous fields can be explicitly constructed as follows. First, the nodal displacements $v_\beta$ in each particular Bravais sub-lattice $b \ni \beta$ are interpolated by smooth functions $v_b^s(X_\beta)$ of their undeformed position $X_\beta$: a set of $n_b$ such interpolating functions $v_b^s$ with $b = 1, \ldots, n_b$ are obtained. Next, $u(X)$ is defined as the average $u(X) = \frac{1}{n_b} \sum_{b=1}^{n_b} v_b^s(X)$. Finally, the 'continualized' microscopic fields $\xi_b(X)$ and $\psi_b(X)$ are obtained by interpolating $v_\beta - u(X_\beta)$ and $\theta_\beta - \gamma(X_\beta)$, respectively, over each Bravais sub-lattice separately. The constraint (3.5) is then automatically satisfied.

**Remark 3.3.** Consider the particular case of a rigid-body motion of the lattice made up of a translation $\boldsymbol{u}_0$ and rotation $\boldsymbol{\gamma}_0$, *i.e.*, $(\boldsymbol{v}_\beta, \boldsymbol{\theta}_\beta) = (\boldsymbol{u}_0 + \boldsymbol{\gamma}_0 \times \boldsymbol{X}_\beta, \boldsymbol{\gamma}_0)$ at every node $\beta$. The macroscopic fields are extracted as outlined in Remark 3.2, which yields the macroscopic displacement as $\boldsymbol{u}(\boldsymbol{X}) = \boldsymbol{u}_0 + \boldsymbol{\gamma}_0 \times \boldsymbol{X}$, a constant macroscopic rotation $\boldsymbol{\gamma}(\boldsymbol{X}) = \boldsymbol{\gamma}_0$ given by $(3.8)_2$ and zero microscopic displacement $\boldsymbol{\xi}_b(\boldsymbol{X}) = \boldsymbol{0}_d$ and rotation $\boldsymbol{\psi}_b(\boldsymbol{X}) = \boldsymbol{0}_{n_r}$. By including the term $\boldsymbol{\gamma}(\boldsymbol{X}_\beta)$ in the right-hand side of $(3.4)_2$, we have made the microscopic displacement insensitive to rigid-body motions of the lattice.

## 3.2. Macroscopic strain $\check{\boldsymbol{\varepsilon}}$ and rotation $\boldsymbol{\gamma}$

In linear elasticity, the displacement gradient $\nabla \boldsymbol{u}(\boldsymbol{X})$ is classically decomposed into:

- a symmetric strain *tensor* $\boldsymbol{\varepsilon}(\boldsymbol{X}) = \frac{1}{2}(\nabla \boldsymbol{u}(\boldsymbol{X}) + \nabla \boldsymbol{u}^T(\boldsymbol{X})) \in \mathbb{T}^{(d,d)}$, whose components we arrange into a *vector* $\check{\boldsymbol{\varepsilon}}(\boldsymbol{X}) \in \mathbb{R}^{n_\varepsilon}$, following Mandel's notation,

$$\check{\boldsymbol{\varepsilon}}(\boldsymbol{X}) = \begin{cases} (\varepsilon_{11}(\boldsymbol{X}), \varepsilon_{22}(\boldsymbol{X}), \sqrt{2}\,\varepsilon_{12}(\boldsymbol{X})) & (d=2) \\ (\varepsilon_{11}(\boldsymbol{X}), \varepsilon_{22}(\boldsymbol{X}), \varepsilon_{33}(\boldsymbol{X}), \sqrt{2}\,\varepsilon_{12}(\boldsymbol{X}), \sqrt{2}\,\varepsilon_{23}(\boldsymbol{X}), \sqrt{2}\,\varepsilon_{13}(\boldsymbol{X})) & (d=3), \end{cases} \quad (3.7)$$

where $n_\varepsilon = \frac{d(d+1)}{2}$ is the number of independent strain components,

- and an antisymmetric part $\hat{\boldsymbol{\gamma}}(\boldsymbol{X}) = \frac{1}{2}(\nabla \boldsymbol{u}(\boldsymbol{X}) - \nabla \boldsymbol{u}^T(\boldsymbol{X}))$ representing an infinitesimal rotation.

With the help of the geometric tensors $\mathcal{M} \in \mathbb{T}^{(d,d,n_\varepsilon)}$ and $\mathcal{N} \in \mathbb{T}^{(d,d,n_r)}$, which depend on the space dimension $d$ only and are defined in Equation (B.1) in the Appendix B and Equation (A.6) in the Appendix A.2, respectively, the macroscopic strain $\check{\boldsymbol{\varepsilon}}$ and the pseudo-vector $\boldsymbol{\gamma}$ associated with the infinitesimal rotation $\hat{\boldsymbol{\gamma}}$ are obtained from $\nabla \boldsymbol{u}$ as

$$\check{\boldsymbol{\varepsilon}}(\boldsymbol{X}) = \nabla \boldsymbol{u}(\boldsymbol{X}) : \mathcal{M}, \qquad \boldsymbol{\gamma}(\boldsymbol{X}) = -\frac{1}{2} \nabla \boldsymbol{u}(\boldsymbol{X}) : \mathcal{N}. \quad (3.8)$$

The first equality follows from Equation (B.4) in Appendix B, and from the fact that $\mathcal{M}$ is symmetric with respect to its first pair of indices. The second equality follows by setting $\boldsymbol{\gamma} = \boldsymbol{\theta}$ in $(A.5)_2$ and from the fact that $\mathcal{N}$ is antisymmetric with respect to its first pair of indices.

By the scaling assumptions (3.1), $\nabla \boldsymbol{u} = \mathcal{O}(\eta^0)$, and we have the order-of-magnitude estimates

$$\check{\boldsymbol{\varepsilon}}(\boldsymbol{X}) = \mathcal{O}(\eta^0), \qquad \boldsymbol{\gamma}(\boldsymbol{X}) = \mathcal{O}(\eta^0). \quad (3.9)$$

## 3.3. Macroscopic state vector $\boldsymbol{l}$

Next, we define the *macroscopic degrees of freedom* vector $\boldsymbol{l}(\boldsymbol{X}) \in \mathbb{R}^{n_l}$ by concatenating the macroscopic displacement $\boldsymbol{u}(\boldsymbol{X})$, rotation $\boldsymbol{\gamma}(\boldsymbol{X})$ and strain $\check{\boldsymbol{\varepsilon}}(\boldsymbol{X})$: in block-vector notation,

$$\boldsymbol{l}(\boldsymbol{X}) = (\,\boldsymbol{u}(\boldsymbol{X})\ \ \boldsymbol{\gamma}(\boldsymbol{X})\ \ \check{\boldsymbol{\varepsilon}}(\boldsymbol{X})\,) \in \mathbb{R}^{n_l}, \quad (3.10)$$

where

$$n_l = d + n_r + n_\varepsilon. \quad (3.11)$$

By design, the macroscopic state vector is a quantity of order 1, see (3.1) and (3.8),

$$\boldsymbol{l}(\boldsymbol{X}) = \mathcal{O}(\eta^0). \quad (3.12)$$

The macroscopic state vector $\boldsymbol{l}(\boldsymbol{X})$ collects the quantities that are fixed during homogenization.

As shown in Equation (A.8) in Appendix A.3, the Taylor expansion about a point $\boldsymbol{X}_c$ of the macroscopic contribution $(\boldsymbol{u}, \boldsymbol{\gamma})$ to the nodal degrees of freedom, see (3.4), can be written in terms of the macroscopic state vector $\boldsymbol{l}(\boldsymbol{X})$ as

$$\begin{pmatrix} \boldsymbol{u}(\boldsymbol{X}) \\ \boldsymbol{\gamma}(\boldsymbol{X}) \end{pmatrix} = \mathcal{T}_l(\boldsymbol{X} - \boldsymbol{X}_c) \cdot \boldsymbol{l}(\boldsymbol{X}_c) + \mathcal{T}_{l'}(\boldsymbol{X} - \boldsymbol{X}_c) : \nabla \boldsymbol{l}(\boldsymbol{X}_c) + \mathcal{T}_{l''}(\boldsymbol{X} - \boldsymbol{X}_c) \vdots \nabla^2 \boldsymbol{l}(\boldsymbol{X}_c) + \cdots, \quad (3.13)$$

where the tensors $\mathcal{T}_{l^{(i)}}$ are expressed in block-matrix notation as

$$\begin{aligned} \mathcal{T}_l(\boldsymbol{\Delta}) &= \begin{pmatrix} \boldsymbol{I}_d & -\mathcal{N}^{T_{132}} \cdot \boldsymbol{\Delta} & \mathcal{U}'^{T_{132}} \cdot \boldsymbol{\Delta} \\ \boldsymbol{0}_{n_r \times d} & \boldsymbol{I}_{n_r} & \boldsymbol{0}_{n_r \times n_\varepsilon} \end{pmatrix} & \in \mathbb{T}^{(n_n, n_l)}, \\ \mathcal{T}_{l'}(\boldsymbol{\Delta}) &= \begin{pmatrix} \boldsymbol{0}_{d \times d \times d} & \boldsymbol{0}_{d \times n_r \times d} & \mathcal{U}''^{T_{14523}} : \frac{\boldsymbol{\Delta}^{\otimes 2}}{2} \\ \boldsymbol{0}_{n_r \times d \times d} & \boldsymbol{0}_{n_r \times n_r \times d} & \mathcal{G}'^{T_{1423}} \cdot \boldsymbol{\Delta} \end{pmatrix} & \in \mathbb{T}^{(n_n, n_l, d)}, \\ \mathcal{T}_{l''}(\boldsymbol{\Delta}) &= \begin{pmatrix} \boldsymbol{0}_{d \times d \times d \times d} & \boldsymbol{0}_{d \times n_r \times d \times d} & \mathcal{U}'''^{T_{1567234}} \vdots \frac{\boldsymbol{\Delta}^{\otimes 3}}{6} \\ \boldsymbol{0}_{n_r \times d \times d \times d} & \boldsymbol{0}_{n_r \times n_r \times d \times d} & \mathcal{G}''^{T_{156234}} : \frac{\boldsymbol{\Delta}^{\otimes 2}}{2} \end{pmatrix} & \in \mathbb{T}^{(n_n, n_l, d, d)}, \\ & \qquad \cdots \end{aligned} \quad (3.14)$$

Equations (3.13–3.14) follow from the geometric analysis in Sections A.3 and B.3 in the Appendix, and make use of the geometric tensors $\mathcal{N}, \mathcal{U}', \mathcal{U}'', \mathcal{U}''', \mathcal{G}'$ and $\mathcal{G}''$ defined in Equations (B.7–B.15): they are a compact rewriting of the Taylor expansion (A.8) derived in the Appendix. In the right-hand sides of (3.14), the vertical blocks correspond to the nodal displacement ($\boldsymbol{u}$) versus rotation ($\boldsymbol{\gamma}$), and the horizontal blocks to the $\nabla^i \boldsymbol{u}$, $\nabla^i \boldsymbol{\gamma}$ and $\nabla^i \check{\boldsymbol{\varepsilon}}$ parts of $\nabla^i \boldsymbol{l}$, see (3.10). The generalized-transpose notation is defined in Appendix A.

### 3.4. Microscopic degrees of freedom $y$

Next, we define the microscopic degrees of freedom $y(X)$ as the concatenation into a single vector of those relevant to the different Bravais lattice,

$$y(X) = \left( \begin{array}{ccccccc} \frac{\xi_1(X)}{\eta} & \psi_1(X) & \ldots & \frac{\xi_b(X)}{\eta} & \psi_b(X) & \ldots & \frac{\xi_{n_b}(X)}{\eta} & \psi_{n_b}(X) \end{array} \right) \in \mathbb{R}^{n_y}, \tag{3.15}$$

where $n_y$ is the dimension of $y$,

$$n_y = n_b \times n_n. \tag{3.16}$$

By contrast with $l$, the vector $y$ lists all the degrees of freedom that will be relaxed during homogenization: the discrete energy of the lattice depends on $y$ but the homogenized energy does not. The coefficients $1/\eta$ in (3.15) warrant

$$y(X) = \mathcal{O}(\eta^0), \tag{3.17}$$

see (3.1).

The microscopic displacement and rotation at a particular node can be extracted from $y$ using

$$\begin{pmatrix} \xi_{b(\beta)}(X_\beta) \\ \psi_{b(\beta)}(X_\beta) \end{pmatrix} = \mathcal{B}_{b(\beta)} \cdot y(X_\beta), \tag{3.18}$$

where $\mathcal{B}_b$ is defined as

$$\mathcal{B}_b = \left( \begin{array}{ccc} \mathbf{0}_{n_n \times (b-1)n_n} & \begin{pmatrix} \eta I_d & \mathbf{0} \\ \mathbf{0} & I_{n_r} \end{pmatrix} & \mathbf{0}_{n_n \times (n_b-b)n_n} \end{array} \right) \in \mathbb{T}^{(n_n, n_y)}.$$

We have derived in (3.13) the Taylor expansion of the *macroscopic* contribution to the nodal degrees of freedom about a generic point $X_c$. The counterpart for the *microscopic* contributions is obtained by expanding (3.18) as

$$\begin{pmatrix} \xi_b(X) \\ \psi_b(X) \end{pmatrix} = \mathcal{T}_y^b(X - X_c) \cdot y(X_c) + \mathcal{T}_{y'}^b(X - X_c) : \nabla y(X_c) + \mathcal{T}_{y''}^b(X - X_c) \therefore \nabla^2 y(X_c) + \cdots \tag{3.19}$$

where

$$\begin{aligned} \mathcal{T}_y^b(\Delta) &= \mathcal{B}_b & \in \mathbb{T}^{(n_n, n_y)}, \\ \mathcal{T}_{y'}^b(\Delta) &= \mathcal{B}_b \otimes \Delta & \in \mathbb{T}^{(n_n, n_y, d)}, \\ \mathcal{T}_{y''}^b(\Delta) &= \mathcal{B}_b \otimes \frac{\Delta^{\otimes 2}}{2} & \in \mathbb{T}^{(n_n, n_y, d, d)}, \\ &\cdots \end{aligned} \tag{3.20}$$

### 3.5. Element strain expansion

Combining (2.14), (3.4), (3.13), (3.19), (2.6) and (2.2), one can rewrite the strain $E_\alpha$ in an element $\alpha$ as

$$E_\alpha = E_{\varphi_\alpha}(\eta, X_\alpha^c) \tag{3.21}$$

where $E_\varphi(\eta, X)$ is a smooth function of $X = X_\alpha^c$ depending on the element family $\varphi = \varphi_\alpha$,

$$E_\varphi(\eta, X) = \mathcal{D}_\varphi(\eta, X) : \left( \begin{bmatrix} \sum_{i=0}^\infty \mathcal{T}_{l^{(i)}}(\Delta_\varphi^-(\eta, X)) \cdot_{(i+1)} \nabla^i l(X) \cdots \\ + \sum_{i=0}^\infty \mathcal{T}_{y^{(i)}}^{b_\varphi^-}(\Delta_\varphi^-(\eta, X)) \cdot_{(i+1)} \nabla^i y(X) \end{bmatrix} \begin{bmatrix} \sum_{i=0}^\infty \mathcal{T}_{l^{(i)}}(\Delta_\varphi^+(\eta, X)) \cdot_{(i+1)} \nabla^i l(X) \cdots \\ + \sum_{i=0}^\infty \mathcal{T}_{y^{(i)}}^{b_\varphi^+}(\Delta_\varphi^+(\eta, X)) \cdot_{(i+1)} \nabla^i y(X) \end{bmatrix} \right),$$

and $\cdot_{(i)}$ is a shorthand for the contraction of order $i$ defined in Section A.1 of the Appendix, that is $\cdot_{(1)} = \cdot$, $\cdot_{(2)} = :$, $\cdot_{(3)} = \therefore$, etc.

This can be rewritten in compact form as

$$E_\varphi(\eta, X) = \sum_{i=0}^\infty E_\varphi^{l^{(i)}}(\eta, X) \cdot_{(i+1)} \nabla^i l(X) + \sum_{i=0}^\infty E_\varphi^{y^{(i)}}(\eta, X) \cdot_{(i+1)} \nabla^i y(X), \tag{3.22}$$

where

$$\begin{aligned} E_\varphi^{l^{(i)}}(\eta, X) &= \mathcal{D}_\varphi(\eta, X) : \left( \mathcal{T}_{l^{(i)}}(\Delta_\varphi^-(\eta, X)) \quad \mathcal{T}_{l^{(i)}}(\Delta_\varphi^+(\eta, X)) \right) \\ E_\varphi^{y^{(i)}}(\eta, X) &= \mathcal{D}_\varphi(\eta, X) : \left( \mathcal{T}_{y^{(i)}}^{b_\varphi^-}(\Delta_\varphi^-(\eta, X)) \quad \mathcal{T}_{y^{(i)}}^{b_\varphi^+}(\Delta_\varphi^+(\eta, X)) \right). \end{aligned} \tag{3.23}$$

In (3.21–3.23), the element strain $E_\alpha$ has been obtained as a series expansion about the element center $X = X_\alpha^c$ in terms of the macroscopic state vector $l$, of the microscopic degrees of freedom $y$ and of their successive derivatives. Using explicit formulas for the $\mathcal{T}$'s in (3.14) and (3.20), we can use (3.23) to calculate symbolically and in closed form the coefficients $E_\varphi^l(\eta, X)$, $E_\varphi^{l'}(\eta, X)$, $E_\varphi^{l''}(\eta, X)$, ..., $E_\varphi^y(\eta, X)$, $E_\varphi^{y'}(\eta, X)$, $E_\varphi^{y''}(\eta, X)$, ... entering in the series expansion (3.22) of the element strain. These coefficients depend on both $\eta$ and $X$ when the lattice has non-uniform geometrical properties in space, as happens for instance with the curved lattice shown in Figure 2.1b.

**Remark 3.4.** By (3.10), the first slots in $l$ correspond to a macroscopic rigid-body displacement (translation $u$ and rotation $\gamma$). The corresponding subblocks in $\boldsymbol{E}^l_\varphi(\eta,\boldsymbol{X})$, $\boldsymbol{E}^{l'}_\varphi(\eta,\boldsymbol{X})$ and $\boldsymbol{E}^{l''}_\varphi(\eta,\boldsymbol{X})$ are actually zero, reflecting the fact that the lattice elements (such as beams in the present case) are unaffected by rigid-body displacement:

- because of the zero blocks in (3.14)$_{2,3}$, the subblocks in $\boldsymbol{E}^{l'}_\varphi(\eta,\boldsymbol{X})$ and $\boldsymbol{E}^{l''}_\varphi(\eta,\boldsymbol{X})$ corresponding to $u$ and $\gamma$ are zero, see (3.23);
- in view of (3.14) and (3.23), the $u$ subblock in $\boldsymbol{E}^l_\varphi(\eta,\boldsymbol{X})$ is $\mathcal{D}_\varphi(\eta,\boldsymbol{X}) : \begin{pmatrix} \boldsymbol{I}_d & \boldsymbol{I}_d \\ \boldsymbol{0}_{n_r \times d} & \boldsymbol{0}_{n_r \times d} \end{pmatrix}$, which is zero when the expression of $\mathcal{D}_\varphi$ in (2.15) is used (invariance of the element strain $\boldsymbol{E}_\alpha$ by a rigid translation);
- the $\gamma$ subblock in $\boldsymbol{E}^l_\varphi(\eta,\boldsymbol{X})$ is $\mathcal{D}_\varphi(\eta,\boldsymbol{X}) : \begin{pmatrix} -\mathcal{N}^{T_{132}} \cdot \Delta^-_\varphi & -\mathcal{N}^{T_{132}} \cdot \Delta^+_\varphi \\ \boldsymbol{I}_{n_r} & \boldsymbol{I}_{n_r} \end{pmatrix}$, which can be shown to be zero by a similar argument (invariance of the element strain $\boldsymbol{E}_\alpha$ by a rigid rotation).

The only point of the macroscopic degrees of freedom $l$ being to compute the element strain by Equation (3.22), it is therefore possible to discard entirely the $u$ and $\gamma$ components in $l$, i.e., to amend (3.10) to

$$l(\boldsymbol{X}) = (\ \check{\boldsymbol{\varepsilon}}(\boldsymbol{X})\ ) \qquad \text{(alternate, reduced form of } l\text{)}. \tag{3.24}$$

In this case the length of $n_l$ in (3.11) must obviously be changed to $n_l = n_\varepsilon$. The library is flexible: the definition of $l$ can be selected using an option and we check that the expected subblocks are zero in any case.

**Remark 3.5.** In Equation (3.24), the macroscopic degrees of freedom $l$ coincides with the macroscopic strain $\varepsilon$. We prefer, however, to treat these notions as distinct, as we will occasionally include additional degrees of freedom into $l$, such as the pre-strain magnitude in Equation (7.9). In any case, $l$ collects all the variables on which the homogenized energy depends, so that the energy is always of the form (4.5).

### 3.6. Assembly

Next, we concatenate the elements' strain measures $\boldsymbol{E}_\varphi$ corresponding to the different element families $1 \leqslant \varphi \leqslant n_\varphi$, along with the left-hand side of the constraint (3.5) into a global strain vector $\boldsymbol{E}(\boldsymbol{X})$,

$$\boldsymbol{E}(\eta,\boldsymbol{X}) = \Big(\ \boldsymbol{E}_I(\eta,\boldsymbol{X})\ \ldots\ \boldsymbol{E}_\varphi(\eta,\boldsymbol{X})\ \ldots\ \boldsymbol{E}_{n_\varphi}(\eta,\boldsymbol{X})\ \ \frac{1}{\eta}\sum_{b=1}^{n_b}\boldsymbol{\xi}_b(\boldsymbol{X})\ \Big) \in \mathbb{R}^{n_E}, \tag{3.25}$$

whose dimension is

$$n_E = \left(\sum_{\varphi=I}^{n_\varphi} n_{E_\varphi}\right) + d. \tag{3.26}$$

By (3.1) and (3.6), it scales as

$$\boldsymbol{E}(\eta,\boldsymbol{X}) = O(\eta^0). \tag{3.27}$$

The honeycomb lattice shown in Figure 2.1, for instance, has $\boldsymbol{E} = \big(\ \varepsilon_I\ \kappa_I\ \tau_I\ \varepsilon_{II}\ \kappa_{II}\ \tau_{II}\ \varepsilon_{III}\ \kappa_{III}\ \tau_{III}\ \frac{\xi_1+\xi_2}{\eta} \cdot \boldsymbol{e}_1\ \frac{\xi_1+\xi_2}{\eta} \cdot \boldsymbol{e}_2\ \big)$ and $n_E = 11$.

The global strain vector $\boldsymbol{E}(\boldsymbol{X})$ plays a central role in the formulation of the homogenized problem:

- the strain $\boldsymbol{E}_\alpha$ in a particular element can be found by evaluating $\boldsymbol{E}(\eta,\boldsymbol{X})$ at the element center $\boldsymbol{X} = \boldsymbol{X}^c_\alpha$, and extracting the range of indices corresponding to element family $\varphi_\alpha$, see (3.25) and (3.21),

$$\boldsymbol{E}_\alpha = \Big(\ \boldsymbol{0}_{n_{E_{(\varphi_\alpha)}} \times p_\varphi}\ \ \boldsymbol{I}_{n_{E_{(\varphi_\alpha)}}}\ \ \boldsymbol{0}_{n_{E_{(\varphi_\alpha)}} \times p'_\varphi}\ \Big) \cdot \boldsymbol{E}(\eta,\boldsymbol{X}^c_\alpha) \tag{3.28}$$

where $p_\varphi = \sum_{\varphi'=I}^{\varphi-1} n_{E_{\varphi'}}$ is the number of slots before the $\boldsymbol{E}_\varphi$ block in (3.25) and $p'_\varphi = \sum_{\varphi'=\varphi+1}^{n_\varphi} n_{E_{\varphi'}} + d$ is the number of slots after the $\boldsymbol{E}_\varphi$ block;

- in view of (3.28), the elastic energy in (2.20–2.23) can be rewritten as

$$\Phi_d = \int_\Omega \frac{1}{2} \boldsymbol{E}(\eta,\boldsymbol{X}) \cdot \mathcal{H}_d(\eta,\boldsymbol{X}) \cdot \boldsymbol{E}(\eta,\boldsymbol{X})\, d\boldsymbol{X} \tag{3.29}$$

where $\mathcal{H}_d(\eta,\boldsymbol{X})$ is the global stiffness matrix,

$$\mathcal{H}_d(\eta,\boldsymbol{X}) = \sum_{\varphi=I}^{n_\varphi} \left(\sum_{\substack{\alpha \in \varphi \\ \boldsymbol{X}^c_\alpha \in \Omega}} \delta_D(\boldsymbol{X} - \boldsymbol{X}^c_\alpha)\right) \mathcal{H}^g_\varphi(\eta,\boldsymbol{X}) \in \mathbb{T}^{(n_E, n_E)}, \tag{3.30}$$

which is symmetric, $\delta_D$ denotes the Dirac distribution, and $\mathcal{H}^g_\varphi(\eta,\boldsymbol{X})$ represents the contribution from the element family $\varphi$ to the global stiffness matrix,

$$\mathcal{H}^g_\varphi(\eta,\boldsymbol{X}) = \begin{pmatrix} \boldsymbol{0}_{p_\varphi \times p_\varphi} & \boldsymbol{0}_{p_\varphi \times n_{E_\varphi}} & \boldsymbol{0}_{p_\varphi \times p'_\varphi} \\ \boldsymbol{0}_{n_{E_\varphi} \times p_\varphi} & \mathcal{H}_\varphi(\eta,\boldsymbol{X}) & \boldsymbol{0}_{n_{E_\varphi} \times p'_\varphi} \\ \boldsymbol{0}_{p'_\varphi \times p_\varphi} & \boldsymbol{0}_{p'_\varphi \times n_{E_\varphi}} & \boldsymbol{0}_{p'_\varphi \times p'_\varphi} \end{pmatrix} \in \mathbb{T}^{(n_E, n_E)}; \tag{3.31}$$

- the $n_c = d = 2$ kinematic constraints in (3.5) can be rewritten as

$$\mathcal{Q} \cdot \boldsymbol{E}(\eta,\boldsymbol{X}) = \boldsymbol{0} \quad \forall \boldsymbol{X} \in \Omega \tag{3.32}$$

where the constraint extraction matrix $\mathcal{Q}$ extracts the trailing $n_c$ in the expression (3.25) of $\boldsymbol{E}$,

$$\mathcal{Q} = \begin{pmatrix} \boldsymbol{0}_{(n_E-n_c)\times n_c} & \boldsymbol{I}_{n_c} \end{pmatrix} \in \mathbb{T}^{(n_c, n_E)}; \tag{3.33}$$

- by (3.22), (3.25) the assembled strain vector can be expressed in terms of $\boldsymbol{l}$, $\boldsymbol{y}$ and their gradients as

$$\begin{aligned} \boldsymbol{E}(\eta, \boldsymbol{X}) = {} & \boldsymbol{E}_l(\eta, \boldsymbol{X}) \cdot \boldsymbol{l}(\boldsymbol{X}) + \boldsymbol{E}_l'(\eta, \boldsymbol{X}) : \nabla \boldsymbol{l}(\boldsymbol{X}) + \boldsymbol{E}_l''(\eta, \boldsymbol{X}) \therefore \nabla^2 \boldsymbol{l}(\boldsymbol{X}) + \cdots \\ & + \boldsymbol{E}_y(\eta, \boldsymbol{X}) \cdot \boldsymbol{y}(\boldsymbol{X}) + \boldsymbol{E}_y'(\eta, \boldsymbol{X}) : \nabla \boldsymbol{y}(\boldsymbol{X}) + \boldsymbol{E}_y''(\eta, \boldsymbol{X}) \therefore \nabla^2 \boldsymbol{y}(\boldsymbol{X}) + \cdots \end{aligned} \tag{3.34}$$

where the coefficients $\boldsymbol{E}_l, \boldsymbol{E}_l', \boldsymbol{E}_l'', \boldsymbol{E}_y, \boldsymbol{E}_y'$ and $\boldsymbol{E}_y''$ are obtained by a simple assembly process, see (3.25) and (3.15), as

$$\boldsymbol{E}_l^{(i)}(\eta, \boldsymbol{X}) = \begin{pmatrix} \boldsymbol{E}_I^{l^{(i)}}(\eta, \boldsymbol{X}) \\ \vdots \\ \boldsymbol{E}_{n_\varphi}^{l^{(i)}}(\eta, \boldsymbol{X}) \\ \boldsymbol{0}_{n_c \times n_l \times d^i} \end{pmatrix}, \qquad \boldsymbol{E}_y^{(i)}(\eta, \boldsymbol{X}) = \begin{pmatrix} \boldsymbol{E}_I^{y^{(i)}}(\eta, \boldsymbol{X}) \\ \vdots \\ \boldsymbol{E}_{n_\varphi}^{y^{(i)}}(\eta, \boldsymbol{X}) \\ \begin{cases} (I_d\ \boldsymbol{0}_{d\times n_r}\ \cdots\ I_d\ \boldsymbol{0}_{d\times n_r}) & \text{if } i=0 \\ \boldsymbol{0}_{n_c\times n_y \times d^i} & \text{if } i>0 \end{cases} \end{pmatrix}. \tag{3.35}$$

**Remark 3.6.** Both $\boldsymbol{l}(\boldsymbol{X})$ and $\boldsymbol{y}(\boldsymbol{X})$ are of order $\eta^0$ by (3.12) and (3.17), and so is $\boldsymbol{E}(\eta, \boldsymbol{X})$ as well by (3.27). As a result, the leading-order coefficients in (3.34) are of order $\boldsymbol{E}_l = \mathcal{O}(\eta^0)$ and $\boldsymbol{E}_y = \mathcal{O}(\eta^0)$. A more detailed analysis shows that

$$\begin{aligned} \boldsymbol{E}_l(\eta, \boldsymbol{X}) &= \mathcal{O}(\eta^0), & \boldsymbol{E}_l'(\eta, \boldsymbol{X}) &= \mathcal{O}(\eta^1), & \boldsymbol{E}_l''(\eta, \boldsymbol{X}) &= \mathcal{O}(\eta^2), \\ \boldsymbol{E}_y(\eta, \boldsymbol{X}) &= \mathcal{O}(\eta^0), & \boldsymbol{E}_y'(\eta, \boldsymbol{X}) &= \mathcal{O}(\eta^1), & \boldsymbol{E}_y''(\eta, \boldsymbol{X}) &= \mathcal{O}(\eta^2). \end{aligned} \tag{3.36}$$

The increasing powers of $\eta$ from left to right in this table can be traced back to the presence of increasing powers of $\boldsymbol{\Delta} = \mathcal{O}(\eta)$ in the successive $\mathcal{T}$ operators appearing in (3.14) and (3.20), see (2.6).

### 3.7. Continualized strain energy

The energy $\Phi_d$ has been artificially rewritten as an integral in (3.29) but, because of the Dirac weights in (3.30), it is in fact a discrete sum over the element centers. By the Euler–MacLaurin formula [LP16], this discrete sum can be approximated to any desired order in $\eta$ by a 'genuine' integral plus boundary terms—we work with an error of order $\eta^3$ in the present work,

$$\Phi_d = \Phi + \Phi_{bt} + \mathcal{O}(\eta^3) \tag{3.37}$$

where $\Phi_{bt} = \oint_{\partial \Omega} \ldots \mathrm{d}a$ is the boundary term, and $\Phi$ is the *continualized* energy,

$$\Phi = \int_\Omega \frac{1}{2} \boldsymbol{E}(\eta, \boldsymbol{X}) \cdot \mathcal{H}(\eta, \boldsymbol{X}) \cdot \boldsymbol{E}(\eta, \boldsymbol{X}) \, \mathrm{d}\boldsymbol{X}, \tag{3.38}$$

obtained by taking formally the limit $\eta \to 0$ in (3.29–3.30),

$$\mathcal{H}(\eta, \boldsymbol{X}) = \sum_{\varphi=I}^{n_\varphi} \rho_\varphi(\eta, \boldsymbol{X}) \, \mathcal{H}_\varphi^g(\eta, \boldsymbol{X}). \tag{3.39}$$

Here, $\rho_\varphi(\eta, \boldsymbol{X})$ is the density of element centers for family $\varphi$ in the reference configuration space $\boldsymbol{X}$, see Figure 2.1b. It is given in terms of the known density $\tilde{\rho}_\varphi$ of element centers in the topological lattice space $\tilde{\boldsymbol{X}}$ (see Table 2.1) as

$$\rho_\varphi(\eta, \boldsymbol{X}) = \frac{\tilde{\rho}_\varphi}{\eta^d |J_f(\boldsymbol{f}^{-1}(\boldsymbol{X}))|}, \tag{3.40}$$

where $J_f = \det \nabla \boldsymbol{f}$ is the Jacobian of the diffeomorphism (2.3). Like $\tilde{\rho}_\varphi$, the density $\rho_\varphi$ is expressed per unit area if $d=2$, and per unit volume if $d=3$.

The proof of Equations (3.37–3.40) can be outlined as follows:

1. the energy in (3.29) is a sum by (3.30) of terms concerning a particular family $\varphi$;
2. each of the terms writes as $\sum_{a\in\varphi, \boldsymbol{X}_a^c\in\Omega} \frac{1}{2} \boldsymbol{E}(\eta, \boldsymbol{X}_a^c) \cdot \mathcal{H}_\varphi^g(\eta, \boldsymbol{X}_a^c) \cdot \boldsymbol{E}(\eta, \boldsymbol{X}_a^c)$ and is continualized separately;
3. with the help of the diffeomorphism, each term is rewritten as a sum over the (periodic) underlying topological lattice, namely $\sum_{a\in\varphi, \tilde{\boldsymbol{X}}_a^c\in\eta^{-1}\boldsymbol{f}^{-1}(\Omega)} \frac{1}{2} \boldsymbol{E}(\eta, \boldsymbol{f}(\eta\tilde{\boldsymbol{X}}_a^c)) \cdot \mathcal{H}_\varphi^g(\eta, \boldsymbol{f}(\eta\tilde{\boldsymbol{X}}_a^c)) \cdot \boldsymbol{E}(\eta, \boldsymbol{f}(\eta\tilde{\boldsymbol{X}}_a^c))$;
4. this is now a sum over a large domain in a periodic lattice, whose term is given by a slowly-variable function: by the Euler–MacLaurin formula, see Theorem 2.1(*ii*) by Le Floch and Pelayo [LP16], it can be rewritten as an integral plus boundary terms—this proof is constructive and explicit expressions are available:
   - the integral term thus obtained corresponds to taking the formal limit $\eta \to 0$ in (3.29–3.30) where the Dirac weights are replaced by their density;
   - the boundary terms are available as well but are not needed in the present paper;

5. the proof of Equations (3.37–3.40) follows by summing up the integrals and boundary terms coming from the different families.

This yields the integrand $\frac{1}{2} \boldsymbol{E}(\eta, \boldsymbol{X}) \cdot \mathscr{H}(\eta, \boldsymbol{X}) \cdot \boldsymbol{E}(\eta, \boldsymbol{X})$ announced in (3.38). We do not need the exact expression of the boundary term $\Phi_{\text{bt}}$ in the present paper, and therefore leave to future work both the details of the proof and the expression of $\Phi_{\text{bt}}$. In this future work, we will combine the boundary term $\Phi_{\text{bt}}$ with an analysis of the boundary layer to account for the boundary in an effective way.

Combining (2.21), (3.31), (3.39) and (3.40), we obtain an order-of-magnitude estimate for the continuous stiffness,

$$\mathscr{H}(\eta, \boldsymbol{X}) = \mathcal{O}(\eta^0). \tag{3.41}$$

Since $\boldsymbol{E}(\eta, \boldsymbol{X}) = \mathcal{O}(\eta^0)$ by (3.27), the continualized energy in (3.38) scales as

$$\Phi = \mathcal{O}(\eta^0). \tag{3.42}$$

The above scalings for $\mathscr{H}$ and $\Phi$ are convenient. They motivate the scaling assumption (2.21).

## 3.8. Summary: canonical form

In our previous work, we carried out homogenization of a generic elasticity problem, formulated in a special form called the *canonical form*, see Section 2.1 in [AL23]. In Section 3 above, we have rewritten the equilibrium of the lattice in canonical form, which will allow us to apply the homogenization procedure readily. The canonical form is nothing but

- the expansion (3.34) of the strain $\boldsymbol{E}$ in terms of $\boldsymbol{l}$, $\boldsymbol{y}$ and their gradients, which we can rewrite as $\boldsymbol{E} = \boldsymbol{E}(\eta, \boldsymbol{X}; \boldsymbol{l}(\boldsymbol{X}), \nabla \boldsymbol{l}(\boldsymbol{X}), \ldots; \boldsymbol{y}(\boldsymbol{X}), \nabla \boldsymbol{y}(\boldsymbol{X}), \ldots)$,
- the expression (3.32) of the kinematical constraint,
- the expression (3.38) of the energy.

We will homogenize the continualized energy $\Phi$, ignoring the boundary terms $\Phi_{\text{bt}}$ appearing in (3.37) during homogenization: the latter can only be interpreted in combination with an analysis of the boundary layers that form at edges of the lattice, which we leave for future work.

## 4. SECOND-ORDER HOMOGENIZATION

In this section, we summarize the homogenization procedure described in [AL23] which applies readily to the canonical form of lattice energy $\Phi = \Phi[\eta, \boldsymbol{l}, \boldsymbol{y}]$ derived in Section 3. Here, the square brackets refer to a *functional* dependence on the arguments $\boldsymbol{l}$ and $\boldsymbol{y}$.

## 4.1. Principle of the homogenization method

The macroscopic state vector $\boldsymbol{l}$ is equal to the macroscopic strain $\check{\boldsymbol{\varepsilon}}$, see (3.24) and (3.7). However, to make future extensions easier, we treat $\boldsymbol{l}$ and $\check{\boldsymbol{\varepsilon}}$ as distinct entities, viewing $\boldsymbol{l}$ as the list of degrees of freedom that are fixed during homogenization (which may be more than just the macroscopic strain $\check{\boldsymbol{\varepsilon}}$).

The homogenization procedure described in our previous work [AL23] can be summarized as follows. With $\boldsymbol{l}$ a prescribed function of $\boldsymbol{X}$, we revisited homogenization by working at the energy level and showed that the microscopic degrees of freedom $\boldsymbol{y}(\boldsymbol{X})$ are the solution of a stationary-point problem,

$$\boldsymbol{y}^\star[\eta, \boldsymbol{l}] = \underset{\boldsymbol{y} \text{ such that } \mathcal{Q} \cdot \boldsymbol{E} = \boldsymbol{0} \, \forall \boldsymbol{X}}{\text{stpt}} \Phi[\eta, \boldsymbol{l}, \boldsymbol{y}]. \tag{4.1}$$

This stationary-point problem concerns interior points of the domain $\Omega$ only, and proceeds order by order in the expansion parameter $\eta$, the gradient effect being captured as a higher-order correction in $\eta$. The canonical form derived in Section 3 provided us with the energy functional $\Phi[\eta, \boldsymbol{l}, \boldsymbol{y}]$ which we can now plug into the homogenization engine developed in our previous work [Aud23].

The stationary point $\boldsymbol{y}^\star$ obtained in (4.1) is a functional of $\boldsymbol{l}$, as indicated by the square brackets. It is also a function of $\boldsymbol{X}$ whose values are denoted as $\boldsymbol{y}^\star[\eta, \boldsymbol{l}](\boldsymbol{X})$. Equation (4.1) effectively slaves the microscopic degrees of freedom $\boldsymbol{y}$ to the macroscopic state $\boldsymbol{l}$.

A homogenized energy functional $\Phi^\star$ is obtained by inserting the stationary point $\boldsymbol{y}^\star$ into the original energy,

$$\Phi^\star[\eta, \boldsymbol{l}] = \Phi[\eta, \boldsymbol{l}, \boldsymbol{y}^\star[\eta, \boldsymbol{l}]]. \tag{4.2}$$

The stationary point problem (4.1) is similar to that governing the *equilibrium* of the lattice, except that:

- only $\boldsymbol{y}$ is allowed to vary, $\boldsymbol{l}$ being kept fixed;
- the potential of the external loading $\Psi$ is ignored, only the strain energy $\Phi$ being considered;
- the energy is written in the domain $\Omega$ strictly contained in the physical domain of the lattice, see Figure 2.3, and boundary terms $\Phi_{\text{bc}}$ are ignored.

Once homogenization is complete, the following additional steps should be addressed: the boundary layers taking place outside the domain $\Omega$ must be analyzed, the boundary terms $\Phi_{\mathrm{bc}}$ must obtained, and the total potential energy—including the potential of the external loading $\Psi$—must be optimized with respect to $l$. When this is done, the proper equilibrium of the lattice is recovered, warranting that the homogenization procedure summarized in (4.1–4.2) is asymptotically consistent. This paper focusses on the derivation of the homogenized model $\Phi^\star$: its application to the solution of complete structural problems will be treated in future work.

**Remark 4.1.** The motivation for ignoring the potential of the external loading $\Psi$ during homogenization is as follows. It follows from standard assumptions on the magnitude of the load that the potential $\Psi[\eta, l]$ is a function of $l$ but not of $y$: since $l$ is fixed during homogenization, it is then justified to ignore the external potential $\Psi[\eta, l]$ during homogenization, and to restore it in the final step, when the total energy $\Phi^\star[\eta, l] + \Psi[\eta, l]$ is relaxed with respect to $l$—see also the discussion in Section 3.1 in our previous work [AL23].

The homogenization method treats the scale separation parameter $\eta \ll 1$ as an expansion parameter and solves the stationary point problem (4.1) order by order in $\eta$: the optimal microscopic displacement $y^\star$ is obtained in the form of an expansion,

$$y^\star[\eta, l](X) = Y(\eta, X) \cdot l(X) + Y'(\eta, X) : \nabla l(X) + \mathcal{O}(\eta^2), \tag{4.3}$$

where the localization tensors $Y$ and $Y'$ are an output in symbolic form by the homogenization procedure. They scale as

$$Y(\eta, X) = \mathcal{O}(\eta^0), \qquad Y'(\eta, X) = \mathcal{O}(\eta^1). \tag{4.4}$$

implying that the first term in the right-hand side of (4.3) is the leading-order term, whose order of magnitude agrees with (3.17), while the second term is a corrector capturing the gradient effect.

The homogenized energy $\Phi^\star$ in (4.2) is obtained in the form of an expansion as well, which can be written in compact form as

$$\begin{aligned}\Phi^\star[l] = & \int_\Omega \left( K(\eta, X) : \frac{l(X) \otimes l(X)}{2} + A(\eta, X) \vdots (l(X) \otimes \nabla l(X)) + B(\eta, X) t :: \frac{\nabla l(X) \otimes \nabla l(X)}{2} \right) dX + \\ & \oint_{\partial\Omega} \left[ k(\eta, X) \vdots \left( \frac{l(X) \otimes l(X)}{2} \otimes n(X) \right) + a(\eta, X) t :: (l(X) \otimes \nabla l(X) \otimes n(X)) \right] da + \mathcal{O}(\eta^3). \end{aligned} \tag{4.5}$$

The boundary integral $\oint_{\partial\Omega} \ldots da$ in the second line takes place over the boundary $\partial\Omega$ of the domain, with $n$ denoting the unit outward normal and $da$ the area (if $d=3$) or the length (if $d=2$) of a boundary element, see Figure 2.3. The special integral sign $\oint$ will be used throughout for boundary integrals. Note that the boundary terms appearing in (4.5) as a result of homogenization are different from those that appeared earlier in (3.37) as a result of continualizing the energy: they must be summed up.

The homogenization procedure delivers the tensors characterizing the homogenized energy in symbolic form. Their orders of magnitude are given by

$$K(\eta, X) = \mathcal{O}(\eta^0), \quad A(\eta, X) = \mathcal{O}(\eta^1), \quad B(\eta, X) = \mathcal{O}(\eta^2), \quad k(\eta, X) = \mathcal{O}(\eta^2), \quad a(\eta, X) = \mathcal{O}(\eta^2). \tag{4.6}$$

**Remark 4.2.** When the microscopic degrees of freedom and energy $\Phi^\star$ are truncated to leading-order $\mathcal{O}(\eta^0)$, we obtain the approximations

$$y^\star[\eta, l](X) = Y(\eta, X) \cdot l(X) + \mathcal{O}(\eta^1), \qquad \Phi^\star[\eta, l] = \int_\Omega K(\eta, X) : \frac{l(X) \otimes l(X)}{2} dX + \mathcal{O}(\eta^1)$$

that correspond to classical homogenization, in which the effect of the gradient $\nabla l$ is ignored [CMR06; DO11; CP12; Dav13; LR13].

**Remark 4.3.** The homogenization method producing Equations (4.3–4.5) has been presented in our previous work [AL23] using different scaling conventions: the microscopic cell size, now $\mathcal{O}(\eta^1)$, was then $\mathcal{O}(\eta^0)$ and the lattice size, now $\mathcal{O}(\eta^0)$, was then $\mathcal{O}(\eta^{-1})$. The two approaches are ultimately equivalent. The scaling relations announced in (4.4) and (4.6) are obtained by inserting the estimates (3.36) for $E_l$, $E'_l$, etc., into the homogenization formula from Appendix C in [AL23]: in Equation [C.9] in this Appendix, for instance, $\mathcal{J}^1 = \mathcal{O}(E'_l, E'_y) = \mathcal{O}(\eta^1)$ while $\mathcal{J}^{11} = \mathcal{O}(E''_y) = \mathcal{O}(\eta^2)$ and similarly $\mathcal{J}^2 = \mathcal{O}(E''_l, E''_y) = \mathcal{O}(\eta^2)$. The tensor $\mathcal{A}$ in [C.19] capturing the first-order gradient effect is then $\mathcal{O}(\eta)$, like the quantity $A_0 \sim A$ that derives from it, see [C.21]. A similar analysis shows that the quantities derived at second order, such as $Y'$, $B$, etc., are all $\mathcal{O}(\eta^2)$.

### 4.2. Connecting with the discrete homogenization library

The homogenization of elastic lattices is implemented and distributed as an extension of the general-purpose homogenization library shoal [Aud23] described in Section 4.1. Both components of the library are written in the symbolic calculation language Wolfram Mathematica. The extension is implemented as follows.

In terms of the parameters describing the lattice listed in Table 4.1, we calculate the following quantities in symbolic form using Wolfram Mathematica [21]:

- the tensors $\mathcal{M}$, $\mathcal{N}$, $\mathcal{U}'$, $\mathcal{U}''$, $\mathcal{U}'''$, $\mathcal{G}'$ and $\mathcal{G}''$ depending on space dimension $d$, from Appendix B;

- the tensors $\mathcal{T}_{l^{(i)}}$, $\mathcal{T}_{y^{(i)}}^b$ and $\mathcal{D}_\varphi$ using (3.14), (3.20) and (2.15), respectively;

- the tensors $\boldsymbol{E}_\varphi^{l^{(i)}}$ and $\boldsymbol{E}_\varphi^{y^{(i)}}$ using (3.23) and then the tensors $\boldsymbol{E}_l^{(i)}$ and $\boldsymbol{E}_y^{(i)}$ using (3.35);

- the element stiffness matrix $\mathcal{H}_\varphi$, its contribution $\mathcal{H}_\varphi^g$ to the global stiffness and the global stiffness $\mathcal{H}$ itself using (2.19), (3.31) and (3.39), respectively.

| | | |
|---|---|---|
| underlying topological lattice (see Table 2.1) | $d$ | space dimension |
| | $n_b$ | number of Bravais sub-lattices |
| | $n_\varphi$ | number of element families |
| | $b_\varphi^\pm$ | elements connectivity |
| reference configuration | $\rho_\varphi(\eta, \boldsymbol{X})$ | density of element family |
| | $\boldsymbol{\Delta}_\varphi^\pm(\eta, \boldsymbol{X})$ | endpoint position with respect to element center |
| elastic properties | $\widehat{EA}_\varphi(\eta, \boldsymbol{X})$ | traction modulus |
| | $\hat{\chi}_\varphi(\eta, \boldsymbol{X})$ | beam aspect-ratio parameter |
| variable material parameters | $\boldsymbol{m}$ | (see Remark 7.1) |

**Table 4.1.** Lattice specification used as an input to the homogenization procedure.

Next, we feed these quantities to the homogenization engine proposed in our previous work. Specifically, we process and wrap up the data provided in Table 4.1 to produce the following list of arguments that are passed to the homogenization engine—we are still considering the generic beam lattice from Figure 2.1 for the purpose of illustration:

- the space dimension $d = 2$, the dimension $n_l = 3$ of the macroscopic state vector, the number $n_y = 2 \times 3 = 6$ of microscopic degrees of freedom, the dimension $n_E = 11$ of the global strain vector and the numbers $n_c = 2$ of kinematical constraint, see (3.11), (3.16), (3.26) and (3.33)$_2$;

- the tensorial functions $\boldsymbol{E}_l(\eta, \boldsymbol{X})$, $\boldsymbol{E}_l'(\eta, \boldsymbol{X})$, $\boldsymbol{E}_l''(\eta, \boldsymbol{X})$, $\boldsymbol{E}_y(\eta, \boldsymbol{X})$, $\boldsymbol{E}_y'(\eta, \boldsymbol{X})$ and $\boldsymbol{E}_y''(\eta, \boldsymbol{X})$ of the strain expansion (3.34);

- the constraint extraction matrix $\mathcal{Q}$ in (3.33)$_1$;

- the global stiffness matrix $\mathcal{H}(\eta, \boldsymbol{X})$ in (3.39), in symbolic form;

- the list $\boldsymbol{m}$ of variable material parameters declared by the user is forwarded to the homogenization engine.

This list $\boldsymbol{m}$ declares all the symbols entering in the expressions of $\mathcal{H}$, $\boldsymbol{E}_l$, $\boldsymbol{E}_l'$, $\boldsymbol{E}_l''$, $\boldsymbol{E}_y$, $\boldsymbol{E}_y'$ or $\boldsymbol{E}_y''$, that vary in space. For lattices having homogeneous properties, $\boldsymbol{m} = \{\}$ is set to be an empty list. For lattices having graded properties, one can either work out the explicit dependence of $\mathcal{H}$, $\boldsymbol{E}_l$, ... on $\boldsymbol{X}$ as in (3.34) and set $\boldsymbol{m} = \{X_1, \ldots, X_d\}$, or use the simpler approach discussed in §7.1–7.3.

**Remark 4.4.** The explicit dependence of the various quantities on the scale-separation parameter $\eta$ was implicit in our previous work [AL23] and is now denoted explicitly by the argument $\eta$. As long as the orders of magnitude are preserved, nothing changes when the input to the homogenization procedure depends on $\eta$.

**Remark 4.5.** The homogenization procedure requires that the energy is positive-definite over the subspace of admissible microscopic degrees of freedom, in the sense that $(\boldsymbol{E}_y(\eta, \boldsymbol{X}) \cdot \boldsymbol{y}) \cdot \mathcal{H}(\eta, \boldsymbol{X}) \cdot (\boldsymbol{E}_y(\eta, \boldsymbol{X}) \cdot \boldsymbol{y}) > 0$ must be strictly positive for any $\boldsymbol{y}$ such that $\boldsymbol{Q} \cdot \boldsymbol{E}_y(\eta, \boldsymbol{X}) \cdot \boldsymbol{y} = \boldsymbol{0}$ (see equation (11) in our previous work [AL23]). We rely on the homogenization engine to check this condition. Whenever present, offending microscopic vectors $\boldsymbol{y}$ are reported as 'soft modes' and the engine aborts with an error message. In future work we will cover how soft modes can be dealt with: this requires treating them as macroscopic degrees of freedom, *i.e.*, incorporating them into the vector $\boldsymbol{l}$.

## 5. SYMBOLIC HOMOGENIZATION OF A PERIODIC HONEYCOMB LATTICE

In this section, a first concrete application of the homogenization scheme is presented. Leaving more advanced examples for the following sections, we start with a simple lattice geometry: we address the *periodic* (non-curved) variant of the honeycomb lattice shown in Figure 2.1, limiting attention to the case of *uniform* elastic properties in space. We derive a simple closed-form expression for its second-order homogenized energy, see Equation (5.9) below. Its leading-order is consistent with existing results from the literature but we also include a higher-order contribution which is novel to the best of our knowledge.

### 5.1. Setting up the lattice in the extensible case

The *periodic* honeycomb lattice is generated using the method described in Section 2, with the diffeomorphism set to $\boldsymbol{f} = \mathbf{id}_2$ (identity in the plane). The properties of topological honeycomb lattice listed in Table 2.1 are still used. The resulting lattice is shown in Figure 5.1.

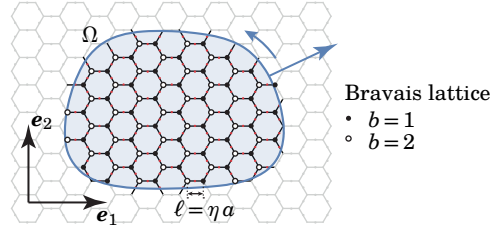

**Figure 5.1.** Undeformed configuration of the periodic honeycomb lattice having uniform properties, used as an input to the homogenization procedure. For this un-curved lattice, the elements centers (red dots) coincide with the midpoints.

Since a contraction factor $\eta$ is applied to the underlying topological lattice (whose beam length is $a = \mathcal{O}(1)$), the generated lattice has beam length $\ell = \eta a \ll 1$, see also Figure 2.1b. In view of the value of $\tilde{\rho}_\varphi = \frac{2}{3\sqrt{3}a^2}$ in Table 2.1 and of Equation (3.40), the element density $\rho_\varphi$ is given, for each one of the three element families $\varphi \in \{I, II, III\}$, by

$$\rho_\varphi = \frac{2}{3\sqrt{3}\,\ell^2}. \tag{5.1}$$

With $\boldsymbol{f} = \mathbf{id}_2$, the element centers $\boldsymbol{X}_a^c$ coincide with the midpoints (red dots in the figure), see (2.5).

The homogenization of the honeycomb lattice is carried out in the companion Mathematica notebook `homogenize-extensible-honeycomb.nb` which is distributed along with the library. The quantities appearing in Table 4.1 are passed on input:

- the integers $d = 2$, $n_b = 2$, $n_\varphi = 3$ and the array $b_\varphi^\pm$ characterizing the underlying topological lattice, see Table 2.1,

- the reference configuration is characterized by $\rho_\varphi$ from (5.1) and by the relative position $\boldsymbol{\Delta}_\varphi^\pm = \pm \frac{\eta \tilde{\boldsymbol{\delta}}_\varphi}{2}$ of the endpoints with respect to the element center (midpoints), where $\tilde{\boldsymbol{\delta}}_\varphi$ is available from Table 2.1,

- each one of the $n_\varphi = 3$ element type is assigned the stiffness matrix $\mathcal{H}_\varphi$ in (2.19) applicable to extensible beams, with *constant* material parameters; indeed, for this particular lattice, the stretching modulus $EA$, the beam length $\ell = \eta a$ and the aspect ratio parameter $\chi$ are all (i) invariant in space and (ii) identical for all three element families $\varphi \in \{I, II, III\}$,

- the variable material parameters $\boldsymbol{m} = \{\}$ is set up as an empty list since the lattice is homogeneous.

The parameters $EA$, $\ell$ and $\chi$ are treated as *symbolic* parameters by the library, and the homogenized energy is delivered in the form of a *function* of $EA$, $\ell$ and $\chi$.

**Remark 5.1.** We noted earlier that the aspect-ratio parameter $\chi \ll 1$ needs to be small for the beam theory to be applicable, see the discussion below Equation (2.17). Mathematically, $\chi$ has to be linked to the small parameter $\eta$, see Remark 2.2. Remarkably, we can carry out homogenization without having to specify how $\chi$ scales with $\eta$: the only assumption we made is $\chi = \mathcal{O}(\eta^0)$ and this does not exclude more specific assumptions such as $\chi = \mathcal{O}(\eta^1)$ or $\chi = \mathcal{O}(\eta^2)$. Delaying the choice of this scaling law until after homogenization is complete enables us to carry out homogenization once for all, and to identify the interesting scaling regimes based on the homogenized energy, see for instance the pantograph application example in Section 7.5. In bending-dominated lattices such as the honeycomb, the inextensible limit can be obtained simply by taking the limit $\chi \to 0$ in the homogenized energy (§5.3).

## 5.2. Homogenization results

This section is a summary of the symbolic homogenization results obtained in the companion Mathematica notebook `homogenize-extensible-honeycomb.nb`. The library yields the homogenized energy $\Phi^\star[\boldsymbol{l}]$ in the form (4.5) that makes use of the strain *vector* $\boldsymbol{l} = \check{\boldsymbol{\varepsilon}}$ in Mandel representation, see (3.7) and (3.24). We systematically rewrite it in terms of the standard 2D strain *tensor* $\boldsymbol{\varepsilon}$ with the help of (B.4).

The homogenization results can be summarized as follows, order by order:

- **Leading order.** The equivalent elastic continuum is isotropic,

$$\begin{aligned}\Phi^\star &= \frac{1}{2}\int_\Omega \frac{1}{2}\check{\boldsymbol{\varepsilon}} \cdot \boldsymbol{K} \cdot \check{\boldsymbol{\varepsilon}}\, \mathrm{d}\boldsymbol{X} + O(\eta)\\ &= \frac{1}{2}\int_\Omega (\lambda \operatorname{tr}^2 \boldsymbol{\varepsilon} + 2\mu\, \boldsymbol{\varepsilon} : \boldsymbol{\varepsilon})\, \mathrm{d}\boldsymbol{X} + O(\eta),\end{aligned} \tag{5.2}$$

and its Lamé parameters are given by

$$\lambda = \frac{\sqrt{3}\, EA}{6\,\ell}\frac{1-12\chi}{1+12\chi}, \qquad \mu = \frac{4\sqrt{3}\, EA}{\ell}\frac{\chi}{1+12\chi}. \tag{5.3}$$

This matches the previous results of [VDP14; GLTK19]. In the slender limit ($\chi \to 0$), we have $\mu = \mathcal{O}(EA\,\chi/\ell) = \mathcal{O}(EI/\ell^3) \ll \lambda = \mathcal{O}(EA/\ell)$, from which we conclude that (i) the lattice is bending-dominated and not stretching-dominated (due to $\mu \propto EI$) and (ii) is incompressible in this limit (due to $\lambda \propto EA$, see (5.8) below).

The microscopic displacement underlying this leading-order energy prediction (5.2–5.3) is obtained from the localization tensor $\boldsymbol{Y}$ appearing in (4.3) as

$$(\boldsymbol{\xi}_b)^\star_{[0]} = (-1)^{b+1} \frac{\ell}{4} \frac{1 - 12\chi}{1 + 12\chi} \boldsymbol{\mathcal{J}} : \boldsymbol{\varepsilon}, \qquad (\psi_b)^\star_{[0]} = 0. \tag{5.4}$$

where $(\boldsymbol{\xi}_b)^\star_{[0]}$ and $(\psi_b)^\star_{[0]}$ are the leading-order microscopic displacement and rotation, respectively, of the Bravais sub-lattice $b \in \{1,2\}$, and $\boldsymbol{\mathcal{J}}$ is a constant tensor associated with the $D_6$ symmetry,

$$\boldsymbol{\mathcal{J}} = \boldsymbol{e}_1 \otimes (-\boldsymbol{e}_1 \otimes \boldsymbol{e}_1 + \boldsymbol{e}_2 \otimes \boldsymbol{e}_2) + \boldsymbol{e}_2 \otimes (\boldsymbol{e}_1 \otimes \boldsymbol{e}_2 + \boldsymbol{e}_2 \otimes \boldsymbol{e}_1). \tag{5.5}$$

In (5.4), $(\psi_b)^\star_{[0]} = 0$ means that the nodal rotation $\theta_b = \gamma$ matches the macroscopic rotation $\gamma = -\frac{1}{2} \nabla \boldsymbol{u} : \boldsymbol{\mathcal{N}}$ set by the macroscopic displacement $\boldsymbol{u}(\boldsymbol{X})$, and not that the node rotation is zero—see (3.4)$_2$.

- **First order.** When pushed to first order, the homogenization procedure delivers the homogenized tensors $\boldsymbol{A} = \boldsymbol{0}$ and $\boldsymbol{k} = \boldsymbol{0}$. In view of (4.5), and as usual with centrosymmetric lattices [ABB10], this implies that the energy has no contribution of order $\eta$. As a result, the leading-order approximation (5.2) is exact up to $\mathcal{O}(\eta^2)$ and not just $\mathcal{O}(\eta)$.

- **Second order.** In the Mathematica notebook, both the bulk term $\boldsymbol{B} t :: \frac{\nabla \boldsymbol{l} \otimes \nabla \boldsymbol{l}}{2}$ and the boundary term $\boldsymbol{a} t :: (\boldsymbol{l} \otimes \nabla \boldsymbol{l} \otimes \boldsymbol{n})$ entering at order $\eta^2$ in the energy (4.5) have been worked out in terms of invariants relevant to the $D_6$ symmetry of the hexagonal network: there are 6 invariants for the surface integral, and 5 for the boundary integral. The full expressions are available in the Mathematica notebook but they are cumbersome and are not included here—their values in the inextensible limit $\chi \to 0$ are nevertheless given in Section 5.3.

  The first-order correction to the microscopic displacement $(\boldsymbol{\xi}_b)^\star_{[1]}$ and $(\psi_b)^\star_{[1]} = \boldsymbol{0}$ that underlies this second-order energy correction is encoded in the localization tensor $\boldsymbol{Y}'$ appearing in (4.3): interpreting the result, we get

$$(\boldsymbol{\xi}_b)^\star_{[1]} = \boldsymbol{0}, \qquad (\psi_b)^\star_{[1]} = -(-1)^{b+1} \frac{3\ell}{4(1+12\chi)} \operatorname{curl}(\boldsymbol{\mathcal{J}} : \boldsymbol{\varepsilon}), \tag{5.6}$$

where $\operatorname{curl} \boldsymbol{a} = \partial a_2/\partial X_1 - \partial a_1/\partial X_2$ denotes the (scalar) curl of a vector field $\boldsymbol{a}(\boldsymbol{X})$ in 2D.

## 5.3. Inextensible limit

We now take the slender limit $\chi \to 0$ in the homogenization results from Section 5.2. Specifically, we address the distinguished limit in which the slenderness parameter $\chi$ and the scale separation parameter $\eta$ both go to zero with

$$(EA)\chi = \mathcal{O}(\eta). \tag{5.7}$$

The equivalent compression modulus $\lambda \sim \frac{EA}{\ell} = \mathcal{O}(\chi^{-1}) \to \infty$ in (5.3) then becomes infinite while the shear modulus $\mu \sim \frac{\chi EA}{\ell} \sim \mathcal{O}(1)$ remains finite. For better legibility of the results, we eliminate $EA$ in favor of $EI$ everywhere using $EA = EI/(\chi \ell^2)$, see (2.17), anticipating that in the limit, the beam elements are effectively inextensible and the lattice deforms by pure bending.

In this limit $\lambda \to \infty$ in (5.3)$_1$, implying that the equivalent Cauchy medium is incompressible,

$$\operatorname{tr} \boldsymbol{\varepsilon}(\boldsymbol{X}) = 0, \quad \forall \boldsymbol{X}. \tag{5.8}$$

The homogenized energy is found by combining the shearing term in leading-order correction (5.2–5.3) with the second-order correction described in words in Section 5.2, taking the kinematic constraint of incompressibility (5.8) into account to remove the apparent divergences that arise when $\chi \to 0$. The result is obtained in the companion Mathematica notebook in the form

$$\begin{aligned}
\Phi^\star = \frac{EI}{\ell^3} \Bigg( & \frac{1}{2} \int_\Omega 2 \times 4\sqrt{3}\, \boldsymbol{\varepsilon} : \boldsymbol{\varepsilon} \, d\boldsymbol{X} \\
& + \frac{\ell^2}{2} \int_\Omega \sqrt{3} \left( -7 \nabla \boldsymbol{\varepsilon} \therefore \nabla \boldsymbol{\varepsilon} + \frac{22}{3} \|\nabla \boldsymbol{\varepsilon} : \boldsymbol{I}_2\|^2 + 2 \left( \operatorname{div}(\boldsymbol{\mathcal{J}} : \boldsymbol{\varepsilon}) \right)^2 \right) d\boldsymbol{X} \\
& + \ell^2 \oint_{\partial \Omega} \frac{\sqrt{3}}{2} \left( 13 \nabla \boldsymbol{\varepsilon} : \boldsymbol{\varepsilon} - \frac{7}{2} \operatorname{div}(\boldsymbol{\mathcal{J}} : \boldsymbol{\varepsilon})\, \boldsymbol{\mathcal{J}} : \boldsymbol{\varepsilon} \right) \cdot \boldsymbol{n} \, dS + \mathcal{O}(\eta^3),
\end{aligned} \tag{5.9}$$

where $\operatorname{div} \boldsymbol{a} = \partial a_1/\partial X_1 + \partial a_2/\partial X_2$ denotes the divergence of a 2D vector field $\boldsymbol{a}(\boldsymbol{X})$. The successive lines in the right-hand side represent the leading-order contribution—of order $EI/\ell^3 \sim EA\chi/\ell = \mathcal{O}(1)$ by (5.7)—, the surface integral for the second-order correction, and the boundary term for the second-order correction, respectively. In (5.9), $\boldsymbol{I}_2$ denotes the identity matrix in dimension 2 and the constant tensor $\boldsymbol{\mathcal{J}}$ has been defined in (5.5). Due to this $\boldsymbol{\mathcal{J}}$ term, the equivalent strain-gradient continuum defined by $\Phi^\star$ is $D_6$-symmetric, like the original honeycomb itself, and the effective isotropy obtained at the leading order is no longer applicable [RA16; RA19].

The analytical, second-order expansion (5.9) of the honeycomb energy is novel to the best of our knowledge.

In the inextensible limit, the microscopic displacement found in (5.4–5.6) becomes

$$\begin{aligned}
\boldsymbol{\xi}^\star_b &= \ell \left( \frac{(-1)^{b+1}}{4} \boldsymbol{\mathcal{J}} : \boldsymbol{\varepsilon}\, \ell^0 + 0\, \ell^1 + \mathcal{O}(\ell^2) \right), \\
\psi^\star_b &= 0\, \ell^0 - \frac{3(-1)^{b+1}}{4} \operatorname{curl}(\boldsymbol{\mathcal{J}} : \boldsymbol{\varepsilon})\, \ell^1 + \mathcal{O}(\ell^2).
\end{aligned} \tag{5.10}$$

**Remark 5.2.** As mentioned earlier, the homogenized energy $\Phi^\star$ approximates only the integral term $\Phi$ that is produced by the continualization of the discrete energy $\Phi_d$ but not its companion boundary term $\Phi_{bt}$, see (3.37). The boundary terms in (5.9) must therefore be added up to $\Phi_{bt}$. This will be covered in future work, along with an analysis of the boundary layers.

**Remark 5.3.** The strain-gradient energy appearing in the second line of (5.9) is not positive, as is often the case in higher-order homogenization. As discussed at the end of Section 4.1 of our previous paper [AL23], this lack of positivity calls for a regularization of the functional, a point which we will address in future work.

## 5.4. Short track: homogenizing the inextensible lattice

In the previous sections (§5.1–5.3), we have homogenized a honeycomb lattice made up of extensible beams first, and taken the inextensible limit $\chi \to 0$ in the homogenized energy next. Here, we show how it is possible to directly homogenize a honeycomb lattice made up of inextensible beams. This yields the same results more elegantly.

The inextensible case is treated in a second Mathematica notebook, also distributed with the library, named `homogenize-inextensible-honeycomb.nb`. The input to the homogenization is identical to that for the extensible lattice (§5.1), except that:

- the beam elements are allocated with the keyword "`inextensibleBeam`" (as opposed to "`beam`" in the extensible case) and make use of the properties $EI$ and $\ell$ only (no $EA$ or $\chi$ parameter),
- an option `rankDeficiency=1` is passed to the library.

If the `rankDeficiency` option is overlooked, the solution of the leading-order problem fails as the underlying system of linear equations is deficient (with rank 1) in the inextensible case: the library then stops with an error, identifies the incompressibility condition (5.8) as the root of the deficiency, and suggests that the procedure is run again with `rankDeficiency=1`.

With the option `rankDeficiency=1`, the library extends the vector $\boldsymbol{l}$ representing the macroscopic degrees of freedom with a fourth slot $p = l_4$: using block-matrix notation, we have now

$$\boldsymbol{l} = (\,\check{\boldsymbol{\varepsilon}}\ \ p\,), \tag{5.11}$$

where $\check{\boldsymbol{\varepsilon}}$ are the usual, 3 components of the 2D macroscopic strain tensor $\boldsymbol{\varepsilon}$ arranged in a vector, see (3.7). The homogenized energy (4.5) produced by the library must be interpreted in the light of this new definition of $\boldsymbol{l}$. The new variable $p$ is the Lagrange multiplier associated with the incompressibility condition (5.8) and can be interpreted as a pressure.

The output for the inextensible honeycomb lattice, as worked out in the notebook, can be summarized as follows:

- A solvability condition on $\boldsymbol{l}$ is issued by the homogenization engine, which, upon examination, is nothing but the incompressibility condition (5.8). This solvability condition warrants that the right-hand sides of the linear problems involving the rank-deficient matrix are indeed part of its image, so that the inversion of the rank-deficient problem is mathematically possible: see Equation [E.11] in [AL23].
- The elasticity tensors $\boldsymbol{K}$, $\boldsymbol{A}$, $\boldsymbol{B}$ and $\boldsymbol{k}$ and localization tensors $\boldsymbol{Y}$, $\boldsymbol{Y}'$ are calculated. When interpreted with the help of (4.3) and (4.5), they yield the same energy (5.9) and microscopic displacement (5.10) as those from Section 5.3.

In the homogenization library, inextensible lattices are implemented as follows:

- For each element family $\varphi \in \{I, II, III\}$, the inextensibility constraint $\varepsilon_\varphi = 0$ is handled by adding one row to the constraint matrix $\mathcal{Q}$ appearing in (3.32–3.33). Each one of these new rows is filled with 0's, except for a single entry equal to 1 in the column whose index corresponds to the position of the element strain $\varepsilon_\varphi$ in $\boldsymbol{E}$, warranting that the generic form of the kinematical constraint $\mathcal{Q} \cdot \boldsymbol{E} = 0$ used throughout now includes the inextensibility constraints $\varepsilon_\varphi = 0$.
- The element stiffness in (2.19) is changed to

$$\mathcal{H}_\varphi(\eta, \boldsymbol{X}) = \frac{\widehat{EI}_\varphi(\eta, \boldsymbol{X})}{\hat{\ell}_\varphi(\eta, \boldsymbol{X})} \mathrm{diag}(0, 1, 12) \in \mathbb{T}^{(3,3)} \qquad \text{(2D inextensible beam)}, \tag{5.12}$$

where $\widehat{EI}_\varphi(\eta, \boldsymbol{X})$ is the prescribed map of elastic (bending) modulus—we set $\widehat{EI}_\varphi(\eta, \boldsymbol{X}) = (EI)$ in the present case of a homogeneous lattice. The value (0) appearing in first position of the diagonal matrix in the right-hand side is irrelevant: this is the modulus of the stretching mode which is inhibited by the inextensibility constraint anyway.

The rest of the homogenization procedure is unchanged. Inextensible lattices make use the ability of the homogenization engine to deal with rank-deficient linear problems, as documented in Appendix E of our previous paper [AL23].

## 6. NUMERICAL VERIFICATION

In this section, we set up a numerical procedure to test both the asymptotic validity of our homogenization scheme in the limit of vanishingly small cell size, and its accuracy for small but fixed cell size. We start in this section with honeycomb lattices subjected to various boundary conditions and loading types. More advanced illustrations will be considered in Section 7.

Near the physical edges of the lattice, boundary layers form and the homogenization breaks down [Dum86; LM18]. Until we properly address these layers, they prevent us from making accurate *global* predictions of the equilibrium solution based on the homogenized theory (4.5)—the analysis of the boundary layers is a significant piece of work and we leave it to a follow-up paper. In the meantime, we give up on making predictions based on the stand-alone homogenized theory (4.5) and keep it tethered to numerical simulations of the full lattice. Our verification method proceeds by extracting the microscopic fluctuations from full numerical simulations of the lattice, and by comparing them to the predictions of the homogenization procedure. This approach is purely *local*, warranting that the systematic errors in the boundary layers do not interfere with the verification in the bulk. Concretely, our verification method produces color maps of the lattice showing where the homogenization method is accurate and where it is not: it is considered to be valid if the accuracy is good everywhere but in a neighborhood of the boundaries, in the sense that the residual error goes sufficiently rapidly to zero with the cell size (see the convergence plot in Figure 6.4).

The verification procedure builds up on the explicit construction of the *continuous* fields proposed in Remark 3.2. We start from the nodal displacement and rotation $(\bar{\boldsymbol{v}}_\beta, \bar{\boldsymbol{\theta}}_\beta) \in \mathbb{R}^{n_n}$ obtained numerically by running discrete lattice simulations based on the energy (2.23). The procedure outlined in Remark 3.2 is implemented through a series of interpolation and averaging steps detailed in Section 6.1 below, delivering interpolations for the macroscopic displacement and the microscopic degrees of freedom $\bar{\boldsymbol{y}}(\boldsymbol{X}) = (\bar{\bar{\boldsymbol{\xi}}}_1(\boldsymbol{X})/\eta, \bar{\boldsymbol{\psi}}_1(\boldsymbol{X}), \dots)$ in the form of *continuous* functions of $\boldsymbol{X}$. Here, the overbar marks quantities extracted from the numerical simulation. By differentiating $\bar{\boldsymbol{u}}(\boldsymbol{X})$ symbolically, one can obtain the macroscopic strain $\bar{\boldsymbol{l}}(\boldsymbol{X})$ and its gradient $\nabla \bar{\boldsymbol{l}}(\boldsymbol{X})$. The accuracy of the homogenization procedure is assessed by comparing locally the microscopic degrees of freedom $\bar{\boldsymbol{y}}(\boldsymbol{X})$ that are directly extracted from the simulations, to the prediction $\boldsymbol{y} = \boldsymbol{Y} \cdot \bar{\boldsymbol{l}} + \boldsymbol{Y}' \cdot \nabla \bar{\boldsymbol{l}}$ in (4.3) based on the localization tensors $\boldsymbol{Y}$ and $\boldsymbol{Y}'$ obtained via homogenization, combined with the interpolated macroscopic strain $\bar{\boldsymbol{l}}(\boldsymbol{X})$ and its gradient $\nabla \bar{\boldsymbol{l}}(\boldsymbol{X})$.

The verification is therefore focussed on the prediction (4.3) of the microscopic displacement $\boldsymbol{y}$. This is indeed the key prediction of the homogenization method as it slaves the microscopic displacement $\boldsymbol{y}$ to the macroscopic degrees of freedom. The expression of the homogenized energy (4.5) follows straightforwardly by inserting this expression of $\boldsymbol{y}$ into the original energy.

## 6.1. Verification procedure

As a first illustration of the verification procedure, we consider a 2D rectangular strip made of a honeycomb lattice subject to compression, as shown in Figure 6.1. The length of the strip is $L = 1$ and its width is $\hat{\delta} = 0.36$. The displacements of the nodes at the left vertical boundary are blocked, $\boldsymbol{v}_{\beta \in \text{lb}} = \boldsymbol{0}$. Those at the right boundary are displaced by an amount $\Delta$, $\boldsymbol{v}_{\beta \in \text{rb}} = -\Delta \boldsymbol{e}_1$, where $\boldsymbol{e}_1$ is a unit vector along the long axis of the strip. The rotations are blocked along both short edges, $\theta_{\beta \in \text{lb}} = \theta_{\beta \in \text{rb}} = 0$. We check convergence for finer and finer lattices, when the beam length $\ell = \mathcal{O}(\eta)$ goes to zero ($\ell \to 0$).

The lattice is made up of extensible 2D beams, modeled with the discrete energy (2.19–2.20). All of them are assigned a bending modulus $EI = 1$ and a stretching modulus $EA = 10^{12}$: this warrants that the aspect-ratio parameter $\chi = \frac{EI}{EA \ell^2}$ remains at most of order $10^{-10}$ in all our simulations, so that all beams are effectively in the inextensible regime. The simulations results are compared to the *inextensible* homogenization theory (§5.3–5.4).

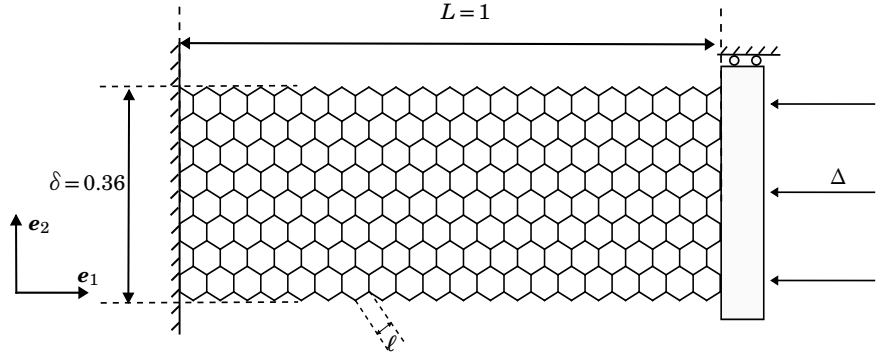

**Figure 6.1.** Honeycomb strip in compression: the lattice is loaded by bringing the short edges closer to one another by a distance $\Delta$. Sliding and rotation is blocked on both short edges. We check convergence of the discrete numerical solution when the beam length goes to zero, $\ell \to 0$.

For a given beam length $\ell$, the discrete simulation yields a set of nodal displacements $\bar{\boldsymbol{v}}_\beta$ and rotations $\bar{\boldsymbol{\theta}}_\beta$. For each beam $\alpha$ in the lattice and for every Bravais sub-lattice $b \in \{1, 2\}$, we pick $\mathcal{N}_b^\alpha = 6$ nodes from the sub-lattice $b$ in the neighborhood of the beam $\alpha$, and use the corresponding nodal values $\bar{\boldsymbol{v}}_\beta$ and $\bar{\boldsymbol{\theta}}_\beta$ to build second-order interpolations $\bar{\boldsymbol{v}}_b^\alpha(\boldsymbol{X})$ and $\bar{\boldsymbol{\theta}}_b^\alpha(\boldsymbol{X})$ (in general, we need $\mathcal{N}_b^\alpha = 1 + d + d(d+1)/2 = (d+1)(d+2)/2$ neighbors to build the second-order interpolations in dimension $d$). These interpolations applied to the sub-lattice $b$ are valid in the neighborhood of the element $\alpha$. In Figure 6.2b, we plot the interpolation of the longitudinal displacement $\bar{u}_b^\alpha(X_1) = \bar{\boldsymbol{v}}_b^\alpha(X_1) \cdot \boldsymbol{e}_1$ for the particular case of simple compression (uniaxial stress): the interpolations $\bar{u}_1^\alpha(X_1)$ and $\bar{u}_2^a(X_1)$ on the two Bravais sub-lattices are different (dashed lines), which points to the existence of a microscopic displacement.

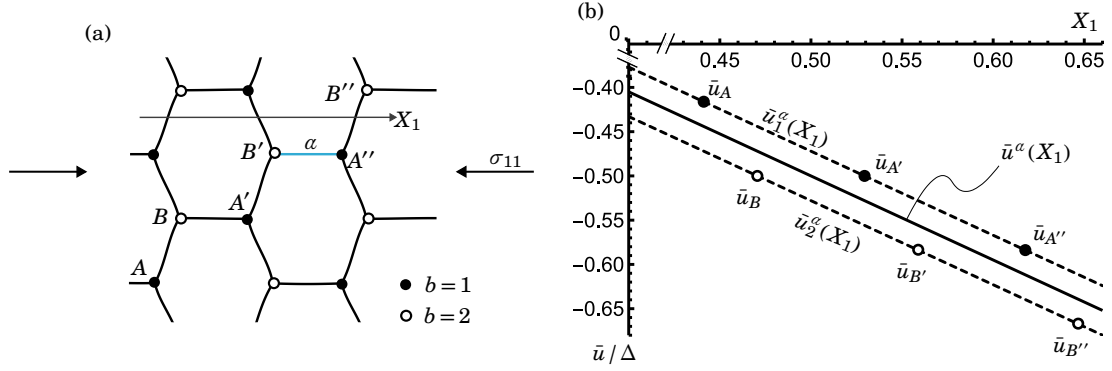

**Figure 6.2.** Local interpolations $\bar{u}_b^a(X) = \bar{v}_b^a(X) \cdot e_1$ of the longitudinal displacement, illustrated in the particular case of uni­axial compression of a honeycomb lattice. The state of uniaxial compression is obtained by modifying the boundary conditions on the short edges of the lattice shown in Figure 6.1 to let the nodes to slide freely in the transverse direction. (a) Schematic view of the deformed lattice. The link $a$ about which the interpolations are constructed is highlighted in blue. Only the $\mathcal{N}_b^a = 6$ neighboring nodes from each Bravais sub-lattice $b \in \{1,2\}$ that enter into the interpolation are shown. (b) Longitudinal displacement from discrete simulation ($\bar{u}_\alpha$, open and filled disks), interpolated in each Bravais sub-lattice ($\bar{u}_b^a(X_1)$, dashed lines) and macroscopic displacement ($\bar{u}^a(X_1)$, solid line). In this simple illustration, the $X_2$ direction is an invariance direction and all displacements are functions of $X_1$ only.

Following Remark 3.2, we obtain a local interpolation $\bar{\boldsymbol{u}}^a(\boldsymbol{X})$ of the *macroscopic* displacement by averaging over the Bravais sub-lattices,

$$\bar{\boldsymbol{u}}^a(\boldsymbol{X}) = \frac{1}{n_b} \sum_{b=1}^{n_b} \bar{\boldsymbol{v}}_b^a(\boldsymbol{X}). \tag{6.1}$$

Differentiating symbolically with respect to $\boldsymbol{X}$, we obtain the displacement gradient $\nabla \bar{\boldsymbol{u}}^a(\boldsymbol{X})$, from which we extract the local interpolations for the macroscopic rotation $\bar{\boldsymbol{\gamma}}^a(\boldsymbol{X}) = -\frac{1}{2}\nabla \bar{\boldsymbol{u}}^a(\boldsymbol{X}) : \mathcal{N}$ and the macroscopic strain $\check{\bar{\boldsymbol{\varepsilon}}}^a(\boldsymbol{X}) = \nabla \bar{\boldsymbol{u}}^a(\boldsymbol{X}) : \mathcal{M}$ using (3.8). The macroscopic state vector is then defined as $\bar{\boldsymbol{l}}^a(\boldsymbol{X}) = \begin{pmatrix} \bar{\boldsymbol{u}}^a(\boldsymbol{X}) & \bar{\boldsymbol{\gamma}}^a(\boldsymbol{X}) & \check{\bar{\boldsymbol{\varepsilon}}}^a(\boldsymbol{X}) \end{pmatrix}$ or $\bar{\boldsymbol{l}}^a(\boldsymbol{X}) = \begin{pmatrix} \check{\bar{\boldsymbol{\varepsilon}}}^a(\boldsymbol{X}) \end{pmatrix}$ depending on whether we use the standard definition (3.10) or the alternate one in (3.24).

Next, the local microscopic displacement and rotation associated to each Bravais sub-lattice $b$ ($1 \leqslant b \leqslant n_b$) are extracted using (3.4) as

$$\bar{\boldsymbol{\xi}}_b^a(\boldsymbol{X}) = \bar{\boldsymbol{v}}_b^a(\boldsymbol{X}) - \bar{\boldsymbol{u}}^a(\boldsymbol{X}) \qquad \bar{\boldsymbol{\psi}}_b^a = \bar{\boldsymbol{\theta}}_b^a(\boldsymbol{X}) - \bar{\boldsymbol{\gamma}}^a(\boldsymbol{X}). \tag{6.2}$$

We finally collect all the microscopic degrees of freedom following (3.15), and evaluate the interpolation at the center $\boldsymbol{X}_a^c$ of the element being considered,

$$\bar{\boldsymbol{y}}^a = \left( \frac{\bar{\boldsymbol{\xi}}_1^a(\boldsymbol{X}_a^c)}{\eta}, \bar{\boldsymbol{\psi}}_1^a(\boldsymbol{X}_a^c), \ldots, \frac{\bar{\boldsymbol{\xi}}_{n_b}^a(\boldsymbol{X}_a^c)}{\eta}, \bar{\boldsymbol{\psi}}_{n_b}^a(\boldsymbol{X}_a^c) \right). \tag{6.3}$$

We compare this to the leading order $(\boldsymbol{y}^\star)_{[0]}^a$ and first correction $(\boldsymbol{y}^\star)_{[1]}^a$ predicted by the homogenized results (5.10) for an inextensible lattice,

$$\begin{aligned}
(\boldsymbol{y}^\star)_{[0]}^a &= \frac{a}{4}(\ +\mathcal{J}:\bar{\boldsymbol{\varepsilon}}^a \quad 0 \quad -\mathcal{J}:\bar{\boldsymbol{\varepsilon}}^a \quad 0\ ), \\
(\boldsymbol{y}^\star)_{[1]}^a &= -\frac{3 a \eta}{4}(\ \boldsymbol{0}_2 \quad \operatorname{curl}(\mathcal{J}:\bar{\boldsymbol{\varepsilon}}^a) \quad \boldsymbol{0}_2 \quad -\operatorname{curl}(\mathcal{J}:\bar{\boldsymbol{\varepsilon}}^a)\ ).
\end{aligned} \tag{6.4}$$

In the right-hand sides, we use the estimate $\bar{\boldsymbol{\varepsilon}}^a$ of the strain, and $\nabla \bar{\boldsymbol{\varepsilon}}^a$ of the strain gradient at the element center $\boldsymbol{X}_a^c$, as obtained from the discrete numerical simulation: these values are extracted from the interpolations $\bar{\boldsymbol{l}}^a(\boldsymbol{X}_a^c)$ and $\nabla \bar{\boldsymbol{l}}^a(\boldsymbol{X}_a^c)$, based on the definition (3.10) or (3.24) of $\boldsymbol{l}$ in terms of the strain $\check{\boldsymbol{\varepsilon}}$ and the Mandel convention (3.7) for representing the strain $\boldsymbol{\varepsilon}$ as a flat vector $\check{\boldsymbol{\varepsilon}}$.

Specifically, we compute at every element $a$ the error $e_{[0]}^a$ on the microscopic displacement based on leading-order homogenization, the error $e_{[1]}^a$ based on second-order homogenization, as well as its magnitude $e_{[\times]}^a$ in the raw output of numerical simulations as

$$\begin{aligned}
e_{[\times]}^a &= \left\| \frac{\bar{\boldsymbol{y}}^a}{a\,\Delta} \right\| \\
e_{[0]}^a &= \left\| \frac{\bar{\boldsymbol{y}}^a - (\boldsymbol{y}^\star)_{[0]}^a}{a\,\Delta} \right\| \\
e_{[1]}^a &= \left\| \frac{\bar{\boldsymbol{y}}^a - ((\boldsymbol{y}^\star)_{[0]}^a + (\boldsymbol{y}^\star)_{[1]}^a)}{a\,\Delta} \right\|,
\end{aligned} \tag{6.5}$$

where $a = \mathcal{O}(\eta^0)$ is the conventional length of the beams in the topological lattice and $\Delta$ is the conventional magni-tude of the applied load in the numerical simulations. The denominators in right-hand sides in (6.5) make the errors $e_{[\cdot]}^\alpha$'s independent of both these conventions. An interpolation present in the right-hand sides is evaluated at the center $\boldsymbol{X}_\alpha^c$ of the element $\alpha$.

In Figure 6.3, these errors are painted on the deformed lattice, for two different beam lengths $\ell$, using a logarithmic colormap. We observe a significant and consistent reduction on the magnitude of the error when we go from no homogenization ($e_{[\times]}^\alpha$) to leading-order homogenization ($e_{[0]}^\alpha$), and then to second-order homogenization ($e_{[1]}^\alpha$). As anticipated earlier, this improvement takes place everywhere in the lattice, except in a layer that is a few cells thick at the boundaries, where it does not decrease noticeably.

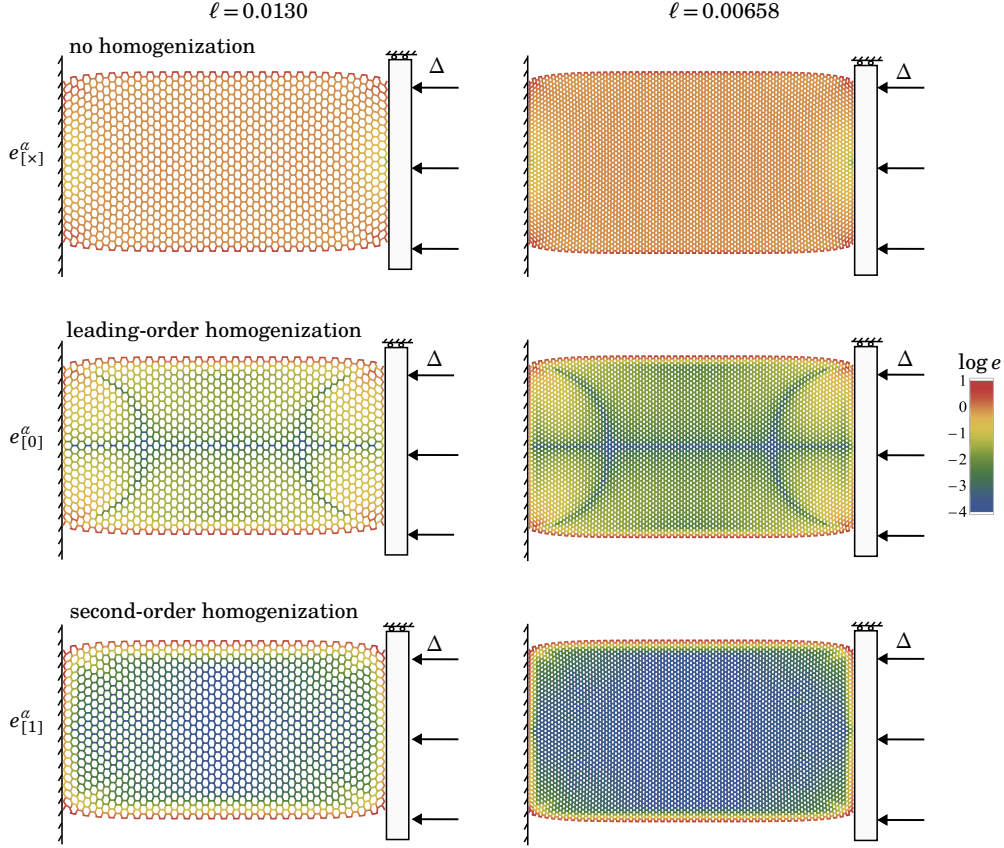

**Figure 6.3.** Honeycomb strip in compression: error $e_{[i]}^\alpha$ on the microscopic displacement predicted by the leading-order ($i = 0$) or second-order ($i = 1$) homogenization, compared with its magnitude ($i = \times$) in the raw numerical solution, for two different lattices (coarser one in the left column, finer one in the right column). The error is painted over the deformed lattice as a base-10-logarithmic colormap. An arbitrary magnification factor is applied on the displacement for plotting the deformed lattice. The numerical simulations were produced with an imposed displacement $\Delta = 0.1$ but the results are rescaled in such a way that this value of $\Delta$ doest not affect the colors. The consistent decrease of the error in logarithmic scale with the order of homogenization is a sign that the homogenization is correct.

This first visual indication that the homogenization is correct can be confirmed by a convergence test with respect to the beam length ($\ell \to 0$). We select a point in the interior of the rectangular domain, shown by the blue cross in the inset in Figure 6.4, and we compute the local errors $e_{[0]}^\alpha$ and $e_{[1]}^\alpha$ at the nearest element center $\boldsymbol{X}_\alpha^c$ for finer and finer lattices. The numerical results, shown in Figure 6.4, reveal the following scaling laws: $e^\alpha_{[0]} = \mathcal{O}(\ell^1)$ and $e^\alpha_{[1]} = \mathcal{O}(\ell^2)$ for $\ell \to 0$. The fact that the order of the error increases from 1 to 2 when we switch from leading-order to second-order homogenization is a numerical evidence that the latter is asymptotically correct. A similar behavior as in Figure 6.4 is obtained when a different target point is chosen (data not shown). For points close to the boundary, the scaling regime seen in Figure 6.4 does not start until $\ell$ becomes significantly smaller than the distance to boundary (data not shown).

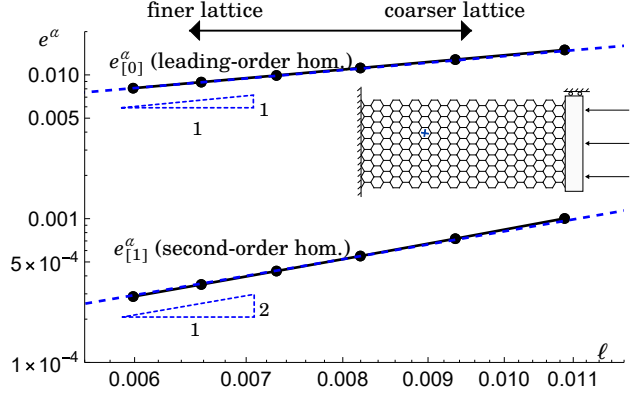

**Figure 6.4.** Honeycomb strip in compression: log-log plot of the homogenization errors $e^\alpha_{[0]}$ and $e^\alpha_{[1]}$ computed at the nearest element $\alpha$ to a fixed target (blue cross in the inset) as a function of the 'microscopic' element length $\ell$. The blue dashed lines are linear fits with slope 1 and 2, respectively.

## 6.2. Sheared strip, with or without a hole

Next, we test the rectangular strip when subjected to a shearing displacement $\Delta$, as shown in Figure 6.5. The displacement is still blocked along the short boundary on the left-hand side ($\boldsymbol{v}_{\beta \in \text{lb}} = \boldsymbol{0}$) and the nodes along the short boundary on the right-hand side are moved vertically ($\boldsymbol{v}_{\beta \in \text{rb}} = +\Delta \, \boldsymbol{e}_2$). In addition, the nodes along both short boundaries can now rotate freely.

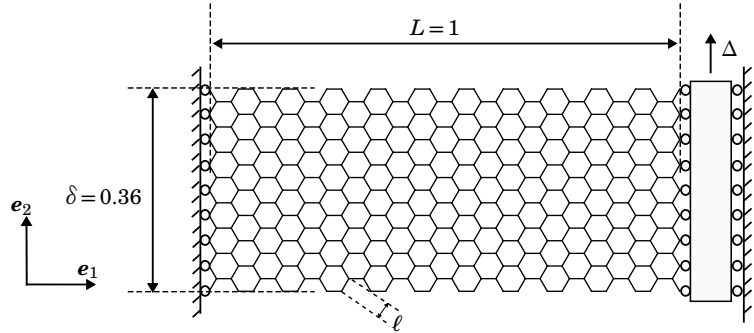

**Figure 6.5.** Honeycomb strip subjected shear: loading geometry. The nodes on both short edges can now rotate freely.

The same testing procedure is applied. The colormaps of the homogenization errors are shown in logarithmic scale in Figure 6.6 for two different beam lengths (a base-10 logarithm is used throughout in the paper). Here again, the higher-order correction significantly increases the accuracy of the prediction, except in a small layer along the boundaries.

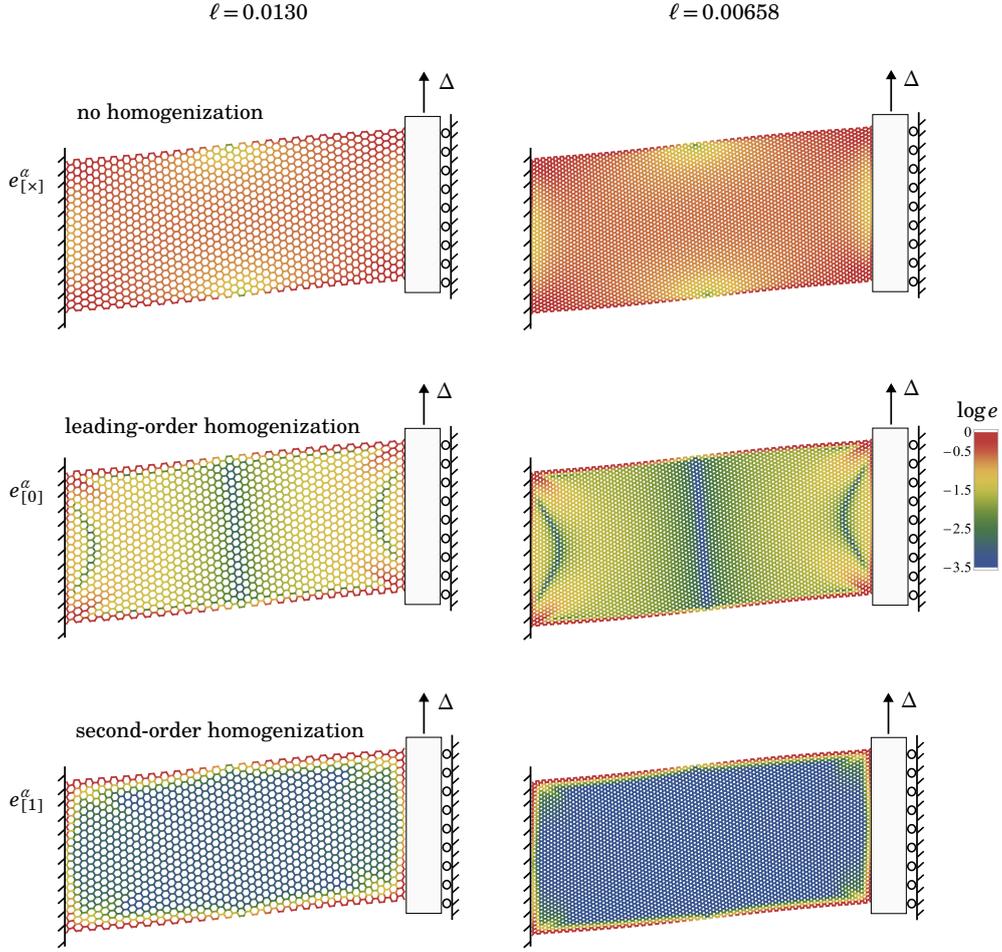

**Figure 6.6.** Honeycomb strip subjected to shear: homogenization error for two different lattices (columns), using the same conventions as in Figure 6.3. The boundary conditions are represented in a simplified way, the reader is referred to Figure 6.5 for details.

In the next test, we punch the lattice by removing a disk-like region with radius $r = 0.04$ in its center. The boundary conditions are identical to those in Figure 6.5. This relatively small hole causes stress concentration, visible on the raw output of the numerical simulations in the first row in Figure 6.7. The typical flower-like stress concentration pattern survives to leading-order homogenization, see the second row. When higher-order homogenization is used (third row), however, it almost disappears from the error map, indicating that second-order homogenization accurately captures the fast variations of the stress in the vicinity of the hole: the error goes to zero quickly away from the hole. The existence of a 'safe zone' surrounding the hole (extended blue region in this third row, where the relative error is as small as $\sim 10^{-3}$) suggests that the punched strip could be very accurately represented by combining effective boundary conditions along the edge of the hole with a higher-order continuum model for the strip. This is probably not possible with leading-order homogenization (second row) as the region of influence of the hole stretches all the way to the lateral boundaries.

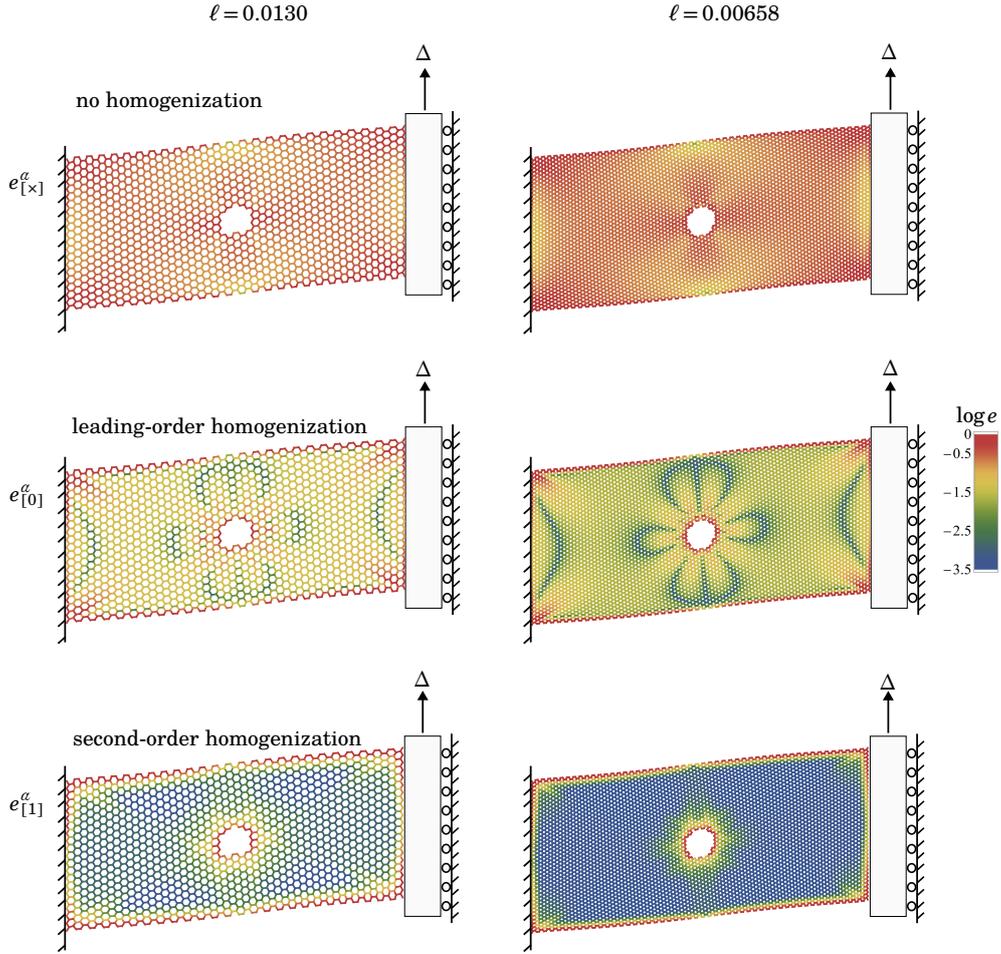

**Figure 6.7.** Honeycomb strip with a hole, subjected to shear: homogenization error for two different lattices (columns), using the same conventions as in Figure 6.3 and 6.5. Same boundary conditions as in Figure 6.5.

## 6.3. Cracked strip in tension

Before moving on to more complex examples, we consider the cracked version of the rectangular honeycomb strip shown in Figure 6.8, having an aspect-ratio $\delta/L = 0.78$. The crack is obtained by removing the beams that intersect the segment with length $a$, making an angle $\pi/6$ with the direction $\boldsymbol{e}_1$, and whose midpoint is at the center of the lattice. Crack propagation is not considered. The same boundary conditions are used as for the shearing test, but a *longitudinal* displacement $\boldsymbol{v}_{\beta \in \mathrm{rb}} = \Delta\, \boldsymbol{e}_1$ is imposed on the right-hand-side boundary.

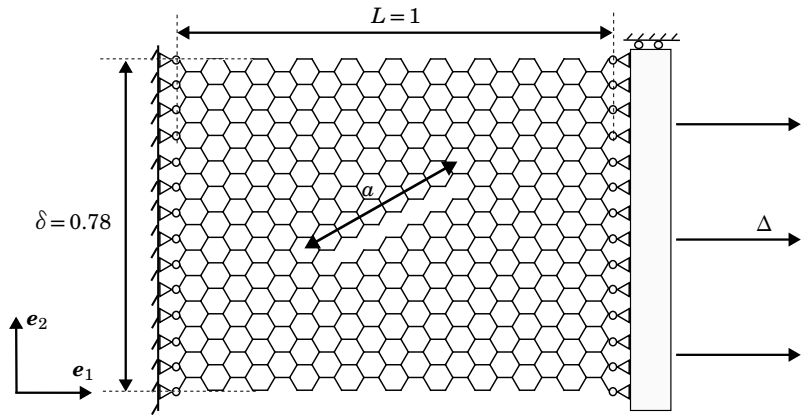

**Figure 6.8.** Cracked honeycomb strip in tension.

Colormaps of the homogenization error are shown in Figure 6.9 for $\ell = 0.0125$. Similar to what has been obtained for a punched lattice, some stress concentration is observed at both crack tips. When second-order homogenization is used, not only does the homogenization error drop significantly, it is also very small except in the intermediate neighborhood of the crack tips and lips and of the edges of the rectangular region. The fact that the perturbation at the crack tips are disconnected from each other suggests that this geometry can be accurately represented by a higher-order continuum model endowed with effective boundary conditions for the crack tips, crack lips and free boundaries.

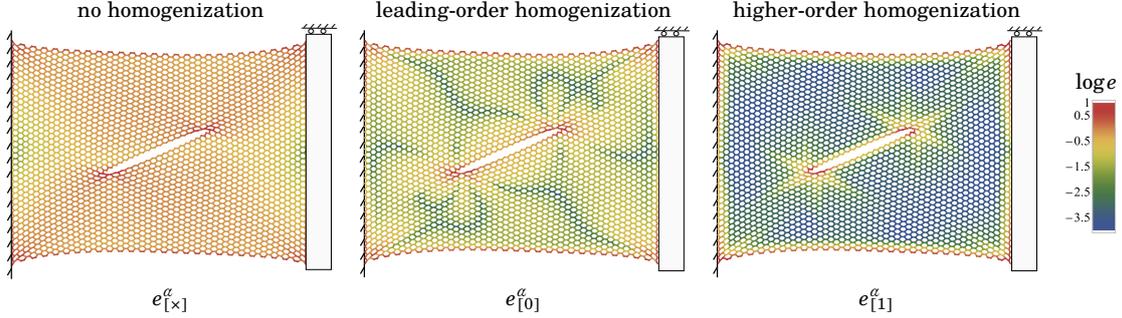

**Figure 6.9.** Cracked honeycomb strip in tension: homogenization errors for $\ell = 0.0125$, using the same conventions as in Figure 6.3. The representation of the boundary conditions is simplified, the reader is referred to Figure 6.8 for details.

## 7. EXTENSIONS AND FURTHER ILLUSTRATIONS

This section illustrates *(i)* more advanced features of the homogenization method, namely non-uniform elastic properties (§7.1), non-uniform geometric properties (§7.3) and pre-strain (§7.2), as well as *(ii)* its versatility, by exploring different lattice topologies (2D Kagome lattice in §7.2), different types of elements (springs §7.3–7.5), different space dimensions (3D truss in §7.4, 1D truss in §7.5) as well the case of a large elastic contrast (§7.5).

### 7.1. Inhomogeneous elastic properties

We consider an extension of the honeycomb compression test done in Section 6.1, now using a lattice whose bending rigidity is graded in space: a beam $\alpha$ with midpoint position $\boldsymbol{X}_\alpha^c$ is assigned a stiffness $EI(\boldsymbol{X}_\alpha^c)$, where

$$EI(X_1, X_2) = \frac{1}{6}\left(1 + 5\tanh^2\left(6\,\frac{2\,X_2^{\,3} - X_1}{1 + 4\,X_2^{\,2}}\right)\right), \tag{7.1}$$

for $-1/2 \leqslant X_1 \leqslant +1/2$ and $-\delta/2 \leqslant X_2 \leqslant +\delta/2$, with the origin of the coordinate axes at the center of the lattice and $\delta = 0.84$. The distribution (7.1) of bending stiffness is shown in Figure 7.1. The lattice is six times softer along an S-shaped region that extends to opposite corners. As earlier, we work in the inextensible regime by using a large stretching modulus $EA = 10^{12}$ for all beams. The detailed boundary conditions are shown in Figure 7.1.

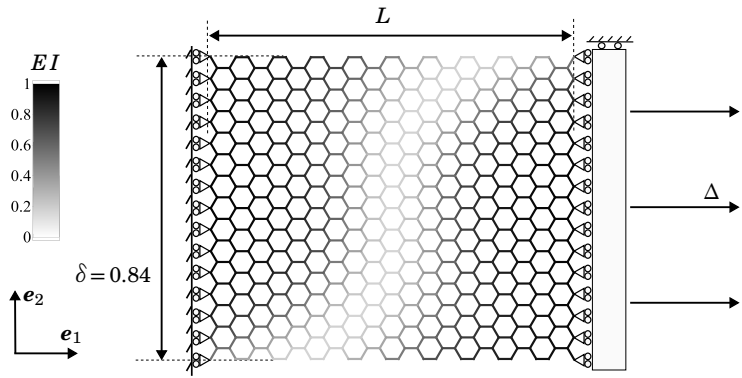

**Figure 7.1.** Honeycomb lattice with a graded bending modulus $EI(\boldsymbol{X})$ given by (7.1). The bending modulus $EI$ is constant in every beam and is found by evaluating the function $EI(\boldsymbol{X})$ at the beam midpoint $\boldsymbol{X}_\alpha^c$. The boundary conditions are depicted using the same conventions as earlier.

The homogenization is based on the inextensible beam model, as earlier. Compared to the homogeneous case (§5.4), the only change is that the symbolic parameter $EI$ must now be included in the list of *variable* material parameters,

$$\boldsymbol{m} = \{EI\}. \tag{7.2}$$

This is done by changing a single line of code: `m={}` is replaced with `m={EI}`, see the companion notebook `homogenize-honeycomb-inextensible-variable.nb`. The library delivers the microscopic displacement in the form

$$\begin{aligned}\boldsymbol{\xi}_b^\star &= \ell\left(\frac{(-1)^{b+1}}{4}\mathcal{J}:\boldsymbol{\varepsilon}\,\ell^0 + 0\,\ell^1 + \mathcal{O}(\ell^2)\right),\\ \psi_b^\star &= 0\,\ell^0 + \frac{3\,(-1)^{b+1}}{4}\left(-\mathrm{curl}\,(\mathcal{J}:\boldsymbol{\varepsilon}) + \frac{(\mathcal{J}:\boldsymbol{\varepsilon})\wedge\nabla(EI)}{EI}\right)\ell^1 + \mathcal{O}(\ell^2),\end{aligned} \quad (7.3)$$

where the constant tensor $\mathcal{J}$ has been defined in (5.5), the curl operator has been introduced in (5.6), and $\wedge$ denotes the 2D wedge product, $\boldsymbol{a}\wedge\boldsymbol{b} = a_1 b_2 - a_2 b_2$. The only difference with the homogeneous case (5.10) is the presence of the gradient term $\nabla(EI)$ in the microscopic rotation $\psi_b$. The gradient effect has now two contributions, one associated with the strain gradient $\nabla\boldsymbol{\varepsilon}$ and the other one with the gradient of elastic properties $\nabla(EI)$.

**Remark 7.1.** The statement (7.2) informs the library that the parameter $EI$ entering in the stiffness $\mathcal{H}$ varies in space, implying that the gradient must be calculated by the chain rule as $\nabla\mathcal{H} = \frac{\mathrm{d}\mathcal{H}}{\mathrm{d}(EI)}\cdot\nabla EI$. The library does not know about the specific profile $EI(\boldsymbol{X})$ in (7.1) and treats the gradient $\nabla(EI)$ symbolically: the explicit expression of $\nabla(EI)$ must be inserted by the user into the output (7.3) of the library.

The validity of the predictions (7.3) are tested using the same procedure as earlier. The results are shown in Figure 7.2. In the discrete simulations, the deformation is concentrated in the soft part of the lattice; by a Poisson's effect, the lattice forms a neck in the center. The rapid variations in bending stiffness creates strong gradients that make the predictions of the leading-order homogenization relatively inaccurate, see Figure 7.2(b). Taking this gradient effect into account improves the accuracy by approximately one order of magnitude, see Figure 7.2(c). Both terms in $(7.3)_2$ are equally important: dropping either one deteriorates the accuracy significantly (plots not shown).

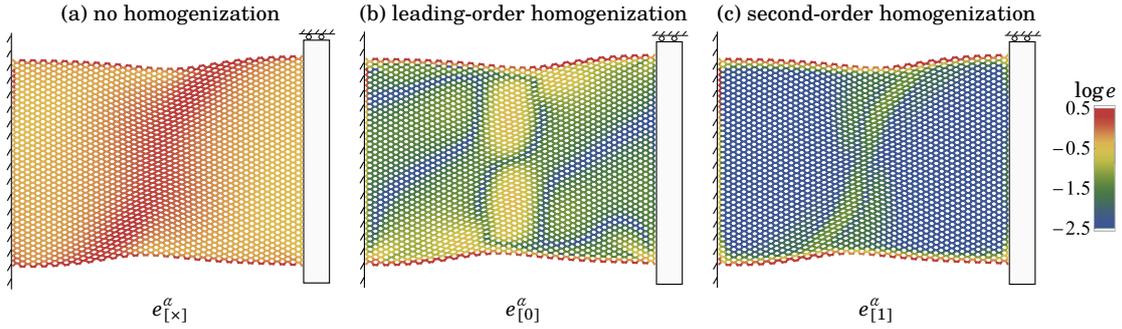

**Figure 7.2.** Honeycomb lattice with graded bending modulus: accuracy of the prediction (7.3) for the microscopic displacement. All beams have length $\ell = 0.0125$.

## 7.2. Kagome lattice with pre-strain

In this section, we change the lattice topology to *Kagome* and analyze the effect of a non-uniform pre-strain, as produced typically by thermal expansion in the presence of a localized heat source. The rectangular lattice is shown in Figure 7.3a: it is pinned at two adjacent corners and there are no other loading than the pre-strain. It is made up of beams with length $\ell = 0.006579$, aspect-ratio parameter $\chi = 0.002$, stretching modulus $EA = 1$ and bending modulus $EI = EA\,\ell^2\,\chi$.

In the discrete simulations, the extensional pre-strain $p(\boldsymbol{X})$ is accounted for by changing the strain energy (2.20) of the beams to

$$w_a(\boldsymbol{E}_a) = \frac{1}{2}\left(\boldsymbol{E}_a - \boldsymbol{E}_0(p(\boldsymbol{X}_a^c))\right)\cdot\mathcal{H}_{\varphi_a}\cdot\left(\boldsymbol{E}_a - \boldsymbol{E}_0(p(\boldsymbol{X}_a^c))\right) \quad \text{where } \boldsymbol{E}_0(p) = (p,0,0). \quad (7.4)$$

The pre-strain vector $\boldsymbol{E}_0$ reflects the ordering convention in (2.12): the pre-strain $p$ being extensional, it offsets the *first* slot of $\boldsymbol{E}_a$ representing the stretching strain $\varepsilon_a$. In (7.4), the value of the pre-strain $p$ in any particular beam $a$ is picked based on a pre-strain map $p(\boldsymbol{X})$ evaluated at the midpoint $\boldsymbol{X}_a^c$: in the simulations, we use the map

$$p(\boldsymbol{X}) = \Delta\,\exp\left[-\left(\tfrac{\|\boldsymbol{X}\|}{0.2}\right)^2\right] \quad (7.5)$$

plotted in Figure 7.3a, which has a maximum at the center $\boldsymbol{X} = \boldsymbol{0}$ of the upper boundary, and decreases exponentially away from this point. The pre-strain magnitude is set to $\Delta = 0.3$ in the simulations but this value is a matter of convention as all our results are rescaled by $\Delta$. The results of the discrete simulations are shown in Figure 2.20 (deformed lattice) along with the homogenization results (colors).

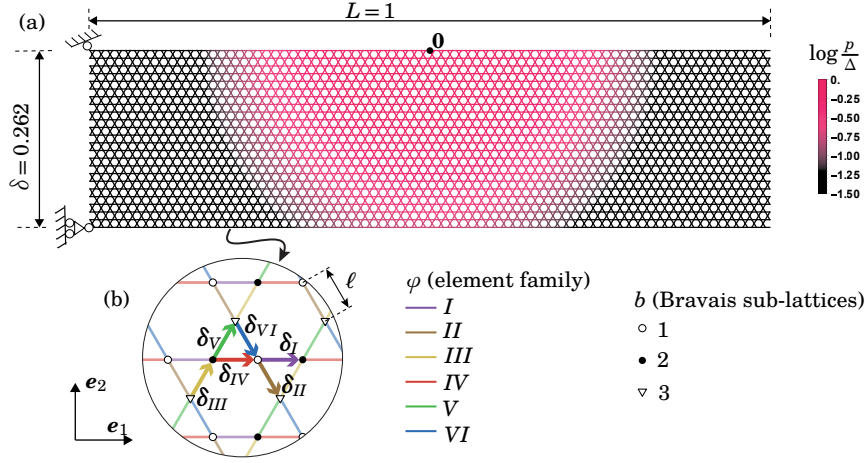

**Figure 7.3.** Kagome lattice with pre-strain. (a) Imposed pre-strain map $p(\boldsymbol{X})$ and boundary conditions. (b) Underlying topological lattice, made up of $n_\varphi = 6$ beam families and $n_b = 3$ Bravais sub-lattices. In each family, a particular beam is shown using a thick line, with the arrow denoting the conventional orientation. All beams have equal length $\ell = 0.006579$, stretching modulus $EA = 1$ and aspect-ratio parameter $\chi = 0.002$.

We proceed to homogenize the lattice. The underlying topological Kagome lattice makes use of the conventions shown in Figure 7.3b. It is specified in the homogenization code based on the properties listed in Table 7.1.

| $d$ | $n_b$ | $n_\varphi$ | $\tilde\rho_\varphi$ |
|---|---|---|---|
| 2 | 3 | 6 | $\frac{1}{2\sqrt{3}\ell^2}$ |

| $\varphi$ | $I$ | $II$ | $III$ | $IV$ | $V$ | $VI$ |
|---|---|---|---|---|---|---|
| $b_\varphi^-$ | 1 | 1 | 3 | 2 | 2 | 3 |
| $b_\varphi^+$ | 2 | 3 | 2 | 1 | 3 | 1 |
| $\boldsymbol{\delta}_\varphi$ | $\ell\boldsymbol{e}_1$ | $\ell\left(\frac{\boldsymbol{e}_1}{2}-\frac{\sqrt{3}}{2}\boldsymbol{e}_2\right)$ | $\ell\left(\frac{\boldsymbol{e}_1}{2}+\frac{\sqrt{3}}{2}\boldsymbol{e}_2\right)$ | $\ell\boldsymbol{e}_1$ | $\ell\left(\frac{\boldsymbol{e}_1}{2}+\frac{\sqrt{3}}{2}\boldsymbol{e}_2\right)$ | $\ell\left(\frac{\boldsymbol{e}_1}{2}-\frac{\sqrt{3}}{2}\boldsymbol{e}_2\right)$ |

**Table 7.1.** Properties of the topological Kagome lattice shown in Figure 7.3b, used as an input to the homogenization code: space dimension $d$, number of Bravais sub-lattices $n_b$, number of element families $n_\varphi$, density $\tilde\rho_\varphi$ per unit area of element belonging to any particular family $\varphi$, connectivity matrix $b_\varphi^\pm$ and end-to-end vector $\boldsymbol{\delta}_\varphi = \boldsymbol{\Delta}_\varphi^+ - \boldsymbol{\Delta}_\varphi^-$ for a beam of type $\varphi$ in reference configuration.

In the homogenization library, the pre-strain is taken into account by expanding (7.4) as

$$w_\alpha(\boldsymbol{E}_\alpha) = \frac{1}{2}\boldsymbol{E}_\alpha \cdot [\mathcal{H}_{\varphi_\alpha}] \cdot \boldsymbol{E}_\alpha + 1\left[-\mathcal{H}_{\varphi_\alpha} \cdot \boldsymbol{E}_0(p(\boldsymbol{X}_\alpha^c))\right] \cdot \boldsymbol{E}_\alpha + \frac{1}{2}1\left[\boldsymbol{E}_0(p(\boldsymbol{X}_\alpha^c)) \cdot \mathcal{H}_{\varphi_\alpha} \cdot \boldsymbol{E}_0(p(\boldsymbol{X}_\alpha^c))\right]1. \quad (7.6)$$

Contrary to what was assumed earlier in (3.29), the element energy (7.6) is no longer homogeneous of degree 2 with respect to the unknown strain $\boldsymbol{E}_\alpha$. To work around this, we incorporate the 1's appearing in (7.6) into the macroscopic strain $\boldsymbol{E}$, amending its definition (3.25) as—mind the trailing 1,

$$\boldsymbol{E} = \left(\varepsilon_I, \kappa_I, \tau_I, \ldots, \varepsilon_{n_\varphi}, \kappa_{n_\varphi}, \tau_{n_\varphi}, \eta^{-1}\sum_{b=1}^{n_b}(\xi_b)_1, \ldots, \eta^{-1}\sum_{b=1}^{n_b}(\xi_b)_d, 1\right). \quad (7.7)$$

Equation (7.6) can then be implemented by incorporating the square brackets appearing in (7.6) into the element contribution $\mathcal{H}_\varphi^g$ (3.31), which we now define as

$$\mathcal{H}_\varphi^g(\eta, \boldsymbol{X}) = \begin{pmatrix} \boldsymbol{0}_{p_\varphi \times p_\varphi} & \boldsymbol{0}_{p_\varphi \times n_{E_\varphi}} & \boldsymbol{0}_{p_\varphi \times (p'_\varphi - 1)} & \boldsymbol{0}_{p_\varphi \times 1} \\ \boldsymbol{0}_{n_{E_\varphi} \times p_\varphi} & [\mathcal{H}_\varphi] & \boldsymbol{0}_{n_{E_\varphi} \times (p'_\varphi - 1)} & \left[-\mathcal{H}_\varphi \cdot \boldsymbol{E}_0(p(\boldsymbol{X}))\right] \\ \boldsymbol{0}_{(p'_\varphi - 1) \times p_\varphi} & \boldsymbol{0}_{(p'_\varphi - 1) \times n_{E_\varphi}} & \boldsymbol{0}_{(p'_\varphi - 1) \times (p'_\varphi - 1)} & \boldsymbol{0}_{(p'_\varphi - 1) \times 1} \\ \boldsymbol{0}_{1 \times p_\varphi} & \left[-\mathcal{H}_\varphi \cdot \boldsymbol{E}_0(p(\boldsymbol{X}))\right]^T & \boldsymbol{0}_{1 \times (p'_\varphi - 1)} & \left[\boldsymbol{E}_0(p(\boldsymbol{X})) \cdot \mathcal{H}_\varphi \cdot \boldsymbol{E}_0(p(\boldsymbol{X}))\right] \end{pmatrix}. \quad (7.8)$$

Now, since $\boldsymbol{E}$ is required to depend linearly on $\boldsymbol{l}, \boldsymbol{y}$ and their gradients by (3.34), the updated definition of $\boldsymbol{E}$ in (7.7) requires that we append a trailing entry 1 in the macroscopic strain $\boldsymbol{l}$ as well, modifying (3.24) into

$$\boldsymbol{l} = \left(\check\varepsilon_1 = \varepsilon_{11}, \check\varepsilon_2 = \varepsilon_{22}, \check\varepsilon_3 = \sqrt{2}\,\varepsilon_{12}, 1\right), \quad (7.9)$$

and that we propagate it to $\boldsymbol{E}$: to do so, we extend the dimension of the tensors $\boldsymbol{E}_l, \ldots, \boldsymbol{E}''_y$ appropriately and fill the new entries with 0's, except that in the lower-right corner of $\boldsymbol{E}_l$ which is set to 1. In the library, the content of the vectors $\boldsymbol{l}, \boldsymbol{y}, \boldsymbol{E}$, etc. is not fixed once for all but can instead be chosen in a flexible way, as documented in the Implementation notes that are distributed along with the library: this makes the implementation of the above changes straightforward.

From the user's perspective, the pre-strain is dealt with by including an option `includeUnitStrain=True` that sets up the trailing 1's in both $\boldsymbol{E}$ and $\boldsymbol{l}$, and by providing an optional argument `prestrain={"ε":p}` to the constructor for beam elements. The pre-strain is represented by a pure symbol p that is treated in a similar way as the variable stiffness in section 7.1: it is declared as a *variable* material parameter by the assignment m={p}, and the homogenization results are given symbolically in terms of $\boldsymbol{m} = (p)$ and its gradient $\nabla \boldsymbol{m} = (\nabla p)$, as well as on $\boldsymbol{l}$ and its gradients $\nabla \boldsymbol{l}$, etc. To interpret these results, the user must take into account the fact that $(\nabla l)_{4i} = \partial(l_4)/\partial X_i = \partial 1/\partial X_i = 0$ in view of (7.9).

In the limit $EI \to 0$ of perfectly flexible beams, the Kagome lattice possesses a number of zero-energy modes [HF06]. One of these zero modes has infinite wavelength and therefore survives homogenization: it involves a rigid rotation of adjacent triangles in opposite directions and translates the Bravais sub-lattice having index $b$ by a vector $\mathcal{W}_b$,

$$\mathcal{W}_1 = \frac{-\boldsymbol{e}_1 - \sqrt{3}\,\boldsymbol{e}_2}{2}, \qquad \mathcal{W}_2 = \frac{-\boldsymbol{e}_1 + \sqrt{3}\,\boldsymbol{e}_2}{2}, \qquad \mathcal{W}_3 = \boldsymbol{e}_1. \tag{7.10}$$

Indeed, the corresponding microscopic displacement $\boldsymbol{y} = (\mathcal{W}_1\ 0\ \mathcal{W}_2\ 0\ \mathcal{W}_3\ 0)$ makes all the *extensional* degrees of freedom in $\boldsymbol{E}_l \cdot \boldsymbol{y}$ vanish.

The Kagome lattice is homogenized in the Mathematica notebook homogenize-kagome-prestress.nb distributed along with the library. The lattice made up of extensible beams, corresponding to the element stiffness matrix in (2.19). The microscopic displacement computed by the library takes the form

$$\boldsymbol{\xi}_b^\star = \ell \left( \boldsymbol{0}\,\ell^0 + \begin{pmatrix} \dfrac{c_1(\chi)}{\chi}\mathrm{div}\,(\mathcal{J}\!:\!\boldsymbol{\varepsilon})\,\mathcal{W}_b \\ + [\mathcal{W}_b \cdot \mathcal{J}] \cdot \bigl( c_2(\chi)\,(\nabla\,\mathrm{tr}\,\boldsymbol{\varepsilon} - 2\,\nabla p) + c_3(\chi)\,\mathcal{J}\!:\!\nabla(\mathcal{J}\!:\!\boldsymbol{\varepsilon}) \bigr) \\ + c_4(\chi)\,\mathrm{curl}\,(\mathcal{J}\!:\!\boldsymbol{\varepsilon})\,[\mathcal{A}^T \cdot \mathcal{W}_b] \end{pmatrix} \ell + \mathcal{O}(\ell^2) \right)$$

$$\psi_b^\star = -\tfrac{1}{2}\,\mathcal{W}_b \wedge (\mathcal{J}\!:\!\boldsymbol{\varepsilon})\,\ell^0 + 0\,\ell^1 + \mathcal{O}(\ell^2), \tag{7.11}$$

where $\mathcal{A}$ is the Levi-Civita (purely anti-symmetric) symbol in dimension 2, $\mathcal{A} = \boldsymbol{e}_1 \otimes \boldsymbol{e}_2 - \boldsymbol{e}_2 \otimes \boldsymbol{e}_1$ and the coefficients $c_i$ are rational functions of $\chi$ that are bounded for $\chi \to 0$,

$$\begin{aligned} c_1(\chi) &= \frac{1 + 18\,\chi}{288} & c_2(\chi) &= \frac{1}{12\,(1 + 12\,\chi)} \\ c_3(\chi) &= \frac{1 - 24\,\chi}{24\,(1 + 12\,\chi)} & c_4(\chi) &= -\chi/4. \end{aligned} \tag{7.12}$$

The compact tensorial expressions in (7.11) has been obtained by applying to the raw output of the homogenization code a procedure similar to that yielding the irreducible form of elasticity tensors in the presence of symmetries [AKO17].

In Figure 7.4, the expression (7.11) of the microscopic displacement is verified using the usual procedure. The leading-order homogenization fails to give accurate predictions in regions where the gradient of pre-strain is important and the improvement brought about by second-order homogenization is very significant there. If the correction proportional to $\nabla p$ is dropped in the second-order homogenization results (7.11)$_2$, the agreement with the numerical solution degrades noticeably (data not shown).

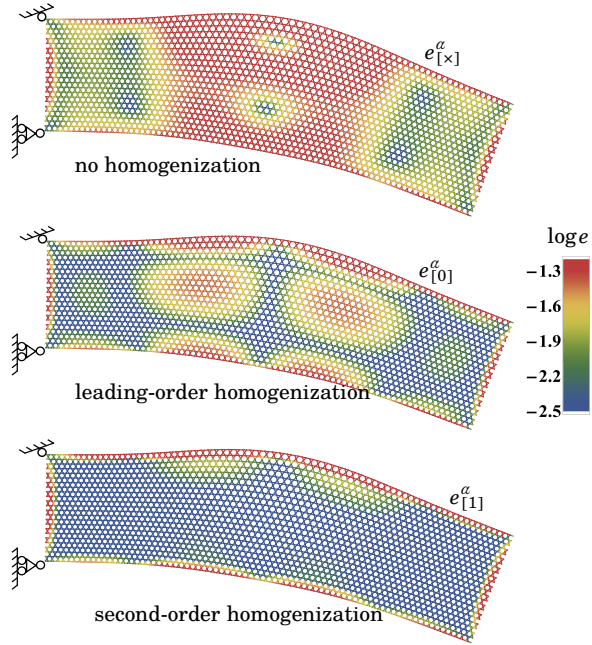

**Figure 7.4.** Kagome lattice with spatially graded pre-strain: verification of the microscopic displacement (7.11) based on the discrete numerical solution. Same lattice as in Figure 7.3.

**Remark 7.2.** We have considered a Kagome lattice made up of *extensible beams*. Indeed, it is neither possible to use *springs* $EI = 0$ as the lattice then possess a zero-energy mode—the library aborts, printing out the zero mode (7.10)—, nor *inextensible beams* ($\chi = 0$) as this makes the lattice fully rigid—the library aborts, reporting a compatibility condition in the form $\varepsilon(X) = 0$ for all $X$.

**Remark 7.3.** The $\chi^{-1}$ coefficient in factor of $c_1$ in the right-hand side of (7.11)$_1$ implies that the amplitude of the soft mode $\mathcal{W}_b$ becomes larger and larger in the limit of inextensible beams, $\chi \to 0$. This is not surprising as, by definition, this mode has a vanishing elastic energy in the limit. A detailed analysis of the inextensible limit of the Kagome lattice is beyond the scope of this paper—in principle, a limit energy can be obtained by analyzing the energy of the discrete lattice computed by the library in the limit $\chi \to 0$, similar to what we did earlier in Section 5.3.

### 7.3. Circular arch

Our next example is an half-annular lattice made up of elastic springs ($EI = 0$), sketched in Figure 7.5. This lattice features *spatial inhomogeneity*: both the orientation and the natural length of the springs vary across the annular domain.

The discrete arch is generated as follows. We pick two integers $n_1$ and $n_2$ and define

$$\eta = \frac{\pi}{n_2} \quad \text{and} \quad \delta = \pi \frac{n_1}{n_2}. \tag{7.13}$$

In the simulations shown in Figure 7.7, we used $(n_1, n_2) = (13, 90)$. The nodes are indexed by a pair of integers $(i, j)$ with $-n_1 \leq i \leq +n_1$ and $0 \leq j \leq n_2$ and their position $X_{(i,j)}$ in reference configuration is set to

$$X_{(i,j)} = \exp(\eta i) \, e_r(\eta j) \quad \text{where } e_r(\theta) = \cos\theta \, e_1 + \sin\theta \, e_2. \tag{7.14}$$

The inner and outer radii of the arch are therefore given as $\exp(\pm \eta n_1) = \exp(\pm \delta)$. Each node is assigned a Bravais index $b$ with $b = 1$ if $i + j$ is even (black nodes in the figure) and $b = 2$ if it is odd (white nodes). As shown in the figure, we set up elastic springs radially between *all pairs* of adjacent nodes, azimuthally between *all pairs* of adjacent nodes, and diagonally between pairs of adjacent *black* nodes ($b = 1$).

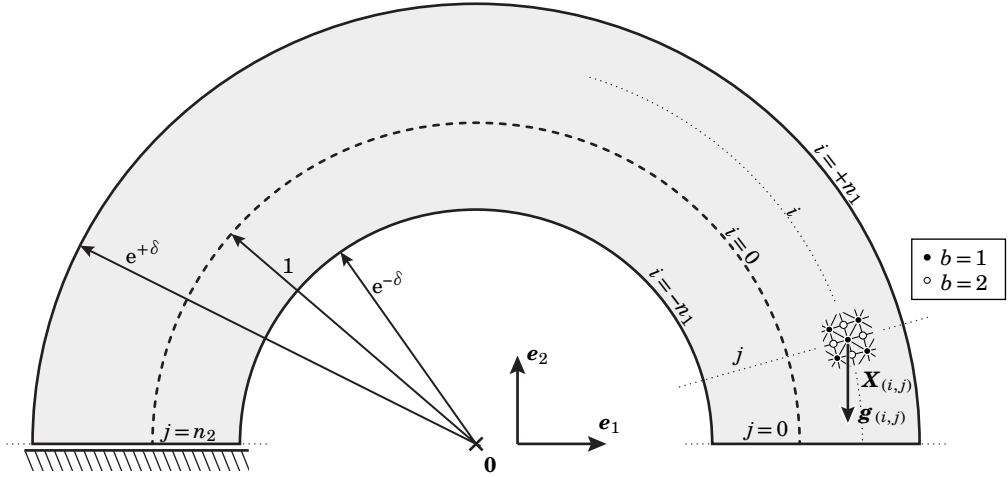

**Figure 7.5.** Circular arch. This annular lattice is made up of *spring* elements connecting nodes $X_{(i,j)}$ indexed by two integers, a radial coordinate $-n_1 \leq i \leq +n_1$ and an azimuthal coordinate $0 \leq j \leq n_2$. The short edge $j = n_2$ is blocked and the other edges are free. A weight-like vertical force $g_{(i,j)}$ is applied on the nodes, with uniform density $\Delta$ per unit area.

The elasticity of the spring (elements) is set up as follows. Considering a spring $\alpha$ having adjacent nodes ($i_\alpha^-$, $j_\alpha^-$) and ($i_\alpha^+, j_\alpha^+$), we define the undeformed end-to-end vector as $\delta_\alpha = X_{(i_\alpha^+, j_\alpha^+)} - X_{(i_\alpha^-, j_\alpha^-)}$, the natural spring length as $\ell_\alpha = \|\delta_\alpha\|$, the undeformed unit tangent $t_\alpha = \delta_\alpha / \ell_\alpha$ and the linearized spring elongation $z_\alpha$ as

$$z_\alpha = t_\alpha \cdot (v_{(i_\alpha^+, j_\alpha^+)} - v_{(i_\alpha^-, j_\alpha^-)}). \tag{7.15}$$

The spring is assigned an elastic energy $w_\alpha = \frac{1}{2} k_\alpha z_\alpha^2$, where $k_\alpha$ is the spring constant which we set up as

$$k_\alpha = \frac{EA}{\ell_\alpha} \tag{7.16}$$

to represent elastic bars having all the same traction modulus $EA$. The short edge $j = n_2$ is blocked and a force

$$g_{(i,j)} = -\Delta \, (\eta r_i)^2 \, e_2 \tag{7.17}$$

is applied on the node $(i, j)$, representing a weight-like force having a constant density $(-\Delta e_2)$ per unit area, where $r_i = \exp(\eta i)$ is the radial coordinate of the node. The nodal displacements $v_\beta$ at equilibrium are proportional to $\Delta$ and independent of the global elasticity constant $EA$. The invariable forces $g_{(i,j)}$ are taken into account through their potential energy $-\sum_{i=-n_1}^{+n_1} \sum_{j=0}^{n_2} g_{(i,j)} \cdot v_{(i,j)}$.

Typical results of the discrete simulations are shown in Figure 7.7 (deformed lattice), along with a comparison to the predictions of homogenization (colors) which we proceed to derive now.

We use homogenization to characterize the limit where both discretization parameters $n_1$ and $n_2$ are large: the scale separation parameter $\eta = \mathcal{O}(1/n_2)$ is then small, $\eta \ll 1$. We limit attention to the case of a stubby arch, and not a slender one, by assuming that the aspect-ratio parameter $\delta = \mathcal{O}(n_1/n_2)$ remains finite.

The periodic topological lattice underlying the arch is shown in Figure 7.6a. Its nodes are positioned on a square lattice spanning a rectangular domain, $\tilde{X}_{(i,j)} = i\,e_1 + j\,e_2$, with $-n_1 \leqslant i \leqslant +n_1$ and $0 \leqslant j \leqslant n_2$. There are $n_\varphi = 6$ element families (colors in Figure 7.6a) whose properties are listed in Table 7.2. In this imaginary configuration, the area density of the elements belonging to any particular family $\varphi$ is $\tilde{\rho}_\varphi = 1/2$.

| $d$ | $n_{\mathrm{b}}$ | $n_\varphi$ | $\tilde{\rho}_\varphi$ |
|---|---|---|---|
| 2 | 2 | 6 | 1/2 |

| $\varphi$ | I | II | III | IV | V | VI |
|---|---|---|---|---|---|---|
| $b_\varphi^-$ | 1 | 2 | 2 | 1 | 1 | 1 |
| $b_\varphi^+$ | 2 | 1 | 1 | 2 | 1 | 1 |
| $\tilde{\boldsymbol{\delta}}_\varphi$ | $e_2$ | $e_2$ | $e_1$ | $e_1$ | $e_1 + e_2$ | $e_1 - e_2$ |

**Table 7.2.** Properties of the topological square lattice underlying the arch shown Figure 7.6a: space dimension $d$, number $n_{\mathrm{b}}$ of Bravais sub-lattices, number $n_\varphi$ of element (spring) families, element density $\tilde{\rho}_\varphi$, connectivity $b_\varphi^\pm$, and end-to-end vectors $\tilde{\boldsymbol{\delta}}_\varphi$.

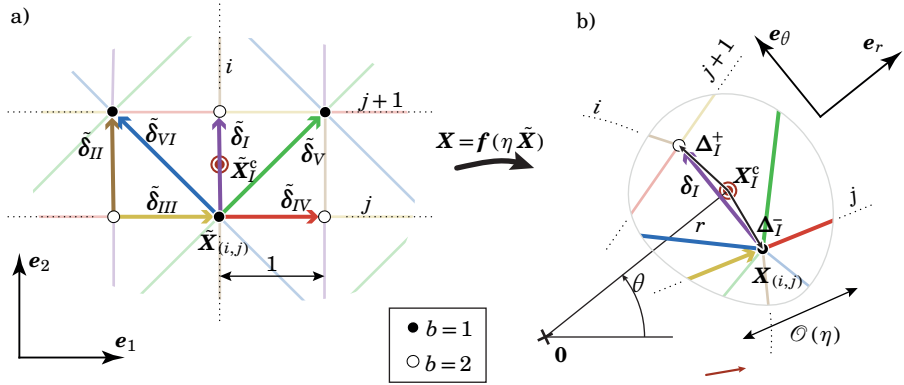

**Figure 7.6.** Circular arch: homogenization conventions. (a) Underlying periodic topological lattice, with the different element families represented by colors. (b) Close-up view of particular element of type $\varphi = I$ in reference configuration, drawn with a large value of the cell size $\sim \eta$ for the sake of legibility. Note that the element center $X_I^c$ (target-like symbol, defined as the image of the element midpoint $\tilde{X}_I^c$ by the diffeomorphism $f$) is distinct from the spring midpoint.

The node positions $X_{(i,j)}$ in (7.14) are produced by applying the map $X_{(i,j)} = f(\eta\,\tilde{X}_{(i,j)})$ in (2.3) to the topological lattice $\tilde{X}_{(i,j)}$ using an exponential-map diffeomorphism $f$ similar to (2.4),

$$f(\tilde{x}) = \exp(\tilde{x}_1)\,e_r(\tilde{x}_2). \tag{7.18}$$

The arch therefore complies with the lattice generation framework presented in Section 2. The Jacobian can be calculated as $J_f(\tilde{x}) = \exp(2\,\tilde{x}_1)$, and Equation (3.40) yields the density of springs in a given family $\varphi$ as

$$\rho_\varphi(r,\theta) = \frac{1}{2\,(\eta\,r)^2}, \tag{7.19}$$

where $(r,\theta)$ are the polar coordinates in reference configuration, $X = r\,e_r(\theta)$.

As shown by the target-like symbols in Figures 2.1b' and 7.6b, the link centers $X_\alpha^c$, conventionally defined as the image by the diffeomorphism of the edge midpoints, see (2.5), do not coincide with the midpoints. The vectors $\boldsymbol{\Delta}_\varphi^\pm$ yielding the endpoints $\pm$ relative to the center $X_\alpha^c$ (thin black arrows in Figure 7.6b) are given by Equation (2.7) in terms of the family index $\varphi$, the scale separation parameter $\eta$ and of the polar coordinates $(r,\theta)$ of the center $X_\alpha^c$ as

$$\boldsymbol{\Delta}_\varphi^\pm(\eta,r,\theta) = r\exp\left(\pm\frac{\eta}{2}\tilde{\boldsymbol{\delta}}_\varphi \cdot e_1\right) e_r\left(\theta \pm \frac{\eta}{2}\tilde{\boldsymbol{\delta}}_\varphi \cdot e_2\right) - r\,e_r(\theta), \tag{7.20}$$

where the edge vectors $\tilde{\boldsymbol{\delta}}_\varphi$ of the topological lattice are listed in Table 7.2.

In the homogenization code, the lattice is initialized with the information listed in Table 7.2, along with the expressions of $\rho_\varphi$ in (7.19) and $\boldsymbol{\Delta}_\varphi^\pm$ in (7.20).

The following changes are made to the code to handle *spring* elements, as opposed to the *beam* elements that have been used so far. The rotational degrees of freedom $\boldsymbol{\psi}_b$ are dropped. There is a *single* strain ($n_{E_\varphi} = 1$) attached to a spring $\alpha$, namely the elongation $z_\alpha$: the element strain in (2.12) becomes $E_\alpha = (z_\alpha)$. In view of the expression (7.15) of the elongation, we change the displacement–strain matrix in (2.14–2.15) to

$$\mathscr{D}_\varphi(\eta,r,\theta) = \begin{pmatrix} -\hat{\boldsymbol{t}}_\varphi(\eta,r,\theta) \\ +\hat{\boldsymbol{t}}_\varphi(\eta,r,\theta) \end{pmatrix} \in \mathbb{T}^{(1,2,2)} \qquad \text{(2D springs)}, \tag{7.21}$$

where $\hat{\boldsymbol{t}}_\varphi$ is still defined by (2.13). The choice of the spring constants in (7.16) is captured by setting up the element stiffness matrix as

$$\mathcal{H}_\varphi(\eta, r, \theta) = \left( \frac{EA}{\hat{\ell}_\varphi(\eta, r, \theta)} \right) \in \mathbb{T}^{(1,1)} \qquad \text{(springs)}, \tag{7.22}$$

so that (2.20) yields the elastic energy of the springs.

As anticipated in Section 2, the properties (such as $\mathcal{H}_\varphi$, $\rho_\varphi$, etc.) of this curved lattice depend on position $\boldsymbol{X}$ through the polar coordinates $(r, \theta)$. To inform the library that the symbols $r$ and $\theta$ vary in space, we set up the list of variable material parameters as

$$\boldsymbol{m} = (r, \theta). \tag{7.23}$$

The library furnishes the homogenized tensors in terms $\boldsymbol{m}$ and $\nabla \boldsymbol{m}$, treated as symbolic quantities: in the companion Mathematica notebook homogenize-circulararch.nb, we insert the explicit expression of the gradient $\nabla \boldsymbol{m} = (\boldsymbol{e}_r(\theta), \boldsymbol{e}_\theta(\theta)/r)$ in this symbolic output, where $\boldsymbol{e}_\theta(\theta) = -\sin\theta\,\boldsymbol{e}_1 + \cos\theta\,\boldsymbol{e}_2$ is the azimuthal unit vector.

The homogenization proceeds automatically as usual, except that the symbolic expressions of $\boldsymbol{E}_l(\eta; r, \theta)$, …, $\boldsymbol{E}''_y(\eta; r, \theta)$ obtained by (3.23) are so long that they make the procedure hang: to work around this, we replace $\boldsymbol{E}_l(\eta; r, \theta)$, …, $\boldsymbol{E}''_y(\eta; r, \theta)$ by their Taylor expansions with respect to $\eta$ to order $\eta^3$, which are considerably simpler. These quantities being all of order $\eta$, expanding to order $\eta^3$ warrants a *relative* accuracy of order $\eta^2$ is preserved, which is consistent with our second-order homogenization scheme.

The second-order homogenization procedure delivers the microscopic displacement in the form

$$\boldsymbol{\xi}^\star_b = (-1)^{b+1} r \eta \left( 0\,\eta^0 + \frac{\eta^1}{8\sqrt{2}} \left[ \left( -\mathrm{tr}\,\boldsymbol{\varepsilon} + r\,\frac{\partial \mathrm{tr}\,\boldsymbol{\varepsilon}}{\partial r} + 2\,\frac{\partial \varepsilon_{r\theta}}{\partial \theta} \right) \boldsymbol{e}_r + \left( 2\varepsilon_{r\theta} + \frac{\partial \mathrm{tr}\,\boldsymbol{\varepsilon}}{\partial \theta} + 2r\,\frac{\partial \varepsilon_{r\theta}}{\partial r} \right) \boldsymbol{e}_\theta \right] + \mathcal{O}(\eta^2) \right). \tag{7.24}$$

where $\mathrm{tr}\,\boldsymbol{\varepsilon} = \varepsilon_{rr} + \varepsilon_{\theta\theta}$, and $\varepsilon_{rr}$, $\varepsilon_{\theta\theta}$ and $\varepsilon_{r\theta}$ are the components of the macroscopic strain in the polar basis $(\boldsymbol{e}_r, \boldsymbol{e}_\theta)$. This prediction is verified by comparing to discrete simulations following the usual procedure in Figure 7.7. The vanishing contribution $0\,\eta^0$ in the parenthesis in (7.24) points to the fact that the leading-order microscopic displacement is zero for this particular truss: it does not help improving the error, $e^\alpha_{[0]} = e^\alpha_{[\times]}$ (Figure 7.7, top). The higher-order term in (7.24) improves the error substantially and uniformly in the interior (Figure 7.7, bottom).

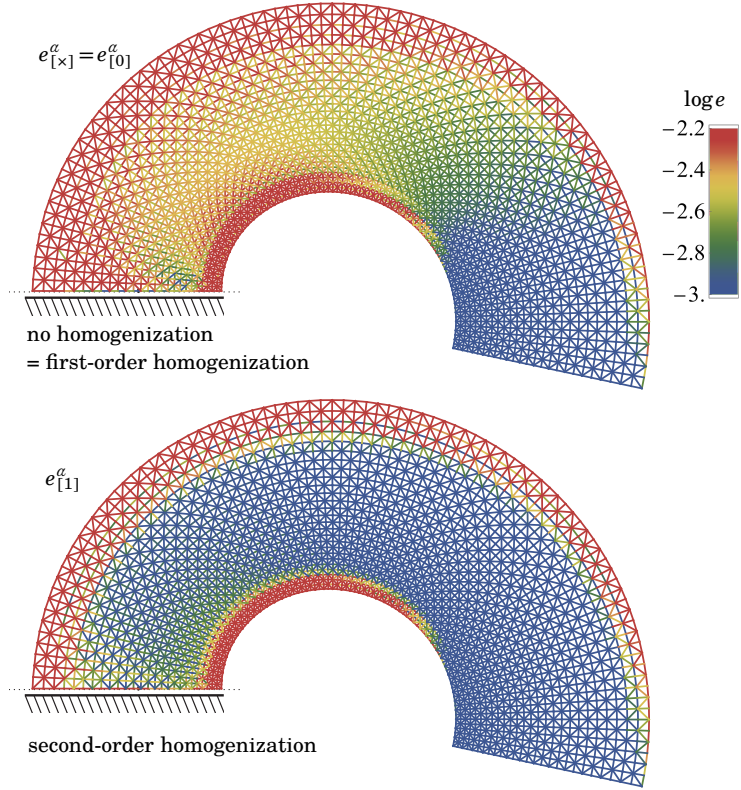

**Figure 7.7.** Circular arch subjected to a weight-like force $-\Delta\,\boldsymbol{e}_2$ per unit area, with discretization parameters $(n_1, n_2) = (13, 90)$ and loading intensity $\Delta = 0.1$. The logarithmic color maps represents the error on the microscopic displacement: the first non-zero prediction from homogenization in (7.24) comes from second-order (bottom).

**Remark 7.4.** The microscopic displacement in (7.24) can be rewritten as

$$\boldsymbol{\xi}^\star_b = (-1)^{b+1} r \eta \left( 0\,\eta^0 + \frac{\eta^1}{8\sqrt{2}} \left[ (\mathcal{J}_p + \boldsymbol{e}_r \otimes (\mathcal{J}_r - \boldsymbol{I}_2)) : \boldsymbol{\varepsilon} + r\,(\nabla \mathrm{tr}\,\boldsymbol{\varepsilon} + \mathcal{J}_\theta \cdot \nabla (\mathcal{J}_\theta : \boldsymbol{\varepsilon})) \right] + \mathcal{O}(\eta^2) \right), \tag{7.25}$$

where $\mathcal{J}_p$ is the polar version of the tensor $\mathcal{J}$ introduced in (5.5), $\mathcal{J}_p = e_r \otimes \mathcal{J}_r + e_\theta \otimes \mathcal{J}_\theta$ with $\mathcal{J}_r = -e_r \otimes e_r + e_\theta \otimes e_\theta$ and $\mathcal{J}_\theta = e_r \otimes e_\theta + e_\theta \otimes e_r$. In the presence of variable properties, the corrective displacement not only includes a term proportional to $\nabla \varepsilon$ but also one proportional to $\varepsilon$ (see the first term in the square bracket above). The latter is produced by inserting $\nabla m = (e_r(\theta), e_\theta(\theta)/r)$ into the corrective $\nabla m \otimes \varepsilon$ term returned by the library in symbolic form.

### 7.4. Shearing and twisting of a 3D elastic truss

We now address an example in dimension $d = 3$, namely an elastic lattice made up of springs whose unit cell is shown in Figure 7.8. The unit cell dimensions is $\ell \times \ell \times \sqrt{3}\,\ell$. There are $n_\varphi = 11$ different types of bars: their length is $\ell_\alpha = \ell$ for types $\varphi \in \{I, VI\}$, $\ell_\alpha = \sqrt{3}\,\ell$ for type $\varphi = XI$, and $\ell_\alpha = \sqrt{5}\,\ell/2$ for the other types $\varphi$. The spring constants $k_\alpha$ in Equation (7.16) are used, representing elastic bars have all the same stretching modulus $EA$. As earlier in 2D, the spring elongation $z_\alpha$ is given by (7.15) and their energy is of the form $w_\alpha = k_\alpha z_\alpha^2/2$.

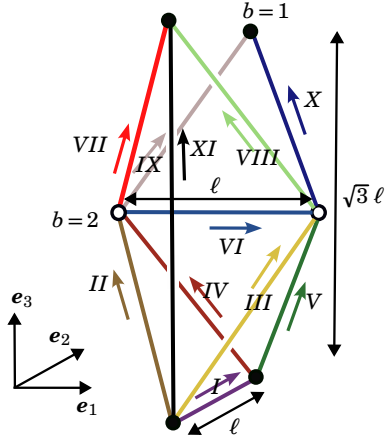

**Figure 7.8.** Unit cell of a 3D lattice made up of springs, with box dimensions $\ell \times \ell \times \sqrt{3}\,\ell$. The lattice has $n_b = 2$ Bravais sub-lattices (solid vs. open disks) and $n_\varphi = 11$ element families (colors). The springs are treated as elastic bars having identical stretching modulus $EA$.

In Figure 7.9, a shear test is conducted on a block comprising $20 \times 10 \times 10$ unit cells: the displacement of the nodes belonging to the face shown in the left-side is blocked, whereas that of those belonging to the opposite face is set to $\Delta\,e_3$, with $\Delta = 0.1\,\ell$. In Figure 7.10, a twisting test is conducted on a block comprising $30 \times 15 \times 11$ unit cells: a rotation by an angle $\Delta$ is imposed on the second face, with $\Delta = 0.1$ (only the components of the displacement in the perpendicular plane $(e_2, e_3)$ are prescribed, the longitudinal component along $e_1$ remaining unconstrained).

| $d$ | $n_b$ | $n_\varphi$ | $\rho_\varphi$ |
|---|---|---|---|
| 3 | 2 | 11 | $(\sqrt{3}\,\ell^3)^{-1}$ |

**Table 7.3.** Parameters used to set up the 3D elastic truss in the homogenization code, see Figures 7.8–7.10. The connectivity matrix $b_\varphi^\pm$ and the edge vectors $\delta_\varphi$ are omitted as they can be read off easily from Figure 7.8.

The homogenization of the 3D truss is carried out in the companion notebook `homogenize-3Dtruss.nb`. The lattice is set up using the parameters listed in Table 7.3 (the connectivity matrix $b_\varphi^\pm$ and the edge vectors $\boldsymbol{\delta}_\varphi$ are provided to the library but are not included in the table). To represent the springs, the 3D version of the displacement-to-strain matrix $\mathscr{D}_\varphi$ in (7.21) is used, along with the element stiffness $\mathscr{K}_\varphi$ in (7.22). For this particular truss, the leading-order microscopic displacement vanishes, $\boldsymbol{Y}_0 = \boldsymbol{0}$, and the first-order one is returned by the library in the form

$$\boldsymbol{\xi}_b^\star = (-1)^{b+1} \ell \left( 0\,\ell^0 + \frac{5\sqrt{5}}{64}\ell \left( -\varepsilon_{11,1}\boldsymbol{e}_1 + \varepsilon_{22,2}\boldsymbol{e}_2 + \frac{1}{\sqrt{3}}\varepsilon_{33,3}\boldsymbol{e}_3 \right) + \mathcal{O}(\ell^2) \right). \tag{7.26}$$

This prediction is verified against numerical simulations in Figures 7.9 and 3.4, by the usual procedure. As in all the previous examples, a significant improvement is brought about by second-order homogenization, except in the vicinity of the edges.

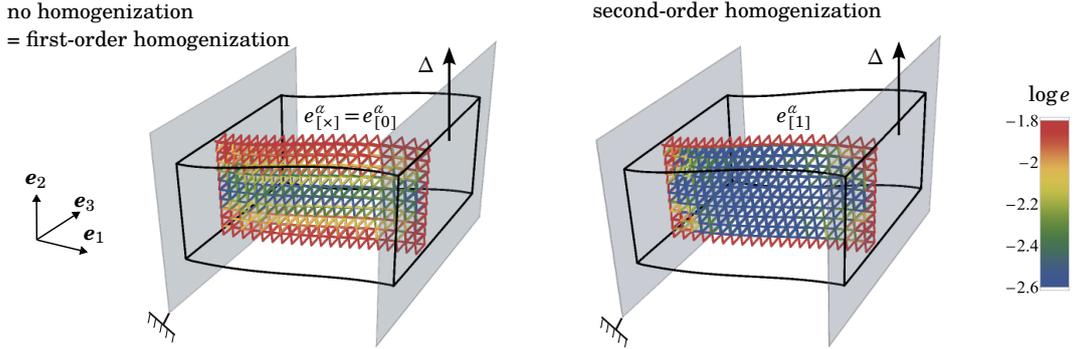

**Figure 7.9.** 3D truss, shearing test. The springs are colored according to the magnitude of the microscopic displacement (no homogenization) or of the error the displacement predicted by first-order or second-order homogenization. To aid visualization, a single layer of springs is shown, the full lattice domain being shown by the black box.

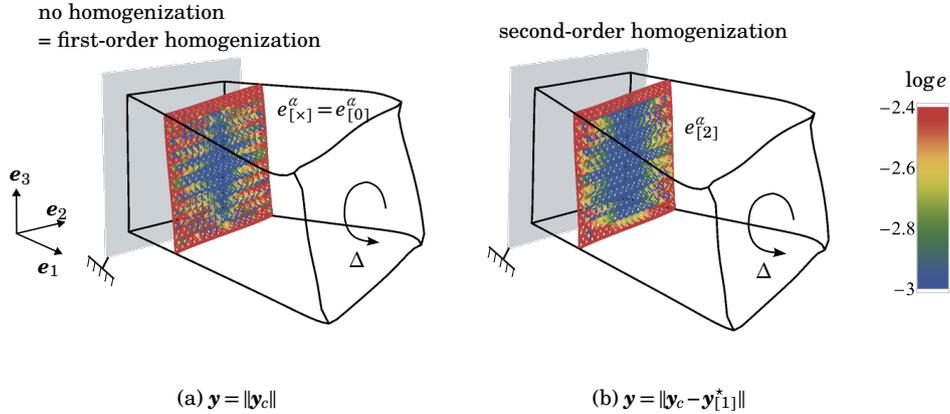

(a) $\boldsymbol{y} = \|\boldsymbol{y}_c\|$          (b) $\boldsymbol{y} = \|\boldsymbol{y}_c - \boldsymbol{y}_{[1]}^\star\|$

**Figure 7.10.** 3D truss, twisting test. Same conventions as in Figure 7.9. On the face shown in foreground, the rotation with angle $\Delta$ is imposed on the $\boldsymbol{e}_2$ and $\boldsymbol{e}_3$ (normal) components of the displacement, the $\boldsymbol{e}_1$ (longitudinal) component remains unconstrained.

## 7.5. A pantograph

Our final example is the 1D elastic truss shown in Figure 7.11, known as a *pantograph*. It is made up of springs having identical elastic constants $k$, except for the vertical springs whose elastic constant is $\chi k \ll k$; we address the case of a small elastic contrast, $\chi \ll 1$. The homogenization of pantograph-like lattices has been studied in a number of work, see for instance [SAd11; AS18b]. More broadly, microstructures featuring large elastic contrast have been extensively studied, see for instance the work of [PS97; Bel17; JS20]. By recovering known results about the pantograph, we demonstrate how our method can handle the special case of a large elastic contrast.

In all the previous illustrations given in this paper, the gradient effect was a *correction* to the leading order effect—our homogenization scheme has indeed been derived with this situation in mind. The pantograph is different as the strain-gradient term $d\varepsilon_{11}/dX_1$ ends up contributing to its energy at the same order as the strain $\varepsilon_{11}$, see Equation (7.29) below. Our homogenization scheme can be revisited to deal with this situation, as we show now.

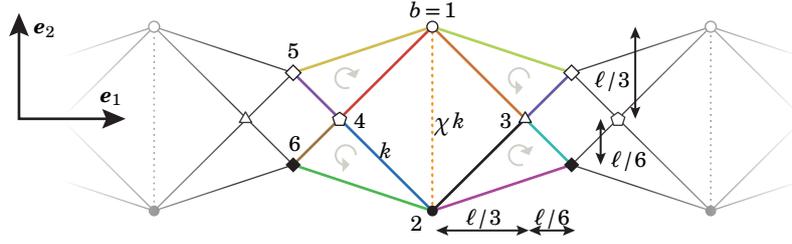

**Figure 7.11.** A pantograph is a 1D truss made up of springs. All springs have the same constant $k$, except for the vertical dashed springs whose spring constant is $\chi k$. We consider the limit of large elastic contrast $\chi \ll 1$. There are $n_b = 6$ Bravais sub-lattices (symbols and integers) and $n_\varphi = 13$ different types of springs: a unit cell is shown using colored segments. In the limit $\chi = 0$ where the dashed vertical springs are absent, the truss possesses a zero-energy mechanism. This is the same lattice as studied in [AS18b], except that beams have been replaced by springs and the weak (dashed vertical) springs have been added.

The pantograph is homogenized in the companion Mathematica notebook `homogenize-pantograph-springs-highcontrast.nb`. In is drawn in the Euclidean plane ($d = 2$), see Figure 7.11, but the limit model is a one-dimensional bar. We could write specialized code handling 2D lattices that are 1D in the limit but a straightforward work-around is to work in the plane and imagine that we are homogenizing an infinite number of replicas of the pantograph obtained by copying the original one, translating it along $e_2$ and pasting it, repeatedly. Concretely, we set up the dimension as $d = 2$ in the library and identify the density per area $\rho_\varphi$ of each family of elements with its density per unit length $\rho_\varphi = 1/\ell$, where $\ell = O(\eta)$ is the size of the unit cell. We also check that the homogenized energy computed by the library does not depend on $\varepsilon_{12}$ and $\varepsilon_{22}$ since the replicas are disconnected.

The homogenized energy returned by the library takes the form

$$\Phi^\star = \frac{k\,\ell}{2}\int_{-\infty}^{+\infty}\left(4\chi c_5(\chi)\,(\varepsilon_{11})^2 + \frac{2}{23}c_6(\chi)\left(\ell\frac{d\varepsilon_{11}}{dX_1}\right)^2 + \mathcal{O}(\eta^3)\right)dX_1, \tag{7.27}$$

where $c_5(\chi) = \left(1 + \frac{7}{2}\chi\right)^{-1}$ and $c_6(\chi) = \left(1 - \frac{19}{54}\chi - \frac{109}{27}\chi^2\right)/\left(1 + \frac{7}{2}\chi\right)^2$ are $\mathscr{C}^\infty$-smooth functions of $\chi$ at $\chi = 0$, with $c_5(0) = c_6(0) = 1$. The underlying microscopic displacement $\boldsymbol{\xi}_b^\star$ is available from the companion notebook. We have not yet specified how $\chi$ scales with the cell size $\eta$, and the above result is correct as long as $\chi$ remains bounded, $\chi = \mathcal{O}(\eta^0)$, see Remark 2.2.

The limit of large elastic contrast is now addressed by taking the *formal* limit $\chi \to 0$ in (7.27). The distinguished limit where both $\chi \to 0$ and $\eta \to 0$ with

$$\chi = \mathcal{O}(\ell^2) = \mathcal{O}(\eta^2) \tag{7.28}$$

is particularly interesting, as it makes the two terms balanced in the integral: at leading order, we then have

$$\Phi^\star = \frac{4k\,\ell\,\chi}{2}\int_{-\infty}^{+\infty}\left((\varepsilon_{11})^2 + \frac{1}{46}\left(\frac{\ell}{\chi^{1/2}}\frac{d\varepsilon_{11}}{dX_1}\right)^2 + \mathcal{O}(\eta)\right)dX_1, \tag{7.29}$$

where $\ell/\chi^{1/2} = \mathcal{O}(1)$ is the macroscopic persistence length of the gradient effect. This energy can be interpreted based on the existence of a soft mode having zero energy in the limit $\chi \to 0$ (vertical dashed springs are removed). This soft mode involves combined rigid rotations of the triangles, as shown by the light gray circular arrows in Figure 7.11, with an amplitude proportional to the macroscopic strain $\varepsilon_{11}$. The first term in the integral in (7.29) can be rewritten as $\frac{1}{2}\int (k\chi)(\ell\varepsilon_{11})^2 dX_1/\ell$, which describes the small energy penalty brought about by the weak springs when the soft mode is activated. Classical homogenization captures only this term, yielding a non-definite energy functional $\frac{1}{2}\int 0 \times (\ell\varepsilon_{11})^2 dX_1/\ell$ in the limit of large elastic contrast $\chi = 0$. The gradient effect solves this issue.

The pantograph example shows that the special behavior of high-contrast microstructures, which has been documented in the literature, can be readily recovered by taking the formal limit $\chi \to 0$ in the output of the homogenization library. The *symbolic* and *second-order* capabilities of the library are instrumental here.

**Remark 7.5.** For consistency with the scaling assumption that the energy is $\Phi^\star = \mathcal{O}(\eta^0)$, one should view the spring constant $k$ as quantity of order $1/\ell^3 = \mathcal{O}(\eta^{-3})$. However, the magnitude of this overall multiplicative constant in the energy is a matter of convention; nothing changes if different scaling assumptions are used.

## 8. DISCUSSION AND CONCLUSION

We have proposed a symbolic, second-order homogenization method for elastic lattices. It is distributed as a library, named `shoal`, implemented in the symbolic calculation language Wolfram Mathematica: given a description of the lattice, the library runs automatically and returns both the localization tensors in (4.3) and the homogenized energy (4.5) in symbolic form.

In the design of the library, emphasis has been placed on versatility: it works in arbitrary dimension $d$ ($1 \leqslant d \leqslant 3$), can handle lattices featuring graded geometric and/or elastic properties, pre-stress or a large elastic contrast. Elastic lattices made up of springs or beams have been demonstrated and new types of elements can be created easily. We are not aware of previous work addressing both graded properties and pre-stress in the context of higher-order, discrete homogenization.

For a variety of lattices, we have derived in closed analytical form both the nodal displacements and rotations—see Equations (5.10), (7.3), (7.11), (7.25), (7.26)—and the homogenized energy—see Equations (5.9), (7.29)—up to second order. Expressions of this kind are novel to the best of our knowledge. Being symbolic, they fully convey the influence of the various lattice parameters. We hope that this kind of results will be used in the future to gain analytical insights into the effective properties of various lattices.

We have proposed a procedure for verifying the correctness (in an asymptotic sense) of the microscopic displacement predicted by the library, based on comparisons with discrete simulations, see Figure 6.4. We have presented a number of validation examples for which we have plotted physical maps of the homogenization error. The improvement in accuracy brought about by second-order homogenization is significant everywhere except in layers forming at free boundaries, where the assumption of scale separation breaks down.

Until these boundary layers have been analyzed and accounted for in the homogenized energy $\Phi^\star$, it is not possible to solve global equilibrium problems for the lattice by making the functional $\Phi^\star$ stationary. This is one main limitation of the present work, which is just a step towards this goal—a rather ambitious one in the context of higher-order homogenization. The derivation of the *other* boundary terms in (4.5)—produced by homogenization and not by boundary layers—is original to the best of our knowledge and will need to be included in the homogenized energy functional predicting equilibrium.

There is a second reason that prevents the homogenized energy $\Phi^\star$ from being used in self-contained simulations, at least in its current form. As noted in Remark 5.3 and in our previous work [AL23], it contains negative strain-gradient moduli, such as the coefficient –7 appearing in the second line of (5.9), and is therefore non-positive. This is a common finding in higher-order homogenization [LM18; DLSS22]. This strong limitation explains the limited popularity of higher-order homogenization so far. In future work, we hope to introduce a regularization strategy that preserves the accuracy of the solution while restoring the desired positivity.

The library is *symbolic* and does *second-order* homogenization. When combined together, these two original features are quite powerful. Consider for instance the challenging class of lattice homogenization problems involving a small dimensionless parameter $\chi \ll 1$ (such as the slenderness ratio $EI/[EA\,\ell^2]$ in beam lattices or the elastic contrast, see §7.5) in addition to the scale separation parameter $\eta \ll 1$. Non-conventional limit behaviors can be obtained in such lattices, which have been approached using specialized methods in the literature. These limit behaviors are often associated with the presence of a large-amplitude, slowly decaying soft mode (Remark 7.3 in §7.2, §7.5). Our approach provides both a unifying perspective and a straightforward approach to these problems: by homogenizing ($\eta \to 0$) with fixed $\chi$ first, and taking a distinguished limit in $\chi$ and $\eta$ in the homogenized energy next, one can identify the limit energy based on simple dimensional analysis (§7.5). This approach involves exchanging the limits on $\chi$ and $\eta$, which is not mathematically rigorous but yields correct results in all the cases we have tried.

Along the same lines, albeit at a more basic (and less original) level, the homogenization procedure handles bending-dominated and stretching-dominated lattices in a unified way: the class of a particular lattice can be deducted from the symbolic expression of the leading-order homogenized energy, see the discussion immediately after Equation (5.3).

The proposed method is subject to the following practical limitations. (*i*) The library solves linear algebra problems in symbolic form, which can require a large amount of computing time especially if there are many microscopic degrees of freedom. When this happens, it is possible to revert to *numeric* homogenization; this is done easily, by assigning numerical values to the lattice parameters. The symbolic computations required to generate the results included in this paper all took from a few seconds to less than a minute on a personal computer. (*ii*) The homogenized energy and microscopic displacement computed by the library have been cast by hand into compact tensor form, shown for instance in Equation (5.9–5.10): this process could in principle be automated by exploiting the symmetries of the lattice. (*iii*) We have homogenized discrete lattices made up of 1D beams of springs, without questioning the validity of these 1D models. To address this question, one needs to revert to the equations of *bulk* elasticity in a periodic domain comprising slender junctions [AB21; DLSS22].

In this work, we have limited attention to *linear elastic* lattices and thus have not considered material or geometrical nonlinearity. When treated incrementally, homogenization problems for lattices having a more general constitutive behavior (such as non-linearly elastic, visco-elastic or elasto-plastic) yield linearized problems featuring both pre-stress and spatially graded properties (such as the tangent elastic stiffness). This is precisely the kind of problems that our approach can solve. In future work, we plan to apply our homogenization scheme incrementally. By doing so, we hope to gain insights into the homogenizing of lattices featuring a wide range of constitutive behaviors.

**Acknowledgements** Yang Ye thanks the *China Scholarship Council* (CSC) for his doctoral funding. This manuscript has been typeset using TeX_MACS, a powerful, multi-platform and freely distributed scientific editor [vdHGG+13].

# BIBLIOGRAPHY

[21]   Wolfram Research, Inc. Mathematica, Version 13.0.0. 2021. Champaign, IL, 2021.

[AB21] Bilen Emek Abali and Emilio Barchiesi. Additive manufacturing introduced substructure and computational determination of metamaterials parameters by means of the asymptotic homogenization. *Continuum Mechanics and Thermodynamics*, 33(4):993–1009, 2021.

[ABB10] Nicolas Auffray, Regis Bouchet, and Y Brechet. Strain gradient elastic homogenization of bidimensional cellular media. *International Journal of Solids and Structures*, 47(13):1698–1710, 2010.

[AG97] Michael F Ashby and Lorna J Gibson. Cellular solids: structure and properties. *Press Syndicate of the University of Cambridge, Cambridge, UK*, pages 175–231, 1997.

[AKO17] N. Auffray, B. Kolev, and M. Olive. Handbook of bi-dimensional tensors: part I: harmonic decomposition and symmetry classes. *Mathematics and Mechanics of Solids*, 9(22):1847–1865, 2017.

[AL23] B. Audoly and C. Lestringant. An energy approach to asymptotic, higher-order, linear homogenization. *Journal of Theoretical, Computational and Applied Mechanics*, 2023.

[AS18a] Houssam Abdoul-Anziz and Pierre Seppecher. Homogenization of periodic graph-based elastic structures. *Journal de l'École polytechnique, Mathématiques*, 5:259–288, 2018.

[AS18b] Houssam Abdoul-Anziz and Pierre Seppecher. Strain gradient and generalized continua obtained by homogenizing frame lattices. *Mathematics and mechanics of complex systems*, 6(3):213–250, 2018.

[ASB19] Houssam Abdoul-Anziz, Pierre Seppecher, and Cédric Bellis. Homogenization of frame lattices leading to second gradient models coupling classical strain and strain-gradient terms. *Mathematics and Mechanics of Solids*, 24(12):3976–3999, 2019.

[Aud23] B. Audoly. The shoal library. https://archive.softwareheritage.org/browse/origin/https://git.renater.fr/anonscm/git/shoal/shoal.git, 2023.

[Bel17] M. Bellieud. Homogenization of stratified elastic composites with high contrast. *SIAM Journal on Mathematical Analysis*, 49(4):2615–2665, 2017.

[Bou96] Claude Boutin. Microstructural effects in elastic composites. *International Journal of Solids and Structures*, 33(7):1023–1051, 1996.

[Bou19] C. Boutin. Homogenization methods and generalized continua in linear elasticity. In H. Altenbach and A. Öchsner, editors, *Encyclopedia of Continuum Mechanics*. Springer, Berlin, Heidelberg, 2019.

[BT94] Scott Bardenhagen and Nicolas Triantafyllidis. Derivation of higher order gradient continuum theories in 2, 3-d non-linear elasticity from periodic lattice models. *Journal of the Mechanics and Physics of Solids*, 42(1):111–139, 1994.

[CMR06] Denis Caillerie, Ayman Mourad, and Annie Raoult. Discrete homogenization in graphene sheet modeling. *Journal of Elasticity*, 84(1):33–68, 2006.

[CP12] Doina Cioranescu and Jeannine Saint Jean Paulin. *Homogenization of reticulated structures*, volume 136. Springer Science & Business Media, 2012.

[Dav13] Cesare Davini. Homogenization of linearly elastic honeycombs. *Mathematics and Mechanics of Solids*, 18(1):3–23, 2013.

[DG12] F Dos Reis and JF Ganghoffer. Construction of micropolar continua from the asymptotic homogenization of beam lattices. *Computers & Structures*, 112:354–363, 2012.

[DLSS22] Baptiste Durand, Arthur Lebée, Pierre Seppecher, and Karam Sab. Predictive strain-gradient homogenization of a pantographic material with compliant junctions. *Journal of the Mechanics and Physics of Solids*, page 104773, 2022.

[DO11] Cesare Davini and Federica Ongaro. A homogenized model for honeycomb cellular materials. *Journal of Elasticity*, 104(1):205–226, 2011.

[Dum86] Hélène Dumontet. Study of a boundary layer problem in elastic composite materials. *ESAIM: Mathematical Modelling and Numerical Analysis*, 20(2):265–286, 1986.

[DYF+20] Bolei Deng, Siqin Yu, Antonio E Forte, Vincent Tournat, and Katia Bertoldi. Characterization, stability, and application of domain walls in flexible mechanical metamaterials. *Proceedings of the National Academy of Sciences*, 117(49):31002–31009, 2020.

[FDA10] Norman A Fleck, Vikram S Deshpande, and Michael F Ashby. Micro-architectured materials: past, present and future. *Proceedings of the Royal Society A: Mathematical, Physical and Engineering Sciences*, 466(2121):2495–2516, 2010.

[GLTK19] Raphaël N Glaesener, Claire Lestringant, Bastian Telgen, and Dennis M Kochmann. Continuum models for stretching- and bending-dominated periodic trusses undergoing finite deformations. *International Journal of Solids and Structures*, 171:117–134, 2019.

[HF06] R. G. Hutchinson and N. A. Fleck. The structural performance of the periodic truss. *Journal of the Mechanics and Physics of Solids*, 54:756–782, 2006.

[JS20] L. Jakabčin and P. Seppecher. On periodic homogenization of highly contrasted elastic structures. *Journal of the Mechanics and Physics of Solids*, 144:104104, 2020.

[KM04] Rajesh S Kumar and David L McDowell. Generalized continuum modeling of 2-d periodic cellular solids. *International Journal of solids and structures*, 41(26):7399–7422, 2004.

[LA20] C. Lestringant and B. Audoly. Asymptotically exact strain-gradient models for nonlinear slender elastic structures: a systematic derivation method. *Journal of the Mechanics and Physics of Solids*, 136:103730, 2020.

[LM18] Duc Trung Le and Jean-Jacques Marigo. Second order homogenization of quasi-periodic structures. *Vietnam Journal of Mechanics*, 40(4):325–348, 2018.

[LP16] Y. Le Floch and À Pelayo. Euler—MacLaurin formulas via differential operators. *Advances in Applied Mathematics*, 73:99–124, 2016.

[LR13] Hervé Le Dret and Annie Raoult. Homogenization of hexagonal lattices. *Networks & Heterogeneous Media*, 8(2):541, 2013.

[MA02] Andrei V Metrikine and Harm Askes. One-dimensional dynamically consistent gradient elasticity models derived from a discrete microstructure: part 1: generic formulation. *European Journal of Mechanics-A/Solids*, 21(4):555–572, 2002.

[MA06] AV Metrikine and H Askes. An isotropic dynamically consistent gradient elasticity model derived from a 2d lattice. *Philosophical Magazine*, 86(21-22):3259–3286, 2006.

[ME68] R. D. Mindlin and N. N. Eshel. On first strain-gradient theories in linear elasticity. *International Journal of Solids and Structures*, 4:109–124, 1968.

[NCH20] Hussein Nassar, Hui Chen, and Guoliang Huang. Microtwist elasticity: a continuum approach to zero modes and topological polarization in kagome lattices. *Journal of the Mechanics and Physics of Solids*, 144:104107, 2020.

[PS97] C. Pideri and P. Seppecher. A second gradient material resulting from the homogenization of an heterogeneous linear elastic medium. *Continuum Mechanics and Thermodynamics*, 9:241–257, 1997.

[PTC22] Paul Plucinsky, Ian Tobasco, and Paolo Celli. Continuum field theory for the deformations of planar kirigami. *Bulletin of the American Physical Society*, 2022.

[RA16] Giuseppe Rosi and Nicolas Auffray. Anisotropic and dispersive wave propagation within strain-gradient framework. *Wave Motion*, 63:120–134, 2016.

[RA19] Giuseppe Rosi and Nicolas Auffray. Continuum modelling of frequency dependent acoustic beam focussing and steering in hexagonal lattices. *European Journal of Mechanics-A/Solids*, 77:103803, 2019.

[RAPG19] Ondrej Rokoš, Maqsood M Ameen, Ron HJ Peerlings, and Mark GD Geers. Micromorphic computational homogenization for mechanical metamaterials with patterning fluctuation fields. *Journal of the Mechanics and Physics of Solids*, 123:119–137, 2019.
[RDK17] Julien Réthoré, Thi Bach Tuyet Dang, and Christine Kaltenbrunner. Anisotropic failure and size effects in periodic honeycomb materials: a gradient-elasticity approach. *Journal of the Mechanics and Physics of Solids*, 99:35–49, 2017.
[RDVB19a] G Rizzi, F Dal Corso, D Veber, and D Bigoni. Identification of second-gradient elastic materials from planar hexagonal lattices. part i: analytical derivation of equivalent constitutive tensors. *International Journal of Solids and Structures*, 176:1–18, 2019.
[RDVB19b] G Rizzi, F Dal Corso, D Veber, and D Bigoni. Identification of second-gradient elastic materials from planar hexagonal lattices. part ii: mechanical characteristics and model validation. *International Journal of Solids and Structures*, 176:19–35, 2019.
[RKD+15] Julien Réthoré, Christine Kaltenbrunner, Thi Bach Tuyet Dang, Philippe Chaudet, and Manuel Kuhn. Gradient-elasticity for honeycomb materials: validation and identification from full-field measurements. *International Journal of Solids and Structures*, 72:108–117, 2015.
[SAd11] P. Seppecher, J.-J. Alibert, and F. dell'Isola. Linear elastic trusses leading to continua with exotic mechanical interactions. *Journal of Physics: Conference Series*, 319:21–25, 2011.
[San80] Évariste Sanchez-Palencia. Non-homogeneous media and vibration theory. *Lecture Note in Physics, Springer-Verlag*, 320:57–65, 1980.
[SC00] Valery P Smyshlyaev and Kirill D Cherednichenko. On rigorous derivation of strain gradient effects in the overall behaviour of periodic heterogeneous media. *Journal of the Mechanics and Physics of Solids*, 48(6-7):1325–1357, 2000.
[SCO+22] Angkur Jyoti Dipanka Shaikeea, Huachen Cui, Mark O'Masta, Xiaoyu Rayne Zheng, and Vikram Sudhir Deshpande. The toughness of mechanical metamaterials. *Nature materials*, 21(3):297–304, 2022.
[TB93] Nicolas Triantafyllidis and S Bardenhagen. On higher order gradient continuum theories in 1-d nonlinear elasticity. derivation from and comparison to the corresponding discrete models. *Journal of Elasticity*, 33(3):259–293, 1993.
[vdHGG+13] J. van der Hoeven, A. Grozin, M. Gubinelli, G. Lecerf, F. Poulain, and D. Raux. GNU TEXmacs: a scientific editing platform. *ACM Communications in Computer Algebra*, 47(1–2):59–61, 2013.
[VDP14] Andrea Vigliotti, Vikram S Deshpande, and Damiano Pasini. Non linear constitutive models for lattice materials. *Journal of the Mechanics and Physics of Solids*, 64:44–60, 2014.
[VP12] Andrea Vigliotti and Damiano Pasini. Linear multiscale analysis and finite element validation of stretching and bending dominated lattice materials. *Mechanics of Materials*, 46:57–68, 2012.

## APPENDIX A. NOTATIONS AND TOOLBOX

### A.1. Tensors

The dimension of the Euclidean space is denoted as $d$. We address both elastic lattices in the plane ($d=2$, see the illustration examples in §6.1–7.3 and §7.5) and in the three-dimensional space ($d=3$, §7.4). The Euclidean space is endowed with an orthonormal Cartesian basis $(\boldsymbol{e}_1,\ldots,\boldsymbol{e}_d)$. A generic point in the Euclidean space is denoted as $\boldsymbol{X} \in \mathbb{R}^d$, see Figure 2.3.

We denote as $\mathbb{T}^{(n_1,n_2,\ldots,n_p)}$ the tensor space $\mathbb{T}^{(n_1,n_2,\ldots,n_p)} = \mathbb{R}^{n_1} \otimes \mathbb{R}^{n_2} \otimes \ldots \otimes \mathbb{R}^{n_p}$ made of tensors $\boldsymbol{G}$ of rank $p$ and dimensions $n_1 \times n_2 \times \cdots \times n_p$. In particular, $\boldsymbol{I}_r \in \mathbb{T}^{(r,r)}$ denotes the identity matrix in dimension $r$ and $\boldsymbol{0}_{n_1 \times \cdots \times n_p} \in \mathbb{T}^{(n_1,n_2,\ldots,n_p)}$ the null tensor with dimensions $n_1 \times \cdots \times n_p$.

Tensors and vectors are denoted using bold symbols, while scalars (including tensor *components*) are denoted using non-bold symbols.

Given two tensors $\boldsymbol{G} \in \mathbb{T}^{(n_1,n_2,\ldots,n_p)}$ and $\boldsymbol{G}' \in \mathbb{T}^{(n'_1,n'_2,\ldots,n'_p)}$, we denote as

- $\boldsymbol{G} \cdot \boldsymbol{G}' \in \mathbb{T}^{(n_1,n_2,\ldots,n_{p-1},n'_2,\ldots,n'_p)}$ their *simple* contraction (whose existence requires $n_p = n'_1$),
- $\boldsymbol{G} : \boldsymbol{G}' \in \mathbb{T}^{(n_1,n_2,\ldots,n_{p-2},n'_3,\ldots,n'_p)}$ their *double* contraction (whose existence requires $n_{p-1} = n'_1$ and $n_p = n'_2$),
- $\boldsymbol{G} \therefore \boldsymbol{G}' \in \mathbb{T}^{(n_1,n_2,\ldots,n_{p-3},n'_4,\ldots,n'_p)}$ their *triple* contraction, (whose existence requires $n_{p-2} = n'_1$, $n_{p-1} = n'_2$ and $n_p = n'_3$),
- etc.

The contracted tensors are given by

$$\begin{aligned}(G \cdot G')_{i_1\ldots i_{p-1}i'_2\ldots i'_p} &= G_{i_1\ldots i_{p-1}j} G'_{ji'_2\ldots i'_p} \\ (G : G')_{i_1\ldots i_{p-2}i'_3\ldots i'_p} &= G_{i_1\ldots i_{p-1}jk} G'_{jki'_2\ldots i'_p} \\ (G \therefore G')_{i_1\ldots i_{p-3}i'_4\ldots i'_p} &= G_{i_1\ldots i_{p-1}jkl} G'_{jkli'_2\ldots i'_p}.\end{aligned} \tag{A.1}$$

Here and elsewhere in the paper, we use Einstein summation whereby any index that is repeated on one side of the equal sign is implicitly summed. Note the ordering of the contracted indices $j$, $k$, $l$, etc. in the right-hand sides.

The action of a matrix $\boldsymbol{G}$ on a vector $\boldsymbol{v}$ is viewed as a special case of the contraction of a tensor of rank 2 with a tensor of rank 1, and is denoted as $\boldsymbol{G} \cdot \boldsymbol{v}$, with a dot.

The outer product of two tensors $\boldsymbol{T}$ and $\boldsymbol{T}'$ is denoted as $\boldsymbol{T} \otimes \boldsymbol{T}'$. In particular, the outer product of two vectors is denoted as $\boldsymbol{v} \otimes \boldsymbol{v}'$. Vector transposition is never used nor is even meaningful.

Given a tensor $\boldsymbol{G} \in \mathbb{T}^{(n_1,n_2,\ldots,n_p)}$, and a permutation $(\sigma_1,\ldots,\sigma_p)$ of the levels $(1,\ldots,p)$ of the tensor, we denote as $\boldsymbol{G}^{T_{\sigma_1\ldots\sigma_p}}$ the generalized transpose of $\boldsymbol{G}$, such that the level $i$ in the original tensor becomes level $\sigma_i$ in the transpose:

$$(G^{T_{\sigma_1\ldots\sigma_p}})_{i_1\ldots i_p} = G_{i_{\sigma_1}\ldots i_{\sigma_p}}. \tag{A.2}$$

For a tensor of rank $p=4$ and the permutation $(\sigma_1,\sigma_2,\sigma_3,\sigma_4) = (1,3,4,2)$, for instance, we have $(G^{T_{1342}})_{ijkl} = G_{iklj}$.

Transposing enables us to reorder the indices of a tensor in any desired order. Suppose that we wish to rewrite an expression $G_{iklj}$ as the component $G'_{ijkl}$ of another tensor whose indices are in *alphabetical* order: $G'$ is a transpose of $G$, the permutation being found by noting that the levels $(1,2,3,4)$ in the original tensor $G$, corresponding to the indices $(i,k,l,j)$, become respectively the levels $(1,3,4,2) = (\sigma_1, \sigma_2, \sigma_3, \sigma_4)$ in $G'$. This yields

$$G_{iklj} = (G^{T_{1342}})_{ijkl}. \tag{A.3}$$

Index reordering using transposition will be routinely used in combination with contractions to remove indices in tensor algebra, as in $G_{iklj} G'_{ijkl} = G^{T_{1324}} t :: G'$. The usual transpose $G^T$ of a *matrix* $G \in \mathbb{T}^{(n_1,n_2)}$ is a particular case of the generalized transpose, $G^T = G^{T_{21}}$.

Given a tensor $G(Y) \in \mathbb{T}^{(n_1,n_2,\ldots,n_p)}$ taking an argument $Y \in \mathbb{R}^s$, we denote as $\nabla G(Y) \in \mathbb{T}^{(n_1,n_2,\ldots,n_p,s)}$ its gradient,

$$\nabla G_{i_1 \ldots i_p j} = \frac{\partial G_{i_1 \ldots i_p}}{\partial Y_j}.$$

By a standard convention, the index $j$ corresponding to differentiation is therefore the *last one* even though the symbol $\nabla$ appears *first* in $\nabla G$.

## A.2. Infinitesimal rotations

In terms of the dimension $d$ of the Euclidean space, we define $n_r = \frac{d(d-1)}{2}$ as the dimension of infinitesimal rotations: $n_r = 1$ in dimension $d = 2$ and $n_r = 3$ in dimension $d = 3$. An infinitesimal rotation is represented by a vector $\boldsymbol{\theta} \in \mathbb{R}^{n_r}$, also known as a *pseudo-vector*. By convention, its indices are written as

$$\boldsymbol{\theta} = \begin{cases} (\theta_3) & (d=2) \\ (\theta_1, \theta_2, \theta_3) & (d=3), \end{cases} \tag{A.4}$$

i.e., the unique index $k$ of $\theta_k$ is taken to be $k = 3$ and not $k = 1$ when $d = 2$, as $\boldsymbol{\theta}$ then lives on the line perpendicular to the Euclidean plane $\mathbb{R}^2$.

The pseudo-vector $\boldsymbol{\theta}$ is uniquely associated with an antisymmetric matrix $\hat{\boldsymbol{\theta}} \in \mathbb{T}^{(d,d)}$ by

$$\hat{\boldsymbol{\theta}} = -\mathcal{N} \cdot \boldsymbol{\theta} \quad \Leftrightarrow \quad \boldsymbol{\theta} = -\frac{1}{2} \hat{\boldsymbol{\theta}} : \mathcal{N}, \tag{A.5}$$

where $\mathcal{N} \in \mathbb{T}^{(d,d,n_r)}$ is a constant rank-3 tensor that is is antisymmetric with respect to its first pair of indices,

$$\mathcal{N}_{ijk} = \begin{cases} \tilde{\mathcal{A}}_{ij3} & (d=2) \\ \tilde{\mathcal{A}}_{ijk} & (d=3). \end{cases} \tag{A.6}$$

Here $\tilde{\mathcal{A}}_{ijk}$ is the alternating symbol in dimension 3: $\tilde{\mathcal{A}}_{ijk} \in \{-1, 0, +1\}$ is 0 if at least two of the indices $(i,j,k)$ are identical, or the signature $\pm 1$ of the permutation if $(i,j,k)$ is a permutation of $(1,2,3)$. In dimension $d = 2$, the last index $k$ of $\mathcal{N}_{ijk}$ in (A.6) is restricted to $k = 3$ as in (A.4), which warrants that the contraction appearing in (A.5)$_1$, as well as the right-most indices on both sides of (A.5)$_2$, are consistent.

The equivalence in (A.5) is established in Appendix B.2.

When represented using an antisymmetric matrix $\hat{\boldsymbol{\theta}}$, the infinitesimal rotation maps a vector $\boldsymbol{x} \in \mathbb{R}^d$ to $\boldsymbol{x}' = (I_d + \hat{\boldsymbol{\theta}}) \cdot \boldsymbol{x}$. When represented as a pseudo-vector $\boldsymbol{\theta}$, it maps $\boldsymbol{x}$ to $\boldsymbol{x}' = \boldsymbol{x} + \theta_3 (\mathcal{A}^T \cdot \boldsymbol{x})$ when $d = 2$ (with $\mathcal{A}$ as the 2D Levi-Civita symbol, defined below Equation (7.11)), or $\boldsymbol{x}' = \boldsymbol{x} + \boldsymbol{\theta} \times \boldsymbol{x}$ when $d = 3$ (with $\times$ as the 3D cross-product). Equation (A.5) warrants that the result is the same.

## A.3. Toolbox for a strain-gradient elasticity

In Appendix B, we show that the higher-order gradients $\nabla^k \boldsymbol{u}$ and $\nabla^k \boldsymbol{\gamma}$ of the displacement and rotation (for $k \geq 1$) can be expressed in terms of $\boldsymbol{\gamma}$ and of the gradients $\nabla^j \check{\boldsymbol{\varepsilon}}$ (with $j \geq 0$) as

$$\begin{aligned}
\nabla \boldsymbol{\gamma}(\boldsymbol{X}) &= \mathcal{G}' : \nabla \check{\boldsymbol{\varepsilon}}(\boldsymbol{X}) \\
\nabla^2 \boldsymbol{\gamma}(\boldsymbol{X}) &= \mathcal{G}'' \mathbin{\vdots} \nabla^2 \check{\boldsymbol{\varepsilon}}(\boldsymbol{X}) \\
\nabla \boldsymbol{u}(\boldsymbol{X}) &= \mathcal{U}' \cdot \check{\boldsymbol{\varepsilon}}(\boldsymbol{X}) - \mathcal{N} \cdot \boldsymbol{\gamma}(\boldsymbol{X}) \\
\nabla^2 \boldsymbol{u}(\boldsymbol{X}) &= \mathcal{U}'' : \nabla \check{\boldsymbol{\varepsilon}}(\boldsymbol{X}) \\
\nabla^3 \boldsymbol{u}(\boldsymbol{X}) &= \mathcal{U}''' \mathbin{\vdots} \nabla^2 \check{\boldsymbol{\varepsilon}}(\boldsymbol{X}).
\end{aligned} \tag{A.7}$$

Note that no gradient of the rotation appears in the right-hand sides above. The constant tensors $\mathcal{G}' \in \mathbb{T}^{(n_r,d,n_\varepsilon,d)}$, $\mathcal{G}'' \in \mathbb{T}^{(n_r,d,d,n_\varepsilon,d,d)}$, $\mathcal{U}' \in \mathbb{T}^{(d,d,n_\varepsilon)}$, $\mathcal{U}'' \in \mathbb{T}^{(d,d,d,n_\varepsilon,d)}$ and $\mathcal{U}''' \in \mathbb{T}^{(d,d,d,d,n_\varepsilon,d,d)}$ are defined in Equations (B.9), (B.11), (B.13) and (B.15) in the Appendix. They are symmetric with respect to any pair of indices associated with second or third gradients, as specified in Appendix B. They depend only on the dimension $d$ of the Euclidean space.

Equation (A.7) is a reformulation of a result by [ME68] who derived expressions for $\nabla \boldsymbol{\gamma}$, $\nabla^2 \boldsymbol{u}$ and their successive gradients in terms of the strain $\check{\boldsymbol{\varepsilon}}$ and its gradients.

Equation (A.7) yields the Taylor expansion of the macroscopic fields $\boldsymbol{u}$ and $\boldsymbol{\gamma}$ about a point $\boldsymbol{X}_c$ as

$$\begin{aligned}
\boldsymbol{u}(\boldsymbol{X}) &= \boldsymbol{u}(\boldsymbol{X}_c) + [\mathcal{U}' \cdot \check{\boldsymbol{\varepsilon}}(\boldsymbol{X}_c) - \mathcal{N} \cdot \boldsymbol{\gamma}(\boldsymbol{X}_c)] \cdot (\boldsymbol{X} - \boldsymbol{X}_c) + [\mathcal{U}'' : \nabla \check{\boldsymbol{\varepsilon}}(\boldsymbol{X}_c)] : \frac{(\boldsymbol{X} - \boldsymbol{X}_c)^{\otimes 2}}{2} \\
&\qquad + [\mathcal{U}''' \mathbin{\vdots} \nabla^2 \check{\boldsymbol{\varepsilon}}(\boldsymbol{X}_c)] \mathbin{\vdots} \frac{(\boldsymbol{X} - \boldsymbol{X}_c)^{\otimes 3}}{6} + \cdots \\
\boldsymbol{\gamma}(\boldsymbol{X}) &= \boldsymbol{\gamma}(\boldsymbol{X}_c) + \quad [\mathcal{G}' : \nabla \check{\boldsymbol{\varepsilon}}(\boldsymbol{X})] \cdot (\boldsymbol{X} - \boldsymbol{X}_c) \quad + [\mathcal{G}'' \mathbin{\vdots} \nabla^2 \check{\boldsymbol{\varepsilon}}(\boldsymbol{X})] : \frac{(\boldsymbol{X} - \boldsymbol{X}_c)^{\otimes 2}}{2} + \cdots
\end{aligned} \tag{A.8}$$

where $\boldsymbol{x}^{\otimes 2} = \boldsymbol{x} \otimes \boldsymbol{x}$, $\boldsymbol{x}^{\otimes 3} = \boldsymbol{x} \otimes \boldsymbol{x} \otimes \boldsymbol{x}$. In Section 3, Equation (A.8) is used to write the strain energy of a particular element, which is expressed in terms of the degrees of freedom at the adjacent nodes, into a Taylor expansion at the center of the element.

## APPENDIX B. PROOF OF THE GEOMETRIC IDENTITIES

### B.1. Mandel representation $\check{\boldsymbol{\varepsilon}}$ of the macroscopic strain

We introduce the following constant tensor $\mathcal{M} \in \mathbb{T}^{(d,d,n_\varepsilon)}$ depending only on the dimension $d$ of the Euclidean space,

$$\mathcal{M} = \sum_{i=1}^{d} [\boldsymbol{i}_i^d \otimes \boldsymbol{i}_i^d] \otimes \boldsymbol{i}_i^{n_\varepsilon} + \sum_{1 \leqslant i < j \leqslant d} \left[ \frac{\boldsymbol{i}_i^d \otimes \boldsymbol{i}_j^d + \boldsymbol{i}_j^d \otimes \boldsymbol{i}_i^d}{\sqrt{2}} \right] \otimes \boldsymbol{i}_{i+(j-i)\left(d-\frac{j-i-1}{2}\right)}^{n_\varepsilon}, \tag{B.1}$$

where $\boldsymbol{i}_a^b \in \mathbb{R}^b$ is the discrete Dirac vector in dimension $b$, whose components are all zero except for the $a$-th one which is equal to 1,

$$(\boldsymbol{i}_a^b)_j = \begin{cases} 0 & \text{if } j \neq a \\ 1 & \text{if } j = a \end{cases}, \quad \text{with } 1 \leqslant j \leqslant b. \tag{B.2}$$

The rationale behind the definition of $\mathcal{M}$ is that the quantities in square brackets form an orthonormal basis of the space of symmetric $d \times d$ matrices, and that the indices $i$ (first term) and $i + (j-i)\left(d - \frac{j-i-1}{2}\right)$ (second term) provide a sequential numbering of the tensors forming this orthonormal basis, from 1 to $n_\varepsilon$. The orthonormal basis of symmetric $d \times d$ matrices can be therefore be extracted as $(\mathcal{M} \cdot \boldsymbol{i}_k^{n_\varepsilon})_{1 \leqslant k \leqslant n_\varepsilon}$.

The following two identities are a consequence of the orthonormal character of the basis appearing in square brackets in (B.1),

$$\mathcal{M}^{T_{231}} : \mathcal{M} = \boldsymbol{I}_{n_\varepsilon} \qquad \mathcal{M} \cdot \mathcal{M}^{T_{231}} = ((\boldsymbol{I}_d \otimes \boldsymbol{I}_d)^{T_{1324}})^{S_{12} \circ S_{34}}, \tag{B.3}$$

where we recall that $\circ$ stands for the composition of symmetrization operations, see (5.1–2.3).

The Mandel representation $\check{\boldsymbol{\varepsilon}} \in \mathbb{R}^{n_\varepsilon}$ of the strain tensor $\boldsymbol{\varepsilon}$ introduced in (3.7) is nothing but the decomposition of $\boldsymbol{\varepsilon}$ in this orthonormal basis,

$$\check{\boldsymbol{\varepsilon}} = \boldsymbol{\varepsilon} : \mathcal{M}. \tag{B.4}$$

Conversely, the symmetric strain tensor $\boldsymbol{\varepsilon}$ can be reconstructed from the vector $\check{\boldsymbol{\varepsilon}}$ by

$$\boldsymbol{\varepsilon} = \mathcal{M} \cdot \check{\boldsymbol{\varepsilon}}, \tag{B.5}$$

as can be shown using (B.3).

### B.2. Properties of the tensor $\mathcal{N}$

The tensor $\mathcal{N}$ introduced in (A.6) is similar to $\mathcal{M}$ as it delivers an orthogonal (but not orthonormal) basis of antisymmetric tensors in $\mathbb{T}^{(d,d)}$ indexed by $k$ in the form $(\mathcal{N} \cdot \boldsymbol{i}_k^{n_r})$ where

- in dimension $d = 2$, the only possible rotation index $k$ is limited to $k \in \{3\}$ and we set $\boldsymbol{i}_3^1 = (1)$,
- in dimension $d = 3$, $1 \leqslant k \leqslant 3$ and the usual definition of $\boldsymbol{i}_k^{n_r}$ in (B.2) applies.

In view of this and of Equation (A.5)$_1$, $\boldsymbol{\theta}$ is nothing but the components of $\hat{\boldsymbol{\theta}}$ in this basis, up to a sign. The reciprocal formula (A.5)$_2$ follows from the identity

$$\frac{1}{2} \mathcal{N}^{T_{231}} : \mathcal{N} = \boldsymbol{I}_{n_r}, \tag{B.6}$$

which expresses itself the orthogonal character of the underlying basis.

### B.3. Elimination of the rotation gradient

In this section, we provide a detailed justification of the identities (A.7), as well as expressions of the geometric tensors $\mathcal{G}'$, $\mathcal{G}''$, $\mathcal{U}'$, $\mathcal{U}''$, $\mathcal{U}'''$ appearing in the right-hand side. These tensors depend on the dimension $d$ of the Euclidean space only.

Combining the decomposition of $\nabla \boldsymbol{u} = \boldsymbol{\varepsilon} + \hat{\boldsymbol{\gamma}}$ into a symmetric and antisymmetric part with (B.5) and (A.5)$_1$, we have $\nabla \boldsymbol{u}(\boldsymbol{X}) = \mathcal{M} \cdot \check{\boldsymbol{\varepsilon}}(\boldsymbol{X}) - \mathcal{N} \cdot \boldsymbol{\gamma}(\boldsymbol{X})$ which is nothing but (A.7)$_3$ when we identify

$$\mathcal{U}' = \mathcal{M}. \tag{B.7}$$

Next, we repeat the classical argument yielding the rotation gradient $\nabla \boldsymbol{\gamma}$ in terms of the strain gradient $\nabla \check{\boldsymbol{\varepsilon}}$. The rotation in (3.8)$_2$ can be written in components as $\gamma_a = -\frac{1}{2} u_{c,d} \mathcal{N}_{cda} = +\frac{1}{2} \mathcal{N}_{dca} u_{c,d}$, where a comma in subscript denotes differentiation. The gradient of rotation thus takes the form $\gamma_{a,b} = \frac{1}{2} \mathcal{N}_{dca} u_{c,db}$. Since the second gradient $u_{c,db}$ is symmetric with respect to the indices $(b,d)$, this can be rewritten as $\gamma_{a,b} = \frac{1}{2} \mathcal{N}_{dca} (u_{c,b})_{,d} = \frac{1}{2} \mathcal{N}_{dca} (u_{c,b} + u_{b,c})_{,d} - \frac{1}{2} \mathcal{N}_{dca} u_{b,cd}$. The last term is zero since it is a contraction over the indices $(c,d)$ of an antisymmetric tensor $\mathcal{N}_{dca}$ and a symmetric tensor $u_{b,cd}$. This yields $\gamma_{a,b} = \mathcal{N}_{dca} \varepsilon_{cb,d} = \mathcal{N}_{dca} \mathcal{M}_{cbe} \check{\varepsilon}_{e,d}$ by (B.5), which we can rewrite in tensor form as

$$\nabla \boldsymbol{\gamma}(\boldsymbol{X}) = \mathcal{G}' : \nabla \check{\boldsymbol{\varepsilon}}(\boldsymbol{X}), \tag{B.8}$$

where
$$\mathscr{G}' = (\mathscr{N}^{T_{231}} \cdot \mathscr{M})^{T_{1423}}, \tag{B.9}$$
as announced in (A.7)$_1$.

Differentiating one more time, we get $\nabla^2 \boldsymbol{\gamma}(\boldsymbol{X}) = \mathscr{G}' : \nabla^2 \check{\boldsymbol{\varepsilon}}(\boldsymbol{X}) = (\mathscr{G}' \otimes \boldsymbol{I}_d)^{T_{124536}} \therefore \nabla^2 \check{\boldsymbol{\varepsilon}}(\boldsymbol{X})$. We enforce the natural symmetries of $\nabla^2 \boldsymbol{\gamma}$ with respect to the pairs of indices $(2,3)$ and $(5,6)$ by writing
$$\nabla^2 \boldsymbol{\gamma}(\boldsymbol{X}) = \mathscr{G}'' \therefore \nabla^2 \check{\boldsymbol{\varepsilon}}(\boldsymbol{X}), \tag{B.10}$$
as announced in (A.7)$_2$, where
$$\mathscr{G}'' = ((\mathscr{G}' \otimes \boldsymbol{I}_d)^{T_{124536}})^{S_{23} \circ S_{56}}, \tag{B.11}$$
and $S_{23} \circ S_{56}$ denotes successive symmetrization with the two pairs of indices.

Differentiating (A.7)$_3$ in components and using (B.8), we obtain the second gradient of displacement as $u_{a,bc} = (\mathscr{U}'_{abe} \check{\varepsilon}_e - \mathscr{N}_{abd} \gamma_d)_{,c} = \mathscr{U}'_{abe} \check{\varepsilon}_{e,c} - \mathscr{N}_{abd} \mathscr{G}'_{dcef} \check{\varepsilon}_{e,f} = (\mathscr{U}'_{abe} \delta_{cf} - \mathscr{N}_{abd} \mathscr{G}'_{dcef}) \check{\varepsilon}_{e,f}$, which we can rewrite as
$$\nabla^2 \boldsymbol{u}(\boldsymbol{X}) = \mathscr{U}'' : \nabla \check{\boldsymbol{\varepsilon}}(\boldsymbol{X}) \tag{B.12}$$
where $\mathscr{U}''$ is given by
$$\mathscr{U}'' = (\mathscr{U}' \otimes \boldsymbol{I}_d)^{T_{12435}} - \mathscr{N} \cdot \mathscr{G}'. \tag{B.13}$$
Since $\nabla^2 \boldsymbol{u}$ is symmetric with respect to its last pair of indices in the left-hand side of (B.12), $\mathscr{U}''$ is automatically symmetric with respect to the pair $(2,3)$. Equation (B.12) proves (A.7)$_4$.

Next, we have $\nabla^3 \boldsymbol{u} = \mathscr{U}'' : \nabla^2 \check{\boldsymbol{\varepsilon}} = (\mathscr{U}'' \otimes \boldsymbol{I}_d)^{T_{1235647}} \therefore \nabla^2 \check{\boldsymbol{\varepsilon}}$. The natural symmetries by any permutation of the indices $(2,3,4)$ on the one hand and $(6,7)$ on the other hand, are enforced by writing
$$\nabla^3 \boldsymbol{u}(\boldsymbol{X}) = \mathscr{U}''' \therefore \nabla^2 \check{\boldsymbol{\varepsilon}}(\boldsymbol{X}) \tag{B.14}$$
as announced in (A.7)$_5$, where
$$\mathscr{U}''' = ((\mathscr{U}'' \otimes \boldsymbol{I}_d)^{T_{1235647}})^{S_{234} \circ S_{67}}, \tag{B.15}$$
and $S_{234} \circ S_{67}$ denotes symmetrization with respect to the set of indices in subscript.

The proof of the identities (A.7) is complete.